\documentclass[a4paper,12pt]{article}
\usepackage{jheppub} 
\usepackage{tcilatex}
\usepackage{amsfonts}
\usepackage{amssymb}
\usepackage{multicol}
\usepackage{graphicx}
\usepackage{float}
\usepackage{caption}
\usepackage{xcolor}
\usepackage[utf8]{inputenc}
\usepackage{amsmath}
\title{Code CFTs and Topological Matter}
\author[1,2,3]{E.H Saidi }
\author[1,2]{, R. Sammani }

\affiliation[1]{ LPHE-MS, Science Faculty, Mohammed V University in Rabat, Morocco}
\affiliation[2]{Hassan II Academy of Science and Technology, Kingdom of Morocco}
\affiliation[3]{Centre of Physics and Mathematics, CPM- Morocco.}
\emailAdd{e.saidi@um5r.ac.ma}
\emailAdd{rajae\_sammani@um5.ac.ma}
\abstract{ In this paper, we propose a novel framework for modeling
topological phases of matter using code-based Narain conformal field
theories (NCFTs). We show that the algebraic structure of the NCFTs
naturally embeds into critical lattice quantum field theory, yielding
emergent topological features characteristic of gapless fermionic systems
with non-zero Chern numbers. We develop a new representation of
construction A of code CFT in terms of root and weight lattices of Lie
algebras, focusing in particular on SU(2) and SU(3). We then derive
the spectrum of particle states occupying the lattice and, with the help of
fermionisation techniques, demonstrate that it hosts fermionic
excitations with Dirac cones akin to tight binding systems namely the
Haldane model of quantum anomalous hall effect on a honeycomb geometry.}

\begin{document}
\notoc
\maketitle
\flushbottom
\newpage
\tableofcontents
\section{Introduction}

\label{sec:intro} Condensed matter systems, especially in their lattice QFT
formulations, have long provided a testing ground for concepts originating
in high energy physics theory \cite{R1, R2}. These models allow the
realizability of theoretical frameworks that are otherwise difficult or even
impossible to test. Beyond the experimental validation, they also offer
insights on testable predictions that may open pathways towards the
development of potential technologies \cite{R3}.\newline
In this work, we introduce a novel approach to modeling topological phases
of matter by proposing code-based conformal field theories (CFTs) \cite{1}-%
\cite{1hA} as a new class of realizations. Exploiting recent constructions
of Narain CFT as a coding theory \cite{1hA}-\cite{1cB} where one can
systematically build and analyze Narain CFTs using self-dual additive codes
defined over discrete abelian groups, we demonstrate that the associated
algebraic data can be leveraged to model topologically non trivial phases of
matter \cite{R4}-\cite{trs2}. We show that the framework of code based
Narain CFT is not only mathematically rich but also physically relevant with
emergent topological properties \cite{AZ}-\cite{DSL} characteristic of a
gapless fermionic system having a non zero Chern number \cite{R6, R7}. In
doing so, we provide a new framework for describing topological matter as a
code based CFT.%
\begin{equation*}
\begin{tabular}{ccccc}
\textbf{AB} \textbf{Chern-Simons\ } &  &  &  &  \\
${\Huge \downarrow }$ & $\quad {\Huge \Leftarrow }\quad $ & \ \textbf{Code
CFT} &  &  \\
\ \textbf{Narain CFT \ } &  &  &  &  \\ \cline{3-5}
${\Huge \updownarrow }$ &  & \multicolumn{1}{|c}{\ \ \textbf{fermionic CFT}}
& $\mathbf{\rightarrow }$ & \multicolumn{1}{c|}{\textbf{Dirac cones \ \ }}
\\
\ \textbf{bosonic CFT \ } & $\quad {\Huge \Leftrightarrow }{\small \quad }$
& \multicolumn{1}{|c}{$\mathbf{\uparrow }$} & ${\Huge \circlearrowright }$ &
\multicolumn{1}{c|}{$\mathbf{\downarrow }$} \\
&  & \multicolumn{1}{|c}{\ \ \textbf{gapless states}} & ${\small \leftarrow }
$ & \multicolumn{1}{c|}{\textbf{lattice model \ \ }} \\ \cline{3-5}
\end{tabular}%
\end{equation*}%
\begin{equation*}
\end{equation*}

For every even self dual code $\mathcal{C}$, one can construct a triplet of
lattices $\mathbf{\Lambda }_{\mathrm{k}}^{\mathrm{r},\mathrm{r}}\subset
\mathbf{\Lambda }_{\mathrm{k}\mathcal{C}}^{\mathrm{r},\mathrm{r}}\subset (%
\mathbf{\Lambda }_{\mathrm{k}}^{\mathrm{r},\mathrm{r}})^{\ast }$\ realizing
the Narain CFT with U(1)$^{\mathrm{c}_{\text{\textsc{l}}}}\times $U(1)$^{%
\mathrm{c}_{\text{\textsc{r}}}}$ where $\mathrm{c}_{\text{\textsc{l}}}=%
\mathrm{c}_{\text{\textsc{r}}}=\mathrm{r}$ \textrm{\cite{1}-\cite{1kA}} that
admits a holographic dual given by the AB Chern-Simons (CS) theory with
coupling constant $\mathrm{k}\in \mathbb{N}^{\ast }$ (CS level) \textrm{\cite%
{1B}; }report to \emph{appendix A} for a comprehensive review. These
constructions allow to perform explicit calculations of generalised Narain
partition functions using code based algorithms \textrm{\cite{1cB}}. The
structure of the even self dual codes $\mathcal{C}$ enables to build
discrete sets $\mathbf{\Lambda }_{\mathrm{k}\mathcal{C}}^{\mathrm{r},\mathrm{%
r}}$ sitting in $\mathbb{R}^{\mathrm{r},\mathrm{r}}$ which in turn\textrm{\ }%
realise 2r dimensional Narain lattices $\mathbf{\Lambda }_{\text{\textsc{%
narain}}}$. In this\textrm{\ }setting, the discrete group $G_{\mathrm{k}}^{%
\mathrm{r},\mathrm{r}}$ is identified with the discriminant $\mathbf{\Lambda
}_{\mathrm{k}}^{\mathrm{r},\mathrm{r}\ast }/\mathbf{\Lambda }_{\mathrm{k}}^{%
\mathrm{r},\mathrm{r}}$ taken in this study to be $\mathbb{Z}_{\mathrm{k}}^{%
\mathrm{r}}\times \mathbb{Z}_{\mathrm{k}}^{\mathrm{r}}$ and can be
interpreted in terms of the symmetries of unit cells of the real lattice $%
\mathbf{\Lambda }_{\mathrm{k}}^{\mathrm{r},\mathrm{r}},$ its dual $\mathbf{%
\Lambda }_{\mathrm{k}}^{\ast \mathrm{r},\mathrm{r}}$ and the even self dual $%
\mathbf{\Lambda }_{\mathrm{k}\mathcal{C}}^{\mathrm{r},\mathrm{r}}.$ This
construction, referred to as Construction A \cite{1jB}, permits to generate
KK and winding states, related via duality, and populating a lattice (a
honeycomb for SU(3) with k=2) that hosts fermionic hoppings.\textrm{\ }The
resulting lattice model is shown to mimic that of Haldane theory with Dirac
cones.\newline
Our investigation unfolds through the following steps:
\vspace*{-3mm} 
\begin{description}
\item[$(\mathbf{a})$] \textbf{the SU(2) construction}: We begin by showing
that construction A of the $\mathrm{c}_{\text{\textsc{l/r}}}=\mathrm{1}$
Narain theory admits a natural interpretation in terms of the root \textsc{R}%
$^{\mathbf{su}_{2}}:=\left. \text{\textsc{R}}_{\mathrm{k}}^{\mathbf{su}%
_{2}}\right\vert _{\mathrm{k=2}}$ and weight \textsc{W}$^{\mathbf{su}%
_{2}}=\left. \text{\textsc{W}}_{\mathrm{k}}^{\mathbf{su}_{2}}\right\vert _{%
\mathrm{k=2}}$ lattices of SU(2) and their fibrations as given by eq(\ref%
{lrw}) which is valid for generic values of k, including the special level
k=2. Recall that the group SU(2) has rank r=1, and for k=2, the associated
discriminant \textsc{W}$^{\mathbf{su}_{2}}/$\textsc{R}$^{\mathbf{su}_{2}}$
is isomorphic to $\mathbb{Z}_{2}$. This implies that the unit cell of
\textsc{W}$^{\mathbf{su}_{2}}$ contains two different types of sites
(represented by a two-colored grid), whereas \textsc{R}$^{\mathbf{su}_{2}}$
has only one type of sites. This distinction is depicted in Figure \textbf{%
\ref{i1}}. In this illustrative image, we graphically represent the lattice
sequence $\mathbf{\Lambda }_{\mathrm{2}}^{\mathrm{1},\mathrm{1}}\subset $ $%
\mathbf{\Lambda }_{\mathrm{2}\mathcal{C}}^{\mathrm{1},\mathrm{1}}\subset
\mathbf{\Lambda }_{\mathrm{2}}^{\ast \mathrm{1},\mathrm{1}}$ using different
color codings\textrm{: }$\left( i\right) $ four colors for the largest
lattice $\mathbf{\Lambda }_{\mathrm{2}}^{\ast \mathrm{1},\mathrm{1}}=\text{%
\textsc{W}}_{\mathrm{2}}^{\mathbf{su}_{2}}\times \text{\textsc{W}}_{\mathrm{2%
}}^{\mathbf{su}_{2}},$ $\left( ii\right) $ two colors for the intermediate $%
\mathbf{\Lambda }_{\mathrm{2}\mathcal{C}}^{\mathrm{1},\mathrm{1}}=$\textsc{R}%
$_{\mathrm{2}}^{\mathbf{su}_{2}}\times $\textsc{W}$_{\mathrm{2}}^{\mathbf{su}%
_{2}}$; and $\left( iii\right) $ one color for the smallest one $\mathbf{%
\Lambda }_{\mathrm{2}}^{\mathrm{1},\mathrm{1}}=$\textsc{R}$_{\mathrm{2}}^{%
\mathbf{su}_{2}}\times $\textsc{W}$_{\mathrm{2}}^{\mathbf{su}_{2}}.$
\begin{figure}[tbph]
\begin{center}
\includegraphics[width=14cm]{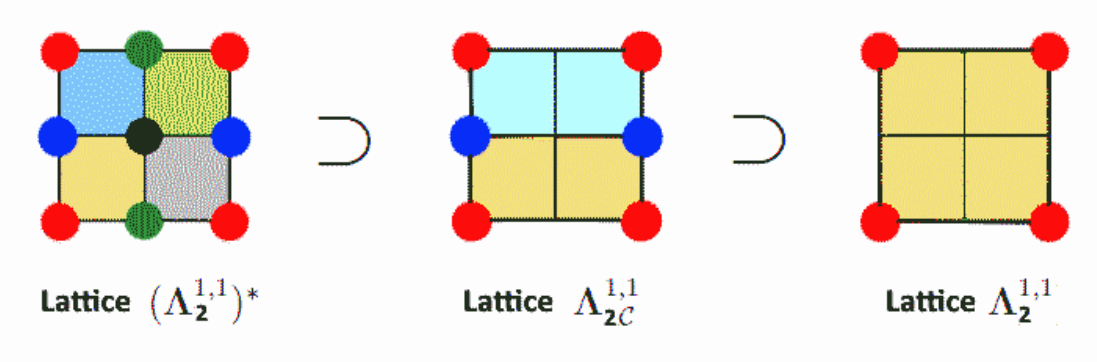}
\end{center}
\par
\vspace{-0.5cm}
\caption{The unit cells of the 2D lattices of construction A of the $%
c_{L/R}=1$ theory. The lattice sites are color coded with the number of
colors reflecting the density of the lattice: k$^{2}$ colors for $\mathbf{%
\Lambda }_{\mathrm{k}}^{\mathrm{1},\mathrm{1}\ast }$, k colors for $\mathbf{%
\Lambda }_{\mathrm{k}\mathcal{C}}^{\mathrm{1},\mathrm{1}}$ and one color for
$\mathbf{\Lambda }_{\mathrm{k}}^{\mathrm{1},\mathrm{1}}.$ The areas of the
unit cells are exhibited in increasing order. On left, $\mathbf{\Lambda }_{%
\mathrm{2}}^{\mathrm{1},\mathrm{1}\ast }$ with 4 unit cells. In middle, the $%
\mathbf{\Lambda }_{\mathrm{2}\mathcal{C}}^{\mathrm{1},\mathrm{1}}$ with 2
unit cells. On right, the unit cell of $\mathbf{\Lambda }_{\mathrm{2}}^{%
\mathrm{1},\mathrm{1}}$. }
\end{figure}
For generic values of the CS level k, the graphical representation of the
triplet $\mathbf{\Lambda }_{\mathrm{k}}^{\mathrm{1},\mathrm{1}}\subset
\mathbf{\Lambda }_{\mathrm{k}\mathcal{C}}^{\mathrm{1},\mathrm{1}}\subset
\mathbf{\Lambda }_{\mathrm{k}}^{\mathrm{1},\mathrm{1}\ast }$ in $\mathbb{R}^{%
\mathrm{1},\mathrm{1}}$ requires \textrm{k}$^{2}$ colored sites. They can be
remarkably realized via the fibrations
\begin{equation}
\begin{tabular}{lll}
$\mathbf{\Lambda }_{\mathrm{k}}^{\mathrm{1},\mathrm{1}}$ & $=$ & $\text{%
\textsc{R}}_{\mathrm{k}}^{\mathbf{su}_{2}}\times \text{\textsc{R}}_{\mathrm{k%
}}^{\mathbf{su}_{2}}$ \\
$\mathbf{\Lambda }_{\mathrm{k}\mathcal{C}}^{\mathrm{1},\mathrm{1}}$ & $=$ & $%
\text{\textsc{R}}_{\mathrm{k}}^{\mathbf{su}_{2}}\times $\textsc{W}$_{\mathrm{%
k}}^{\mathbf{su}_{2}}$ \\
$\mathbf{\Lambda }_{\mathrm{k}}^{\ast \mathrm{1},\mathrm{1}}$ & $=$ &
\textsc{W}$_{\mathrm{k}}^{\mathbf{su}_{2}}\times $\textsc{W}$_{\mathrm{k}}^{%
\mathbf{su}_{2}}$%
\end{tabular}%
\qquad \Leftrightarrow \qquad
\begin{tabular}{lll}
$\mathbf{\Lambda }_{\mathrm{k}}^{\mathbf{su}_{2}}$ & $=$ & $\text{\textsc{R}}%
_{\mathrm{k}}^{\mathbf{su}_{2}}\times \text{\textsc{R}}_{\mathrm{k}}^{%
\mathbf{su}_{2}}$ \\
$\mathbf{\Lambda }_{\mathrm{k}\mathcal{C}}^{\mathbf{su}_{2}}$ & $=$ & $\text{%
\textsc{R}}_{\mathrm{k}}^{\mathbf{su}_{2}}\times $\textsc{W}$_{\mathrm{k}}^{%
\mathbf{su}_{2}}$ \\
$\mathbf{\Lambda }_{\mathrm{k}}^{\ast \mathbf{su}_{2}}$ & $=$ & \textsc{W}$_{%
\mathrm{k}}^{\mathbf{su}_{2}}\times $\textsc{W}$_{\mathrm{k}}^{\mathbf{su}%
_{2}}$%
\end{tabular}%
\end{equation}%
In these expressions, the root lattice\textrm{\ }\textsc{R}$_{\mathrm{k}}^{%
\mathbf{su}_{2}}$ is isomorphic to the 1D lattice $\sqrt{\mathrm{k}}\mathbb{Z%
}$ while the weight lattice \textsc{W}$_{\mathrm{k}}^{\mathbf{su}_{2}}$ is
its dual $\frac{1}{\sqrt{\mathrm{k}}}\mathbb{Z}$. The coset structure
\textsc{W}$_{\mathrm{k}}^{\mathbf{su}_{2}}/$\textsc{R}$_{\mathrm{k}}^{%
\mathbf{su}_{2}}\simeq \mathbb{Z}_{\mathrm{k}}$ shows that \textsc{W}$_{%
\mathrm{k}}^{\mathbf{su}_{2}}$ can be interpreted as a union of k\ (shifted)
sublattices of \textsc{R}$_{\mathrm{k}}^{\mathbf{su}_{2}}$ distant by some
weight vectors as explicitly illustrated in sections 3 and 4. The sites of
these lattices $\mathbf{\Lambda }$ describe quantum particle states $%
\left\vert \psi _{a,n}^{b,m}\right\rangle $ labeled by four indices: ground
configuration indices (a,b), Kaluza $n$ and windings $m$ modes.

\item[$(\mathbf{b})$] \textbf{The SU(3) extension}: We construct novel
realisations of construction A of code CFT, extending beyond the SU(2) based
configurations \cite{HS1}-\cite{HS4} discussed previously. In this
extension, we\ build upon the SU(2) results to generalise the lattice
realisation (\ref{lrw}) to higher dimensional Lie algebras especially those
of rank 2 such as SU(3), SO(5) and G2. Focussing on the interesting case of
SU(3),\textrm{\ }the relevant 4D lattices are embedded in $\mathbb{R}^{%
\mathrm{2},\mathrm{2}}$ with hierarchy $\mathbf{\Lambda }_{\mathrm{k}}^{%
\mathbf{su}_{3}}\subset \mathbf{\Lambda }_{\mathrm{k}\mathcal{C}}^{\mathbf{su%
}_{3}}\subset \mathbf{\Lambda }_{\mathrm{k}}^{\ast \mathbf{su}_{3}}$ and
representation:%
\begin{equation}
\begin{tabular}{lll}
$\mathbf{\Lambda }_{\mathrm{k}}^{\mathbf{su}_{3}}$ & $=$ & $\text{\textsc{R}}%
_{\mathrm{k}}^{\mathbf{su}_{3}}\times \text{\textsc{R}}_{\mathrm{k}}^{%
\mathbf{su}_{3}}$ \\
$\mathbf{\Lambda }_{\mathrm{k}\mathcal{C}}^{\mathbf{su}_{3}}$ & $=$ & $\text{%
\textsc{R}}_{\mathrm{k}}^{\mathbf{su}_{3}}\times $\textsc{W}$_{\mathrm{k}}^{%
\mathbf{su}_{3}}$ \\
$\mathbf{\Lambda }_{\mathrm{k}}^{\ast \mathbf{su}_{3}}$ & $=$ & \textsc{W}$_{%
\mathrm{k}}^{\mathbf{su}_{3}}\times $\textsc{W}$_{\mathrm{k}}^{\mathbf{su}%
_{3}}$%
\end{tabular}%
\end{equation}%
We distinguish three cases according to the value of the Chern-Simons level
\textrm{k} of the holographic dual of the Narain CFT: $\left( \mathbf{i}%
\right) $ \textbf{Case k=3}: A this level, the discriminant group is $%
\mathbb{Z}_{\mathrm{3}}\times \mathbb{Z}_{\mathrm{3}}$ with $\mathbb{Z}_{%
\mathrm{3}}$ being the group centre of SU(3). The 4D lattice $\mathbf{%
\Lambda }_{\mathrm{3}}^{\mathbf{su}_{3}}=$\textsc{R}$_{\mathrm{3}}^{\mathbf{%
su}_{3}}\times $\textsc{R}$_{\mathrm{3}}^{\mathbf{su}_{3}}$ is made out of
the fibration of two hexagonal lattices while its dual $\mathbf{\Lambda }_{%
\mathrm{3}}^{\ast \mathbf{su}_{3}}=$\textsc{W}$_{\mathrm{3}}^{\mathbf{su}%
_{3}}\times $\textsc{W}$_{\mathrm{3}}^{\mathbf{su}_{3}}$ is constructed%
\textrm{\ }from the fibration of two triangular lattices. $\left( \mathbf{ii}%
\right) $ \textbf{Case k\TEXTsymbol{>}3}: the results obtained for k=3
extends naturally to higher values k\TEXTsymbol{>}3 for which the
discriminant satisfies \textsc{W}$_{\mathrm{k}}^{\mathbf{su}_{3}}/$\textsc{R}%
$_{\mathrm{k}}^{\mathbf{su}_{3}}\simeq \mathbb{Z}_{\mathrm{k}}.$ $\left(
\mathbf{iii}\right) $ \textbf{Case k\TEXTsymbol{<}3}: for low levels $%
\mathrm{k}=1,2$, the situation is subtler. While the CS level k=1 is somehow
"exotic" because the condition \textsc{R}$_{\mathrm{k}}^{\mathbf{su}%
_{3}}\subseteq $\textsc{W}$_{\mathrm{k}}^{\mathbf{su}_{3}}$\textrm{, }as we
will discuss, is violated. This bound holds only for k $\geq \sqrt{3}$
therefore ruling out k=1 but allowing k=2 for which the weight lattice
\textsc{W}$_{\mathrm{2}}^{\mathbf{su}_{3}}$ is given by the 2D honeycomb
while the \textsc{R}$_{\mathrm{2}}^{\mathbf{su}_{3}}$ is an hexagonal
lattice.

\item[$(\mathbf{c})$] \textbf{Emergence of topological properties}: After
establishing a new formulation of the framework in terms of algebraic data,
we exploit the new setting to develop a path towards topological phases of
matter by linking Narain CFT to critical lattice QFT. We start by modeling
the particles sitting at the intersection of the lattice $\Lambda _{\mathcal{%
C}}$ grid in terms of KK $\left\vert KK_{\mathbf{m}}\right\rangle $ and
winding states $\left\vert W_{\mathbf{m}}\right\rangle $ akin to
tight-binding systems in condensed matter physics. Our primary focus remains
on SU(2) based code CFTs with central charge giving quantized left/right
momenta for particle states, organizing into SU(2) representations, and
labeled by ground state and excitation quantum numbers. This reveals a
Hilbert space containing k$^{2}$ vacuum states and associated excitations, a
decoupling of energy and momentum sectors, in addition to a globally defined
topological index depending on k.\newline
Proceeding, we analyze the particle content in the SU(3) Chern-Simons (CS)
theory at level k=2. We show that the corresponding weight lattice
decomposes into two sublattices forming a honeycomb structure,
interconnected via a Z$_{\mathrm{2}}$ symmetry. This later also relates the
KK modes residing on one sublattice to the winding modes populating the
other. Using fermionisation techniques \textrm{\cite{ferm1,ferm2,T}}, we
introduce Dirac fermions as propagations on the honeycomb lattice with a
tight binding Hamiltonian that breaks inversion symmetry. We prove that our
lattice model mimics that of Haldane theory \cite{tm,tm1} with Dirac cones
therefore demonstrating that NCFTs can be realised as gapless fermionic
systems having non zero Chern numbers. This in turn closes the loop and
establishes a direct link between code CFT and topological matter.
\end{description}

The organisation is as follows: In \autoref{sec:2}, we present construction
A in code-CFTs realising Narain conformal theories and show how the three
lattices $\left( \mathbf{\Lambda }_{\mathrm{k}},\mathbf{\Lambda }_{\mathrm{k}%
\mathcal{C}},\mathbf{\Lambda }_{\mathrm{k}}^{\ast }\right) $ can be linked
to the root \textsc{R}$^{\mathbf{su}_{2}}$ and the weight \textsc{W}$^{%
\mathbf{su}_{2}}$ lattices of SU(2) and their cross products. New results
for this SU(2) based building are obtained. In \autoref{sec:3}, we extend
the construction to SU(3) and derive realisations of the triplet $(\mathbf{%
\Lambda }_{\mathrm{k}}^{\mathbf{su}_{3}},\mathbf{\Lambda }_{\mathrm{k}%
\mathcal{C}}^{\mathbf{su}_{3}},\mathbf{\Lambda }_{\mathrm{k}}^{\mathbf{su}%
_{3}\ast })$ while distinguishing three intervals k=3, k\TEXTsymbol{>}3 and k%
\TEXTsymbol{<}3. In \autoref{sec:4}, we give applications aiming to
establish a link between the Narain CFTs and lattice quantum matter at Dirac
points. Here, we investigate the properties of particle states in the SU(2)
and the SU(3) based CFTs while in \autoref{sec:5} we draw the way to Chern
matter by giving explicit examples. \autoref{sec:6} is devoted to conclusion
and discussions. \autoref{sec:A}, \autoref{sec:B} and \autoref{sec:C}
contain technical details of our calculations and supplementary material
supporting the main text.

\section{The triplet ($\mathbf{\Lambda }_{\mathrm{k}},\mathbf{\Lambda }_{%
\mathrm{k}\mathcal{C}},\mathbf{\Lambda }_{\mathrm{k}}^{\ast }$) from SU(2)
lattices}

\label{sec:2} In this section, we demonstrate how the three lattices $%
\mathbf{\Lambda },$ $\mathbf{\Lambda }^{\ast }$ and $\mathbf{\Lambda }_{%
\mathcal{C}}$ used in the construction A, \emph{reviewed in appendix B}, can
be linked to the root \textsc{R}$^{\mathbf{su}_{2}}$, the weight \textsc{W}$%
^{\mathbf{su}_{2}}$ lattices of SU(2) and their cross products. This
identification will lay the groundwork foundation for the next section where
we extend the study to a broader class of lattices associated with general
Lie algebras $\mathbf{g}$ going beyond SU(2). This will lead to novel
realisations of the triplet $\left( \mathbf{\Lambda }^{\mathbf{g}},\mathbf{%
\Lambda }_{\mathcal{C}}^{\mathbf{g}},\mathbf{\Lambda }^{\ast \mathbf{g}%
}\right) $ in higher dimensions.

\subsection{Narain CFT with $c_{L/R}=1$}

To start,\textrm{\ }we focus on the case \textrm{r=1} (i.e. central charge $%
c_{\text{\textsc{l}/\textsc{r}}}=1$)\textrm{\ }to clearly present the core
idea of the construction. The extension to r\TEXTsymbol{>}1 follows directly
from (\ref{csa})\textrm{\ }and\textrm{\ }will be briefly commented later. To
that purpose, we define\textrm{\footnote{%
\ \ In general, one may take \textsc{R}$_{\mathrm{k}}=\sqrt{\mathrm{k}}(%
\mathtt{\text{\b{a}}}\mathbb{Z})$ and \textsc{W}$_{\mathrm{k}}=\frac{1}{%
\sqrt{\mathrm{k}}}(\mathbb{Z}/\mathtt{\text{\b{a}}})$ with respective
parameters $\mathtt{\text{\b{a} and 1/\b{a}. }}$}} \textsc{R}$_{\mathrm{k}}=%
\sqrt{\mathrm{k}}\mathbb{Z}$ and \textsc{W}$_{\mathrm{k}}=\frac{1}{\sqrt{%
\mathrm{k}}}\mathbb{Z}$ with integer k\TEXTsymbol{>}1; they naturally
satisfy the inclusion property \textsc{R}$_{\mathrm{k}}\subset $\textsc{W}$_{%
\mathrm{k}}$ and have a discriminant group given by \textsc{W}$_{\mathrm{k}%
}/ $\textsc{R}$_{\mathrm{k}}$, which upon substitution becomes $\frac{1}{%
\sqrt{\mathrm{k}}}\mathbb{Z}/(\sqrt{\mathrm{k}}\mathbb{Z})$ isomorphic to
the abelian group $\mathbb{Z}/(\mathrm{k}\mathbb{Z})\simeq \mathbb{Z}_{%
\mathrm{k}}$ giving therefore
\begin{equation}
\text{\textsc{W}}_{\mathrm{k}}/\text{\textsc{R}}_{\mathrm{k}}\simeq \mathbb{Z%
}_{\mathrm{k}}
\end{equation}%
This coset space will later be given a compelling geometric interpretation.
By viewing \textsc{W}$_{\mathrm{k}}$ \textrm{as }the product $\mathbb{Z}_{%
\mathrm{k}}\times $\textsc{R}$_{\mathrm{k}}$, one may think of it as a
superposition of k sheets of 1D sublattices, each isomorphic to \textsc{R}$_{%
\mathrm{k}}$. Formally, we have%
\begin{equation}
\text{\textsc{W}}_{\mathrm{k}}\simeq \underbrace{\text{\textsc{R}}_{\mathrm{k%
}}\text{ }\cup \text{ \textsc{R}}_{\mathrm{k}}^{[\mathbf{\lambda }]}\text{ }%
\cup \text{ }...\text{ }\cup \text{ \textsc{R}}_{\mathrm{k}}^{[(\mathrm{k-1)}%
\mathbf{\lambda }]}}_{\mathrm{k}\text{ components}}\text{ }\subset \text{ }%
\mathbb{R}
\end{equation}%
where we have set \textsc{R}$_{\mathrm{k}}^{[\varepsilon \mathbf{\lambda }%
]}= $\textsc{R}$_{\mathrm{k}}+\varepsilon \mathbf{\lambda }$ and \textsc{R}$%
_{\mathrm{k}}^{[0]}=$\textsc{R}$_{\mathrm{k}}$. These sheets are related to
one another by the symmetry $\mathbb{Z}_{\mathrm{k}}$, generated by the
algebraic k-cycle (1,2,...,k) which can be represented by the matrix%
\begin{equation}
T=\left(
\begin{array}{cccc}
0 & 1 & \cdots & 0 \\
\vdots & \ddots & \ddots & \vdots \\
0 & 0 & 0 & 1 \\
1 & 0 & 0 & 0%
\end{array}%
\right) _{\mathrm{k\times k}}\qquad ,\qquad T^{\mathrm{k}}=I_{id}
\end{equation}%
For the special case $\mathrm{k=1}$, we have \textsc{R}$_{\mathrm{1}}=%
\mathbb{Z}$ and \textsc{W}$_{\mathrm{1}}=\mathbb{Z}$ so that \textsc{R}$_{%
\mathrm{1}}\simeq $\textsc{W}$_{\mathrm{1}}$; their discriminant $\mathbb{Z}%
_{\mathrm{1}}$ is therefore trivial and given by the identity element $%
I_{id}.$ From the perspective of code CFT, this case $\mathrm{k=1}$
corresponds to the trivial code \cite{1A, 1bB}
\begin{equation}
\mathcal{C}=\left\{ \text{\textsc{c}}=(0,0)\right\} \subseteq \mathbb{Z}_{%
\mathrm{k}}
\end{equation}%
Recall that a generic codeword \textsc{c}, defined by the pair $(a,b)$ with
components taking integer values $a,b=0,...,\mathrm{k}-1$, belongs to an
even self dual code $\mathcal{C}$. In terms of the sets \textsc{R}$_{\mathrm{%
k}}$ and \textsc{W}$_{\mathrm{k}}$, the lattice $\mathbf{\Lambda }_{\mathrm{k%
}}$ is then given by the cross product \textsc{R}$_{\mathrm{k}}\times $%
\textsc{R}$_{\mathrm{k}}$\textrm{\ }while its dual $\mathbf{\Lambda }_{%
\mathrm{k}}^{\ast }$ \textrm{is} \textsc{W}$_{\mathrm{k}}\times $\textsc{W}$%
_{\mathrm{k}}$ thus satisfying the inclusion $\mathbf{\Lambda }_{\mathrm{k}%
}\subset \mathbf{\Lambda }_{\mathrm{k}}^{\ast }$ due to \textsc{R}$_{\mathrm{%
k}}\subset $\textsc{W}$_{\mathrm{k}}.$ The motivation for using \textsc{R}$_{%
\mathrm{k}}$ and \textsc{W}$_{\mathrm{k}}$\ stems from their interesting
interpretation as the root \textsc{R}$_{\mathrm{k}}$ and weight \textsc{W}$_{%
\mathrm{k}}$ lattices of SU(2) which naturally leads to the\textrm{\ }2D
lattices $\mathbf{\Lambda }_{\mathrm{k}}$ and $\mathbf{\Lambda }_{\mathrm{k}%
}^{\ast }$ imagined as follows
\begin{equation}
\begin{tabular}{|c||c|c|}
\hline
Lattice & \qquad SU(2)\qquad\ \  & \qquad SU(2)$\times $SU(2)\qquad \\
\hline\hline
root & \textsc{R}$_{\mathrm{k}}$ & $\mathbf{\Lambda }_{\mathrm{k}}=$\textsc{R%
}$_{\mathrm{k}}\times $\textsc{R}$_{\mathrm{k}}$ \\ \hline
weight & \textsc{W}$_{\mathrm{k}}$ & $\mathbf{\Lambda }_{\mathrm{k}}^{\ast
}= $\textsc{W}$_{\mathrm{k}}\times $\textsc{W}$_{\mathrm{k}}$ \\ \hline\hline
\end{tabular}%
\end{equation}%
To illustrate this structure, we first consider the particular case k=2 for
which \textsc{R}$_{\mathrm{2}}=\sqrt{\mathrm{2}}\mathbb{Z}$ and \textsc{W}$_{%
\mathrm{2}}=\frac{1}{\sqrt{\mathrm{2}}}\mathbb{Z}$. Then, we proceed to the
generic case $\mathrm{k}>2.$

\subsubsection{Models with CS level\emph{\ }k=2}

\begin{figure}[tbph]
\begin{center}
\includegraphics[width=16cm]{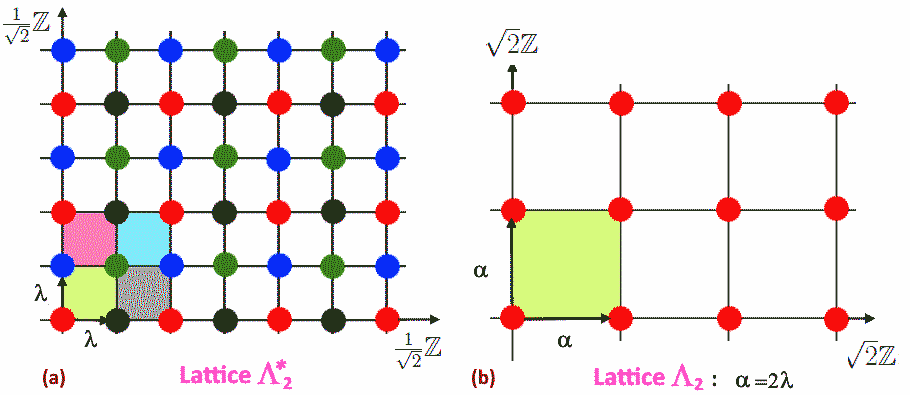}
\end{center}
\par
\vspace{-0.5cm}
\caption{Graphic descriptions of the lattice $\mathbf{\Lambda }_{\mathrm{2}}$
and its dual $\mathbf{\Lambda }_{\mathrm{2}}^{\ast }$. On right, the sites
in $\mathbf{\Lambda }_{\mathrm{2}}$ are painted in red with the unit cell
highlighted in pistachio. In the left, the sites of $\mathbf{\Lambda }_{%
\mathrm{2}}^{\ast }$ are color coded in black, green, blue and red to
illustrate the embedding property $\mathbf{\Lambda _{\mathrm{2}}\subset
\Lambda }_{\mathrm{2}\mathcal{C}}\subset \mathbf{\Lambda }_{\mathrm{2}%
}^{\ast }$. The graphics of $\mathbf{\Lambda }_{\mathrm{2}\mathcal{C}}$\ is
given by the Figure \textbf{\protect\ref{j2}}.}
\label{j1}
\end{figure}

\begin{figure}[tbph]
\begin{center}
\includegraphics[width=10cm]{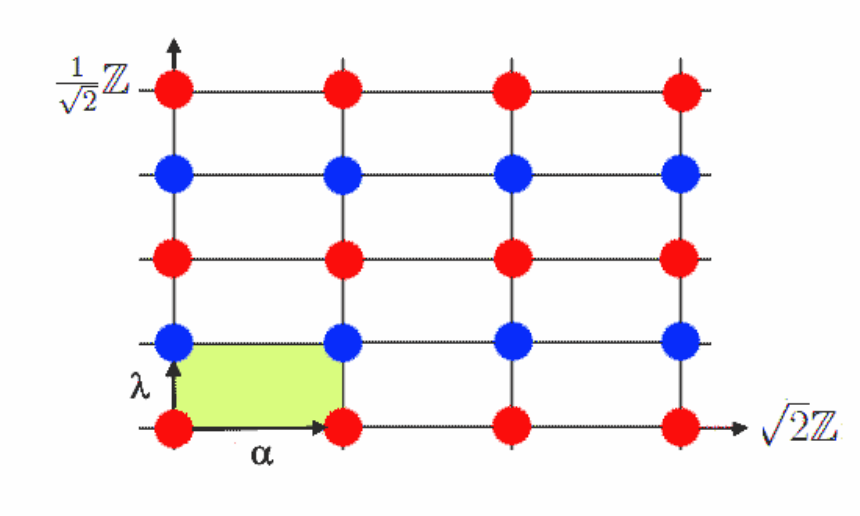}
\end{center}
\par
\vspace{-0.5cm}
\caption{The structure of the lattice $\mathbf{\Lambda }_{\mathrm{2}\mathcal{%
C}}$ with sites shown in two different colors. The even self dual lattice
contains the $\mathbf{\Lambda }_{\mathrm{2}}$ made by red sites. Sites in
blue belong to $\mathbf{\Lambda }_{\mathrm{2}\mathcal{C}}/\mathbf{\Lambda }_{%
\mathrm{2}}\simeq \mathbf{\Lambda }_{\mathrm{2}}+\mathbf{\protect\lambda }$.
Physically, red sites will be thought of as filled by KK particle states and
the blue ones by winding modes; see section 4 for details.}
\label{j2}
\end{figure}

For the particular k=2 case, the discrete spaces $\mathbf{\Lambda }_{\mathrm{%
2}}$, $\mathbf{\Lambda }_{\mathrm{2}\mathcal{C}}$\ and $\mathbf{\Lambda }_{%
\mathrm{2}}^{\ast }$ constructed out of \textsc{R}$_{\mathrm{2}}$ and
\textsc{W}$_{\mathrm{2}}$\ are two dimensional lattices embedded in $\mathbb{%
R}^{1,1}.$ To set the stage, we provide a preview of their graphic
description in the Figures \textbf{\ref{j1}-\ref{j2}}. The $\mathbf{\Lambda }%
_{\mathrm{2}}$ and $\mathbf{\Lambda }_{\mathrm{2}}^{\ast }$ are\textbf{\ }%
given by square lattices characterised, respectively, by parameters $\mathbf{%
\alpha }$ and $\mathbf{\lambda }$ which will be determined later.
In the two\ illustrations of Figure \textbf{\ref{j1}}, we employ distinct
colors to designate the sites of $\mathbf{\Lambda }_{\mathrm{2}}$ and its
dual $\mathbf{\Lambda }_{\mathrm{2}}^{\ast }$. For the real lattice $\mathbf{%
\Lambda }_{\mathrm{2}},$ the sites are painted in red; its unit cell, shaded
in pistachio, has an area equal to 2 (i.e $uc\mathbf{\Lambda }_{\mathrm{2}%
}=2 $). As for the dual lattice $\mathbf{\Lambda }_{\mathrm{2}}^{\ast },$ we
use four colors (black, green, blue and red) to explicitly display the
embedding hierarchy
\begin{equation}
\mathbf{\Lambda }_{\mathrm{2}}\subset \mathbf{\Lambda }_{\mathrm{2}\mathcal{C%
}}\subset \mathbf{\Lambda }_{\mathrm{2}}^{\ast }\qquad \Leftrightarrow
\qquad uc\mathbf{\Lambda }_{\mathrm{2}}>uc\mathbf{\Lambda }_{\mathrm{2}%
\mathcal{C}}>uc\mathbf{\Lambda }_{\mathrm{2}}^{\ast }  \label{em}
\end{equation}%
The area of the unit cell for $\mathbf{\Lambda }_{\mathrm{2}}^{\ast }$ is
equal to 1/2 (i.e $uc\mathbf{\Lambda }_{\mathrm{2}}^{\ast }=1/2=uc\mathbf{%
\Lambda }_{\mathrm{2}}/4$) which is the reciprocal of the unit cell area of
the real $\mathbf{\Lambda }_{\mathrm{2}}$. For completeness, we also give an
anticipatory graphic representation of the even self dual lattice $\mathbf{%
\Lambda }_{\mathrm{2}\mathcal{C}}$ satisfying the embedding property (\ref%
{em}). Its unit cell area is equal to 1 (i.e $uc\mathbf{\Lambda }_{\mathrm{2}%
\mathcal{C}}=1$), a value that we use as \emph{a criterion of self duality}.
The graphic of $\mathbf{\Lambda }_{\mathrm{2}\mathcal{C}}$ is depicted in
Figure \textbf{\ref{j2} }where its lattice sites are distinguished in red
and blue.

\begin{figure}[tbph]
\begin{center}
\includegraphics[width=16cm]{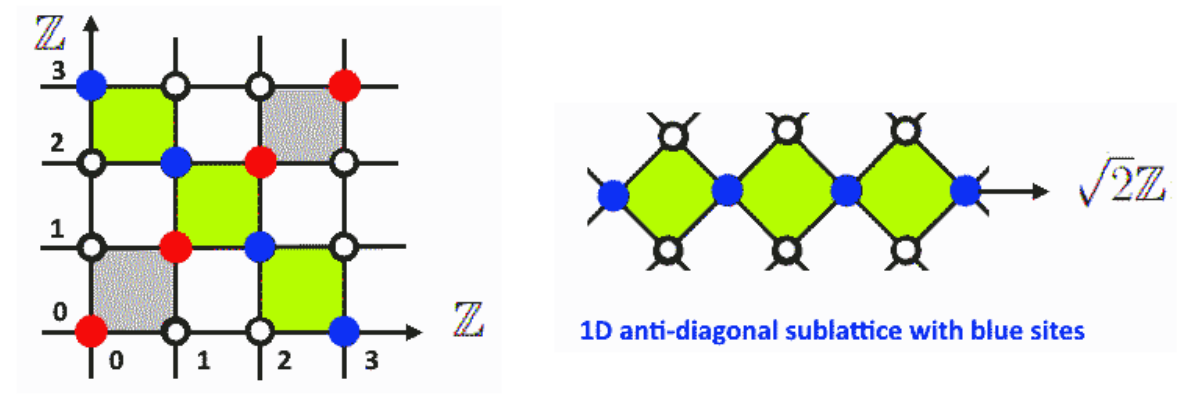}
\end{center}
\par
\vspace{-0.5cm}
\caption{On top, the square lattice $\mathbb{Z\times Z}$. (anti-) diagonal
1D sublattice $\pm \protect\sqrt{\mathrm{2}}\mathbb{Z}$ is given the (blue)
red sites. In bottom, the anti-diagonal 1D sublattice to be used below.}
\label{21a}
\end{figure}

The set of red sites represents the sublattice $\mathbf{\Lambda }_{\mathrm{2}%
}$ contained into $\mathbf{\Lambda }_{\mathrm{2}\mathcal{C}}$ while the blue
sites describe the complement $\mathbf{\Lambda }_{\mathrm{2}\mathcal{C}}/%
\mathbf{\Lambda }_{\mathrm{2}}$ which, up to a global shift $\mathbf{\lambda
,}$ is isomorphic to $\mathbf{\Lambda }_{\mathrm{2}}$. Hence, we have the
decomposition%
\begin{equation}
\begin{tabular}{lll}
$\mathbf{\Lambda }_{\mathrm{2}\mathcal{C}}$ & $=$ & $\mathbf{\Lambda }_{%
\mathrm{2}}+\left( \mathbf{\Lambda }_{\mathrm{2}\mathcal{C}}/\mathbf{\Lambda
}_{\mathrm{2}}\right) $ \\
& $\simeq $ & $\underbrace{\mathbf{\Lambda }_{\mathrm{2}}+\left( \mathbf{%
\Lambda }_{\mathrm{2}}+\mathbf{\lambda }\right) }_{2\text{ superposed }%
\mathbf{\Lambda }_{\mathrm{2}}\text{s }}$%
\end{tabular}%
\end{equation}%
To derive the results illustrated in Figures \textbf{\ref{j1}}-\textbf{\ref%
{j2}}, we start with eq(\ref{3l}) which shows that the lattices $\mathbf{%
\Lambda }_{\mathrm{2}}$ and $\mathbf{\Lambda }_{\mathrm{2}}^{\ast },$
realised in terms of \textsc{R}$_{\mathrm{2}}$ and \textsc{W}$_{\mathrm{2}},$
have the following factorised structure
\begin{equation}
\mathbf{\Lambda }_{\mathrm{2}}=\text{\textsc{R}}_{\mathrm{2}}\times \text{%
\textsc{R}}_{\mathrm{2}}^{\prime }\qquad ,\qquad \mathbf{\Lambda }_{\mathrm{2%
}}^{\ast }=\text{\textsc{W}}_{\mathrm{2}}\times \text{\textsc{W}}_{\mathrm{2}%
}^{\prime }  \label{L2}
\end{equation}%
Since \textsc{W}$_{\mathrm{2}}/$\textsc{R}$_{\mathrm{2}}\simeq \mathbb{Z}_{%
\mathrm{2}},$ the discriminant $\mathbf{\Lambda }_{\mathrm{2}}^{\ast }/%
\mathbf{\Lambda }_{\mathrm{2}}$ is isomorphic to the discrete group $\mathbb{%
Z}_{\mathrm{2}}\times \mathbb{Z}_{\mathrm{2}}$. To elucidate the SU(2)
interpretation of the lattices $\mathbf{\Lambda }_{\mathrm{2}},$ $\mathbf{%
\Lambda }_{\mathrm{2}\mathcal{C}},$ $\mathbf{\Lambda }_{\mathrm{2}}^{\ast },$
we reconsider the discrete sets $\sqrt{\mathrm{2}}\mathbb{Z}$ and $\frac{1}{%
\sqrt{\mathrm{2}}}\mathbb{Z}$ as follows:

$\left( \mathbf{1}\right) $ \textbf{sublattice} $\sqrt{\mathrm{2}}\mathbb{Z}$%
. \newline
The presence of\textrm{\ }$\sqrt{\mathrm{2}}$\textrm{\ }viewed as $\sqrt{%
1^{2}+1^{2}}$, suggest that the discrete set $\sqrt{\mathrm{2}}\mathbb{Z}$
correspond to a 1D sublattice of the ambient square lattice $\mathbb{Z}%
\times \mathbb{Z}$ with canonical basis vectors ($\mathbf{e}_{1},\mathbf{e}%
_{2}$) and lattice\textrm{\ }spacing \texttt{a}$=\mathtt{b}=1$. This
sublattice can be realised in two equivalent ways:\newline
$\left( \mathbf{i}\right) $ As the \emph{diagonal 1D sublattice} with
coordinates ($n,n$) where $n\in \mathbb{Z};$ see red dots in Figure \textbf{%
\ref{21a}}. \newline
$\left( \mathbf{ii}\right) $ As the \emph{anti-diagonal sublattice} with
coordinates ($n,-n$); see the blue dots in the same Figure \textbf{\ref{21a}}%
. These (anti) diagonal lines in $\mathbb{Z}\times \mathbb{Z}$ have slopes $%
\frac{b}{a}=\pm 1$ and spacing $\mathtt{c}=\sqrt{\mathtt{a}^{2}+\mathtt{b}%
^{2}}=\sqrt{2}$.

For our purposes, we adopt the anti-diagonal realisation of \textsc{R}$_{%
\mathrm{2}}\sim \sqrt{2}\mathbb{Z}$ with sites ($n,-n$). This sublattice is
generated by the vector $\mathbf{\alpha }=\mathbf{e}_{1}-\mathbf{e}_{2}$
with components $(1,-1)$ satisfying $\mathbf{\alpha }^{2}=2$ and norm $%
\left\Vert \mathbf{\alpha }\right\Vert =\sqrt{2}.$ We can therefore define
the 1D sublattice like
\begin{equation}
\text{\textsc{R}}_{\mathrm{2}}\simeq \mathbb{Z}\mathbf{\alpha }
\end{equation}%
By using the unit vector $\mathbf{u}=\frac{1}{\sqrt{2}}\left( \mathbf{e}_{1}-%
\mathbf{e}_{2}\right) $ in the direction of \textsc{R}$_{\mathrm{2}}$, we
may also write $\mathbf{\alpha }=\sqrt{2}\mathbf{u}$ and then \textsc{R}$_{%
\mathrm{2}}\simeq (\sqrt{2}\mathbb{Z})\mathbf{u}.$

\begin{figure}[tbph]
\begin{center}
\includegraphics[width=10cm]{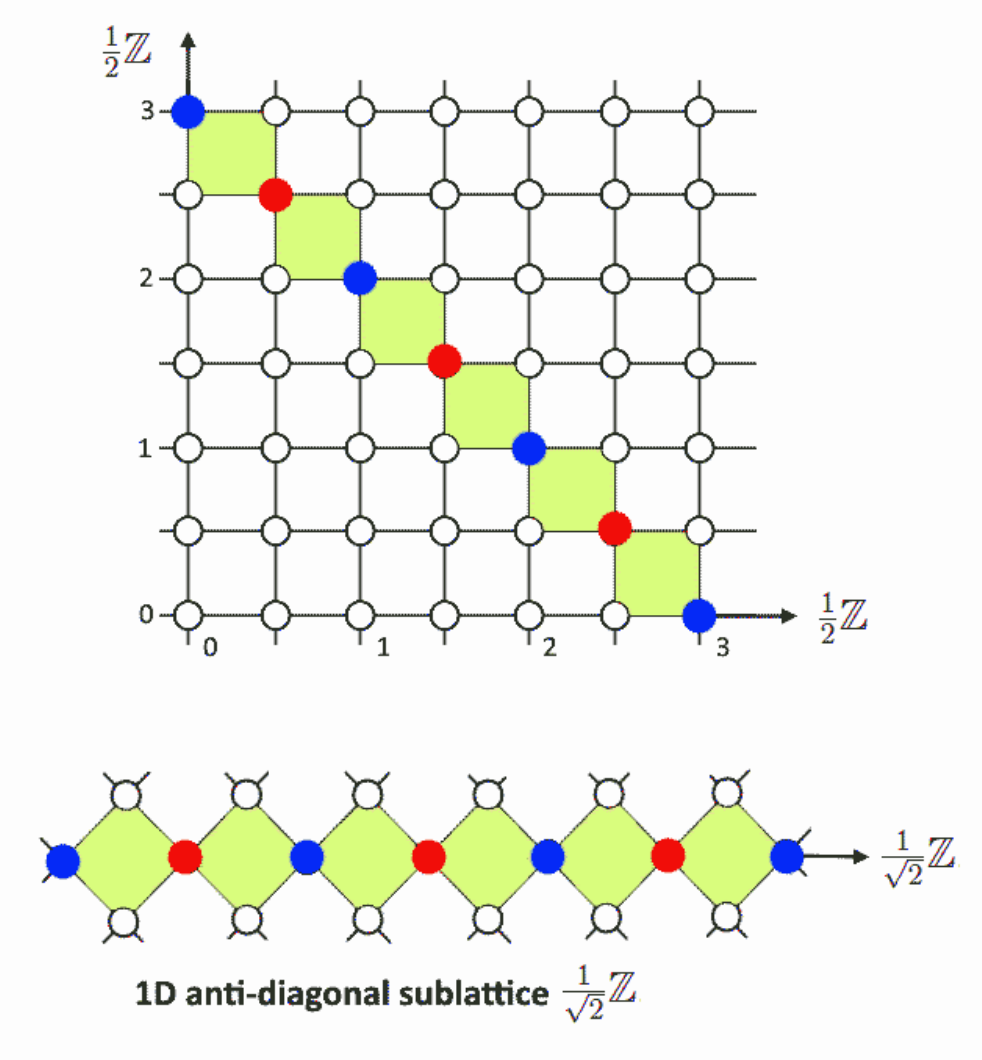}
\end{center}
\par
\vspace{-0.5cm}
\caption{On top, we give the $\frac{1}{2}\mathbb{Z\times }\frac{1}{2}\mathbb{%
Z}$ and anti-diagonal 1D with blue and red sites. In bottom, we find the
anti-diagonal 1D lattice $\frac{1}{\protect\sqrt{\mathrm{2}}}\mathbb{Z}$
containing $\protect\sqrt{\mathrm{2}}\mathbb{Z}$\ as a sublattice. }
\label{22}
\end{figure}

$\left( \mathbf{2}\right) $ \textbf{sublattice} $\frac{1}{\sqrt{\mathrm{2}}}%
\mathbb{Z}$. \newline
This lattice, denoted \textsc{W}$_{\mathrm{2}}=\frac{1}{\sqrt{\mathrm{2}}}%
\mathbb{Z}$ , is\textrm{\ }dual to $\sqrt{\mathrm{2}}\mathbb{Z}.$ It is\
generated by $\mathbf{\lambda }=\frac{1}{2}\mathbf{\alpha }$ with components
$(\frac{1}{2},-\frac{1}{2})$, euclidian length $\mathbf{\lambda }^{2}=1/2$
and norm $\left\Vert \mathbf{\lambda }\right\Vert =1/\sqrt{\mathrm{2}};$
precisely the inverses of the corresponding root quantities $\mathbf{\alpha }%
^{2}$ and $\left\Vert \mathbf{\alpha }\right\Vert .$ The lattice \textsc{W}$%
_{\mathrm{2}}$ sits along \emph{the anti-diagonal line} of the square
lattice $\frac{1}{2}\mathbb{Z}\times \frac{1}{2}\mathbb{Z}$ with lattice
spacing $\mathtt{\tilde{a}}=\frac{1}{2}$ leading to $\mathtt{\tilde{c}}=%
\sqrt{\frac{1}{4}+\frac{1}{4}}=\frac{1}{\sqrt{2}}.$ Thus
\begin{equation}
\text{\textsc{W}}_{\mathrm{2}}\simeq \mathbb{Z}\mathbf{\lambda }
\end{equation}%
Again, in terms of the unit vector $\mathbf{u},$ we also find $\mathbf{%
\lambda }=\frac{1}{\sqrt{2}}\mathbf{u}$ implying \textsc{W}$_{\mathrm{2}%
}\simeq (\frac{1}{\sqrt{2}}\mathbb{Z})\mathbf{u}.$ For illustration, see the
Figure \textbf{\ref{22}}.

In this parametrisation of $\sqrt{\mathrm{2}}\mathbb{Z}$ and $\frac{1}{\sqrt{%
\mathrm{2}}}\mathbb{Z}$, the generators $\mathbf{\alpha }$ and $\mathbf{%
\lambda }$ are very remarkable in the sense they are familiar in Lie algebra
theory. When endowed with the euclidian metric $\delta _{ij}$, $\mathbf{%
\alpha }$ and $\mathbf{\lambda }$ correspond to the simple root and the
fundamental weight vectors of SU(2) with $\mathbf{\alpha }^{2}=2$ and $%
\mathbf{\lambda }^{2}=1/2$ and obeying the duality relation $\mathbf{\alpha
.\lambda }=1$. Therefore, we can accordingly characterise the lattices
\textsc{R}$_{\mathrm{2}}=\sqrt{\mathrm{2}}\mathbb{Z}$ and \textsc{W}$_{%
\mathrm{2}}=\frac{1}{\sqrt{\mathrm{2}}}\mathbb{Z}$\ as follows:
\begin{eqnarray}
\text{\textsc{R}}_{\mathrm{2}} &=&\left\{ \mathbf{x}_{n}=n\mathbf{\alpha }%
\text{ \TEXTsymbol{\vert} }n\in \mathbb{Z}\right\} ,\qquad \left\langle
\mathbf{x}_{\mathbf{n}},\mathbf{x}_{\mathbf{n}}\right\rangle =2n^{2} \\
\text{\textsc{W}}_{\mathrm{2}} &=&\left\{ \mathbf{k}_{m}=m\mathbf{\lambda }%
\text{ \TEXTsymbol{\vert} }m\in \mathbb{Z}\right\} ,\qquad \left\langle
\mathbf{x}_{n},\mathbf{k}_{m}\right\rangle =nm
\end{eqnarray}%
From the relation\textrm{\ }$\mathbf{\lambda }=\frac{1}{2}\mathbf{\alpha ,}$
the dual weight lattice \textsc{W}$_{\mathrm{2}}$ splits into two subsets
\textsc{W}$_{\mathrm{2}}^{{\small even}}$ and \textsc{W}$_{\mathrm{2}}^{%
{\small odd}}$ as described below:

\begin{description}
\item[$\left( \mathbf{i}\right) $] \ the \textsc{W}$_{\mathrm{2}}^{{\small %
even}}$, generated by \emph{even} integers $m=2\mu $, gives vectors $\mathbf{%
k}_{2\mu }=2\mu \mathbf{\lambda }$ which are isomorphic\ to root vectors $%
\mu \mathbf{\alpha }$; as such, we have \textsc{W}$_{\mathrm{2}}^{{\small %
even}}\simeq $\textsc{R}$_{\mathrm{2}}$. This corresponds to the blue lD
lattice in Figure \textbf{\ref{22}}.

\item[$\left( \mathbf{ii}\right) $] \ the subset \textsc{W}$_{\mathrm{2}}^{%
{\small odd}},$ made of \emph{odd}\textrm{\ }vectors $\mathbf{k}_{2\mu
-1}=\left( 2\mu +1\right) \mathbf{\lambda }$ that read also like $\left( \mu
+\frac{1}{2}\right) \mathbf{\alpha ,}$ describes root vectors $\mu \mathbf{%
\alpha }$ shifted by a fundamental weight $\mathbf{\lambda }=\frac{1}{2}%
\mathbf{\alpha }$. Thus \textsc{W}$_{\mathrm{2}}^{{\small odd}}$ is
isomorphic to \textsc{R}$_{\mathrm{2}}+\mathbf{\lambda }$; it corresponds to
the red lD lattice in Figure \textbf{\ref{22}}. Hence, \textsc{W}$_{\mathrm{2%
}}$ decomposes into the superposition
\begin{equation}
\begin{tabular}{lllll}
$\text{\textsc{W}}_{\mathrm{2}}$ & $=$ & $\text{\textsc{W}}_{\mathrm{2}}^{%
{\small even}}$ & $\cup $ & $\text{\textsc{W}}_{\mathrm{2}}^{{\small odd}}$
\\
& $=$ & $\text{\textsc{R}}_{\mathrm{2}}$ & $\cup $ & $\left\{ \text{\textsc{R%
}}_{\mathrm{2}}+\mathbf{\lambda }\right\} $%
\end{tabular}%
\qquad ,\qquad
\begin{tabular}{lll}
$\text{\textsc{W}}_{\mathrm{2}}^{{\small even}}$ & $\simeq $ & $\text{%
\textsc{R}}_{\mathrm{2}}$ \\
$\text{\textsc{W}}_{\mathrm{2}}^{{\small odd}}$ & $\simeq $ & $\text{\textsc{%
R}}_{\mathrm{2}}+\mathbf{\lambda }$%
\end{tabular}%
\end{equation}%
\ in agreement with the discriminant \textsc{W}$_{\mathrm{2}}/$\textsc{R}$_{%
\mathrm{2}}\simeq \mathbb{Z}_{\mathrm{2}}$.
\end{description}

\subsubsection{Lattices of SO(4)}

Here, we use equation (\ref{L2}) to construct the two dimensional lattices $%
\mathbf{\Lambda _{\mathrm{2}},}$ $\mathbf{\Lambda }_{\mathrm{2}\mathcal{C}}$
and $\mathbf{\Lambda }_{\mathrm{2}}^{\ast }$ along with their respective
characteristic matrices represented by the same symbols $\Lambda ,$ $\Lambda
_{\mathcal{C}}$ and $\Lambda ^{\ast }.$

\paragraph{\qquad \textbf{A.} \textbf{the triplet (}$\mathbf{\Lambda _{%
\mathrm{2}},\Lambda }_{\mathrm{2}\mathcal{C}}\mathbf{,\Lambda }_{\mathrm{2}%
}^{\ast }$) \textbf{and} \textbf{SO(4) lattices}}

\ \ \newline
Using the above results, the 2D lattices $\mathbf{\Lambda }_{\mathrm{2}%
}\simeq $\textsc{R}$_{\mathrm{2}}\times $\textsc{R}$_{\mathrm{2}}$ and $%
\mathbf{\Lambda }_{\mathrm{2}}^{\ast }\simeq $\textsc{W}$_{\mathrm{2}}\times
$\textsc{W}$_{\mathrm{2}},$ given by eq(\ref{L2}), can be naturally
interpreted as the 2D root \textsc{R}$_{\mathrm{2}}^{\mathbf{so}_{\mathbf{4}%
}}$ and the 2D weight \textsc{W}$_{\mathrm{2}}^{\mathbf{so}_{\mathbf{4}}}$
lattices of $SU(2)\times SU(2)^{\prime }$. This group is isomorphic to
SO(4), which possesses two orthogonal simple roots ($\mathbf{\alpha ,\alpha }%
^{\prime }$) and two orthogonal fundamental weight vectors ($\mathbf{\lambda
}\mathbf{,\lambda }^{\prime }$). Within this view, the site vectors $\mathbf{%
v}_{n}$ and $\mathbf{w}_{m}$ in the lattices $\mathbf{\Lambda }_{\mathrm{2}}$
and $\mathbf{\Lambda }_{\mathrm{2}}^{\ast }$ are therefore parameterised by
pairs of integers ($n\mathbf{,}n^{\prime }$) and ($m\mathbf{,}m^{\prime }$)
as follows%
\begin{equation}
\begin{tabular}{lll}
$\Lambda _{\mathrm{2}}$ & $=$ & $\left\{ \mathbf{v}_{n}=n\mathbf{\alpha }%
\oplus n^{\prime }\mathbf{\alpha }^{\prime }\right\} $ \\
$\Lambda _{\mathrm{2}}^{\ast }$ & $=$ & $\left\{ \mathbf{w}_{m}=m\mathbf{%
\lambda }\oplus m^{\prime }\mathbf{\lambda }^{\prime }\right\} $%
\end{tabular}%
,\qquad
\begin{tabular}{lll}
$\Lambda _{\mathrm{2}}$ & $=$ & $\{\mathbf{v}_{n}=\sqrt{2}\left( n^{1}%
\mathbf{u}_{1}+n^{2}\mathbf{u}_{2}\right) \}$ \\
$\Lambda _{\mathrm{2}}^{\ast }$ & $=$ & $\{\mathbf{w}_{m}=\frac{1}{\sqrt{2}}%
\left( m^{1}\mathbf{u}_{1}+m^{2}\mathbf{u}_{2}\right) \}$%
\end{tabular}
\label{2l}
\end{equation}%
where $\mathbf{\alpha }=\sqrt{2}\mathbf{u}_{1}$, $\mathbf{\alpha }^{\prime }=%
\sqrt{2}\mathbf{u}_{2}$ and $\mathbf{\lambda }=\frac{1}{\sqrt{2}}\mathbf{u}%
_{1},$ $\mathbf{\lambda }^{\prime }=\frac{1}{\sqrt{2}}\mathbf{u}_{2}$. They
are represented\ in Figure \textbf{\ref{j1}}. Notice that in terms of the
components $\mathbf{u}_{1}=(1,0)$ and $\mathbf{u}_{2}=(0,1)$ with Lorentzian
product $\mathbf{u}_{i}^{T}\eta \mathbf{u}_{j}=\eta _{ij},$ we have
\begin{equation}
v_{n}^{i}=\sqrt{2}u_{j}^{i}n^{j}\qquad ,\qquad w_{m}^{i}=\frac{1}{\sqrt{2}}%
u_{j}^{i}m^{j}  \label{cc}
\end{equation}%
The characteristic matrices follow as $\left( \Lambda _{\mathrm{2}}\right)
_{j}^{i}=\sqrt{2}u_{j}^{i}$ and $\left( \Lambda _{\mathrm{2}}^{\ast }\right)
_{j}^{i}=\frac{1}{\sqrt{2}}u_{j}^{i}.$ From the relationships (\ref{2l}), we
deduce the realisation of $\mathbf{\Lambda }_{\mathrm{2}\mathcal{C}}\simeq $%
\textsc{R}$_{\mathrm{2}}\times $\textsc{W}$_{\mathrm{2}}$ reading like
\begin{equation}
\mathbf{\Lambda }_{\mathrm{2}\mathcal{C}}=\{\mathbf{u}_{n,m}=n\sqrt{2}%
\mathbf{u}_{1}+\frac{m}{\sqrt{2}}\mathbf{u}_{2}\}
\end{equation}%
The lattice picture is depicted by the Figure \textbf{\ref{j2}}.

\paragraph{\qquad \textbf{B.} \textbf{characteristic matrices}}

\ \ \ \newline
Using the components of the vectors $\mathbf{v}_{n}$ and $\mathbf{w}_{m}$
given by (\ref{cc}), the characteristic matrices of the real and dual
lattices are respectively given by $\Lambda _{\mathrm{2}}=\sqrt{2}I_{2\times
2}$ and $\Lambda _{\mathrm{2}}^{\ast }=\frac{1}{\sqrt{2}}I_{2\times 2}.$ To
express the matrix $\Lambda _{\mathrm{2}\mathcal{C}},$ it is convenient to
decompose the identity matrix $I_{2\times 2}$ in terms of the projectors $%
\varrho _{1}$ and $\varrho _{2}$ defined as follows%
\begin{equation}
\varrho _{1}=\left(
\begin{array}{cc}
1 & 0 \\
0 & 0%
\end{array}%
\right) \qquad ,\qquad \varrho _{2}=\left(
\begin{array}{cc}
0 & 0 \\
0 & 1%
\end{array}%
\right)
\end{equation}%
In term of these $\varrho _{i}$'s, the characteristic matrices of the
triplet $(\mathbf{\Lambda }_{\mathrm{2}},\mathbf{\Lambda }_{\mathrm{2}%
\mathcal{C}},\mathbf{\Lambda }_{\mathrm{2}}^{\ast })$ can be written as%
\begin{equation}
\Lambda _{\mathrm{2}}=\sqrt{2}\varrho _{1}+\sqrt{2}\varrho _{2},\qquad
\Lambda _{\mathrm{2}}^{\ast }=\frac{1}{\sqrt{2}}\varrho _{1}+\frac{1}{\sqrt{2%
}}\varrho _{2},\qquad \Lambda _{\mathrm{2}\mathcal{C}}=\sqrt{2}\varrho _{1}+%
\frac{1}{\sqrt{2}}\varrho _{2}
\end{equation}

\subsection{Models with CS level\emph{\ }k\TEXTsymbol{>}2}

\begin{figure}[tbph]
\begin{center}
\includegraphics[width=14cm]{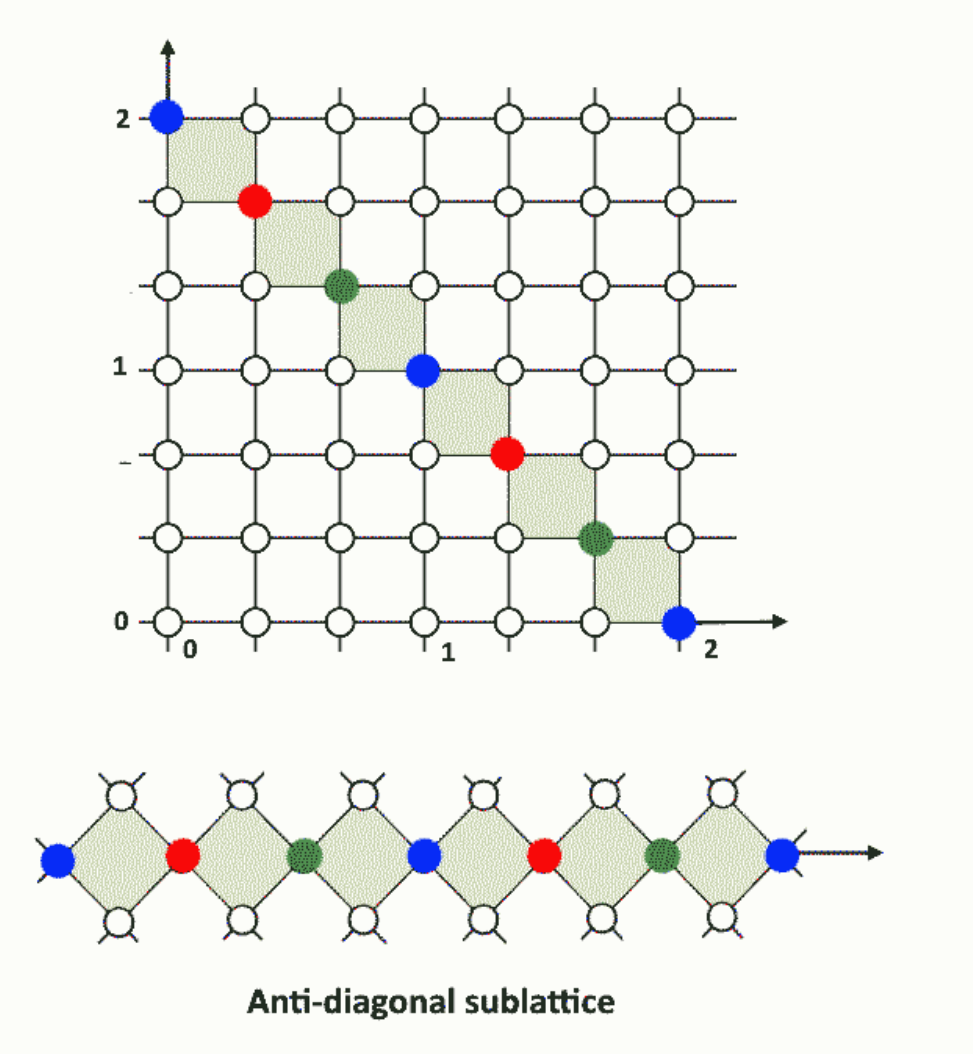}
\end{center}
\par
\vspace{-0.5cm}
\caption{On top, the square lattice $\frac{1}{a}\mathbb{Z\times }\frac{1}{a}%
\mathbb{Z}$ with $a=\protect\sqrt{\frac{\mathrm{k}}{2}}.$ It contains the 1D
weight lattice \textsc{\~{W}}$_{\mathrm{3}}$ given by the superposition of
three 1D sublattices \textsc{R}$_{\mathrm{3}}\cup \left\{ \text{\textsc{R}}_{%
\mathrm{3}}+\mathbf{\protect\lambda }\right\} \cup \left\{ \text{\textsc{R}}%
_{\mathrm{3}}+2\mathbf{\protect\lambda }\right\} $ distinguished by the
colors: blue, red and green. For k=3, the value $2\mathbf{\protect\lambda }$
can be imagined as corresponding to $-\mathbf{\protect\lambda .}$}
\label{23}
\end{figure}

For higher values of the integer k, while keeping the central charge $c_{%
\text{\textsc{l}/\textsc{r}}}=1$, the root $\mathbf{\alpha }$ and the
fundamental weight vector $\mathbf{\lambda }$ are respectively replaced by a
dilated $\mathbf{\tilde{\alpha}}$ and a compressed $\mathbf{\tilde{\lambda}}$
satisfying the euclidian properties:%
\begin{equation}
\mathbf{\tilde{\alpha}}^{2}=\mathrm{k}>2\qquad ,\qquad \mathbf{\tilde{\lambda%
}}^{2}=\frac{1}{\mathrm{k}}<\frac{1}{2}\qquad ,\qquad \mathbf{\tilde{\alpha}}%
.\mathbf{\tilde{\lambda}}=1
\end{equation}%
These newly scaled vectors $\mathbf{\tilde{\alpha}}$ and $\mathbf{\tilde{%
\lambda}}$ generate 1D sublattices \textsc{\~{R}}$_{\mathrm{k}}$ and \textsc{%
\~{W}}$_{\mathrm{k}}$ extending the root and the weight lattices \textsc{R}$%
_{\mathrm{2}}$ and \textsc{W}$_{\mathrm{2}}$ of SU(2).
\begin{figure}[tbph]
\begin{center}
\includegraphics[width=12cm]{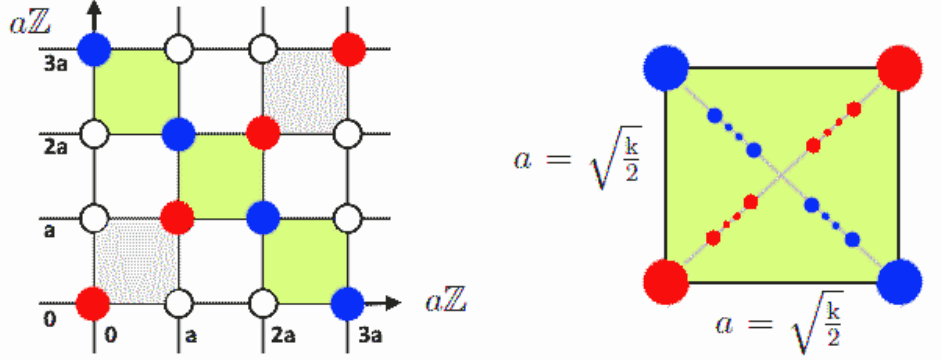}
\end{center}
\par
\vspace{-0.5cm}
\caption{On the left, we represent the square lattice $a\mathbb{Z\times }a%
\mathbb{Z}$ with parameter $a=\protect\sqrt{\frac{\mathrm{k}}{2}}.$ It
contains the 1D root lattice \textsc{\~{R}}$_{\mathrm{k}}$. Here, the 1D
(anti) diagonal lattices are represented in red and blue colors. On the
right, a unit cell having k charges. A similar figure is valid for the
weight lattice \textsc{\~{W}}$_{\mathrm{k}}$ sitting in $\frac{1}{a}\mathbb{Z%
}\times \frac{1}{a}\mathbb{Z};$ see the Figure \textbf{\protect\ref{23} }for
k=3.}
\label{kZ}
\end{figure}
An interesting way to realise this structure is obtained using a square
lattice with scaled spacings $\mathtt{a}_{x}=\mathtt{a}_{y}=\sqrt{\mathrm{k/2%
}}.$ In this case, the vectors $\mathbf{\tilde{\alpha}}$ and $\mathbf{\tilde{%
\lambda}}$ relate to the canonical simple root and the fundamental weight
vector via $\mathbf{\tilde{\alpha}}=\sqrt{\mathrm{k/2}}\mathbf{\alpha }$ and
$\mathbf{\tilde{\lambda}}=\sqrt{\mathrm{2/k}}\mathbf{\lambda }$ and
consequently
\begin{equation}
\mathbf{\tilde{\lambda}}=\frac{1}{\mathrm{k}}\mathbf{\tilde{\alpha}}\qquad
\Rightarrow \qquad \mathbf{\tilde{\alpha}}=\mathrm{k}\mathbf{\tilde{\lambda}}
\end{equation}%
In this realisation, the root lattice \textsc{R}$_{\mathrm{k}}=\sqrt{\mathrm{%
k}}\mathbb{Z}$ is generated by $\left\{ \mathbf{x}_{n}=n\mathbf{\tilde{\alpha%
}}\right\} ,$ depicted in Figure \textbf{\ref{22}}, it is a 1D sublattice of
$\sqrt{\mathrm{k/2}}\mathbb{Z\times }\sqrt{\mathrm{k/2}}\mathbb{Z}$ with
sites%
\begin{equation}
\mathbf{x}_{n}=n\mathbf{\tilde{\alpha}=}\sqrt{\frac{\mathrm{k}}{\mathrm{2}}}n%
\mathbf{\alpha }
\end{equation}%
Setting $\mathtt{a}_{x}=\mathtt{a}_{y}=\mathrm{a}$, the leading prime values
for the CS level correspond to
\begin{equation}
\begin{tabular}{ccccccc}
$\mathrm{a=}$ & 1 & $\frac{3}{2}$ & $\frac{5}{2}$ & $\frac{7}{2}$ & $\frac{11%
}{2}$ & $\frac{13}{2}$ \\
$\mathrm{k=}$ & 2 & 3 & 5 & 7 & 11 & 13%
\end{tabular}%
\end{equation}%
The dual weight lattice \textsc{W}$_{\mathrm{k}}=\frac{1}{\sqrt{\mathrm{k}}}%
\mathbb{Z}$ is defined by $\{\mathbf{k}_{m}=m\mathbf{\tilde{\lambda}\}}$
with $m\in \mathbb{Z};$ it is a 1D sublattice embedded in$\sqrt{\frac{%
\mathrm{2}}{\mathrm{k}}}\mathbb{Z}\times \sqrt{\frac{\mathrm{2}}{\mathrm{k}}}%
\mathbb{Z}.$ By substituting $\mathbf{\tilde{\lambda}}=\frac{1}{\mathrm{k}}%
\mathbf{\tilde{\alpha}},$ we get $\mathbf{k}_{m}=\sqrt{\frac{\mathrm{2}}{%
\mathrm{k}}}m\mathbf{\lambda }$ that reads also like
\begin{equation}
\mathbf{k}_{m}=m\mathbf{\tilde{\lambda}}=\frac{m}{\mathrm{k}}\mathbf{\tilde{%
\alpha}\qquad ,\qquad }m\in \mathbb{Z}
\end{equation}%
By setting $m=\mathrm{k}n+\varepsilon $ with $\varepsilon =0,...,k-1$ $\func{%
mod}k,$ we end up with $\mathbf{k}_{m}=n\mathbf{\tilde{\alpha}}+\varepsilon
\mathbf{\tilde{\lambda}}$ showing that \textsc{W}$_{\mathrm{k}}$ is a\textrm{%
\ }superposition of k sheets labeled by $\varepsilon .$ For odd integers
\textrm{k=2j+1, }the values of\textrm{\ }$\varepsilon \in \left\{
-j,...,j\right\} $\textrm{\ }reflect that the 2j+1 sheets making \textsc{W}$%
_{\mathrm{2j+1}}$ form an irreducible representation of SU(2) with isospin
j. For illustration, see Figure \textbf{\ref{23}} regarding k=3 and \textsc{W%
}$_{\mathrm{3}}/$\textsc{R}$_{\mathrm{3}}\simeq \mathbb{Z}_{3}$ where the
three sheets are colored in blue, red and green.

In conclusion, the two dimensional lattices $(\mathbf{\Lambda }_{\mathrm{2}},%
\mathbf{\Lambda }_{\mathrm{2}\mathcal{C}},\Lambda _{\mathrm{2}}^{\ast })$,
depicted in Figures \textbf{\ref{j1}} and \textbf{\ref{j2},} extend for
Chern-Simons levels $\mathrm{k}>2$ like%
\begin{equation}
\begin{tabular}{lllll}
$\Lambda _{\mathrm{k}}$ & $=$ & $\left\{ \mathbf{v}_{n}=n\mathbf{\tilde{%
\alpha}}\oplus n^{\prime }\mathbf{\tilde{\alpha}}^{\prime }\right\} $ & $=$
& $\{\mathbf{v}_{n}=\sqrt{\mathrm{k}}\left( n\mathbf{u}_{1}+n^{\prime }%
\mathbf{u}_{2}\right) \}$ \\
$\Lambda _{\mathrm{k}}^{\ast }$ & $=$ & $\left\{ \mathbf{w}_{m}=m\mathbf{%
\tilde{\lambda}}\oplus m^{\prime }\mathbf{\tilde{\lambda}}^{\prime }\right\}
$ & $=$ & $\{\mathbf{w}_{m}=\frac{1}{\sqrt{\mathrm{k}}}(m\mathbf{u}%
_{1}+m^{\prime }\mathbf{u}_{2})\}$%
\end{tabular}%
\end{equation}%
where $\mathbf{u}_{i}$ are as previously defined. Substituting $\mathbf{%
\tilde{\lambda}}=\frac{1}{\mathrm{k}}\mathbf{\tilde{\alpha}}$ and $\mathbf{%
\tilde{\lambda}}^{\prime }=\frac{1}{\mathrm{k}}\mathbf{\tilde{\alpha}}%
^{\prime }$ while setting $m=\mathrm{k}n+\varepsilon $ and $m^{\prime }=%
\mathrm{k}n^{\prime }+\varepsilon ^{\prime }$ with $\varepsilon ,\varepsilon
^{\prime }=0,...,k-1,$ we get
\begin{equation}
\mathbf{w}_{m}=\left( n\mathbf{\tilde{\alpha}}+n^{\prime }\mathbf{\tilde{%
\alpha}}^{\prime }\right) +\varepsilon \mathbf{\tilde{\lambda}+}\varepsilon
^{\prime }\mathbf{\tilde{\lambda}}^{\prime }
\end{equation}%
showing that the dual $\Lambda _{\mathrm{k}}^{\ast }$ \textrm{i}s a union of
$\mathrm{k\times k}$ sheets of 2D lattices isomorphic to the real $\Lambda _{%
\mathrm{k}}$. For odd integers \textrm{k=2j+1}, these sheets form a
representation of SO(4) with isospin pair ($j,j$). Moreover, the even self
dual lattice $\mathbf{\Lambda }_{\mathrm{k}\mathcal{C}}$ engendered by the
vectors $n\mathbf{\tilde{\alpha}}+m\mathbf{\tilde{\alpha}}^{\prime
}+\varepsilon ^{\prime }\mathbf{\tilde{\lambda}}^{\prime }$ is composed of k
sheets isomorphic to $\mathbf{\Lambda }_{\mathrm{k}}$.\textrm{\ }It can also
be expressed like%
\begin{equation}
\mathbf{\Lambda }_{\mathrm{k}\mathcal{C}}=\{\mathbf{u}_{n,m}=n\sqrt{\mathrm{k%
}}\mathbf{u}_{1}+\frac{m}{\sqrt{\mathrm{k}}}\mathbf{u}_{2}\}
\end{equation}%
The corresponding characteristic matrix $\left( \Lambda _{\mathrm{k}}\right)
_{\mathcal{C}}$ of the even self dual lattice and the induced metric $\left(
g_{\Lambda _{\mathrm{k}\mathcal{C}}}\right) =\Lambda _{\mathrm{k}\mathcal{C}%
}^{T}\eta \Lambda _{\mathrm{k}\mathcal{C}}$ read as follows
\begin{equation}
\Lambda _{\mathrm{k}\mathcal{C}}=\left(
\begin{array}{cc}
\sqrt{\mathrm{k}} & 0 \\
0 & \frac{1}{\sqrt{\mathrm{k}}}%
\end{array}%
\right) \qquad ,\qquad g_{\Lambda _{\mathrm{k}\mathcal{C}}}=\left(
\begin{array}{cc}
0 & 1 \\
1 & 0%
\end{array}%
\right)
\end{equation}%
with determinant $\det \Lambda _{\mathrm{k}\mathcal{C}}=1.$ \newline
Finally, we note that the generalisation to higher orders r\TEXTsymbol{>}1,
with central charge values $c_{\text{\textsc{l}/\textsc{r}}}=\mathrm{r}$,
follows straightforwardly. A detailed study is omitted here.

\subsection{From scalars in continuum to lattice QFT}

In this subsection, we\ outline the key steps by which the lattices
introduced earlier will be maneuvered in our study. This construction
provides a formal pathway towards a non perturbative description of
topological matter\ that we\textrm{\ }will further develop later on (see
section 5). Focussing on the scalar fields (a fermionic description will be
introduced subsequently via fermionization techniques), the transition from
scalars in continuum to a non perturbative lattice framework can be
summarised as follows:

\subsubsection{Case of one real 2D scalar field\emph{\ }$X=X(\protect\tau ,%
\protect\sigma )$}

We start from the 2D field action $\mathcal{S}\left( X\right) =\int_{%
\mathcal{M}_{2}}d^{2}\xi \mathcal{L}\left( X\right) $ in continuum with
quadratic lagrangian density given by $\frac{1}{2}\left( \partial _{\mu
}X\right) \left( \partial ^{\mu }X\right) +\frac{1}{2}m^{2}X^{2}$ where we
have added the term $m^{2}X^{2}$ as a possible quadratic term. In the
interesting limit $m\rightarrow 0$ (required by conformal invariance), the
field equation of the scalar (bosonic string) field reads like ($\frac{%
\partial ^{2}}{\partial \tau ^{2}}-\frac{\partial ^{2}}{\partial \sigma ^{2}}%
)X=\mathcal{O}\left( m^{2}\right) \simeq 0;$ it remarkably factorises as
follows
\begin{equation}
\left( \frac{\partial }{\partial \tau }-\frac{\partial }{\partial \sigma }%
\right) \left( \frac{\partial }{\partial \tau }+\frac{\partial }{\partial
\sigma }\right) X\simeq 0\qquad \Rightarrow \qquad \left\{
\begin{array}{c}
\frac{\partial }{\partial \xi _{L}}\frac{\partial }{\partial \xi _{R}}%
X\simeq 0 \\
\frac{\partial }{\partial \xi _{R}}\frac{\partial }{\partial \xi _{L}}%
X\simeq 0%
\end{array}%
\right.  \label{eq}
\end{equation}%
and admits solutions in terms of left and right moving modes corresponding
to the usual splitting $X=X_{L}+X_{R}$ with $X_{L}=X\left( \xi _{L}\right) $
and $X_{R}=X\left( \xi _{R}\right) .$ In the euclidian metric, the real
coordinates $\xi _{L/R}$ get replaced by the complex variables $z,\bar{z}$.

\

$\bullet $ \emph{Matrix equation}:\newline
By defining $V_{L}\simeq \partial _{L}X$ and $V_{R}\simeq \partial _{R}X,$
commonly interpreted as the left $J_{L}$ and right $J_{R}$ chiral currents
\textrm{but thought of below as just the components of a 2D "vector" field }$%
\mathcal{V}$ in the limit $m\rightarrow 0,$ the above equations of motion (%
\ref{eq}) become $\partial _{R}V_{L}\simeq 0$ and $\partial _{L}V_{R}\simeq
0 $. For finite m, we have $V_{L/R}=(\partial _{L/R}+\frac{m^{2}}{\partial
_{R/L}})X$. Setting $\mathcal{V}=(V_{L},V_{R})$, one can recast the dynamics
into a 2$\times $2 matrix equation like $\left( \gamma ^{L}\partial
_{L}+\gamma ^{R}\partial _{R}\right) \mathcal{V}\simeq 0$ with $\gamma ^{L}$
and $\gamma ^{R}$ reading from the following expression
\begin{equation}
\left(
\begin{array}{cc}
0 & \partial _{L} \\
\partial _{R} & 0%
\end{array}%
\right) \left(
\begin{array}{c}
V_{L} \\
V_{R}%
\end{array}%
\right) \simeq \left(
\begin{array}{c}
0 \\
0%
\end{array}%
\right)
\end{equation}%
from which we deduce the operators
\begin{equation}
\mathcal{D}\simeq \left(
\begin{array}{cc}
0 & \partial _{L} \\
\partial _{R} & 0%
\end{array}%
\right) \qquad ,\qquad \mathcal{D}^{2}\simeq \left(
\begin{array}{cc}
\partial _{L}\partial _{R} & 0 \\
0 & \partial _{R}\partial _{L}%
\end{array}%
\right)  \label{d}
\end{equation}

$\bullet $ \emph{Fourier transform}:\newline
In the reciprocal space $\partial _{L/R}=\frac{i}{\hbar }q_{L/R}$\ and
where\ the fields $V_{L}$ and $V_{R}$ are expanded in Fourier series\textrm{%
\footnote{%
\ For exact m=0 (CFT), one uses Laurent expansions and the powerful
properties of complex analysis.}} like $\sum_{q}e^{iq.\xi }\tilde{V}_{L}$
and $\sum_{q}e^{iq.\xi }\tilde{V}_{R}$, the above matrix relation can be
brought to the form%
\begin{equation}
\sum_{\left( q_{L},q_{R}\right) }\left(
\begin{array}{cc}
0 & q_{L} \\
q_{R} & 0%
\end{array}%
\right) \left(
\begin{array}{c}
e^{iq.\xi }\tilde{V}_{L} \\
e^{iq.\xi }\tilde{V}_{R}%
\end{array}%
\right) \simeq \left(
\begin{array}{c}
0 \\
0%
\end{array}%
\right)  \label{l}
\end{equation}%
Notice the emergence of the \emph{Dirac-like}\textrm{\footnote{%
\ By using ferminoisation ideas, the real bosoinc components $V_{L/R}$ can
be expressed in terms of complex fermionic fields as $\psi _{L/R}\psi
_{L/R}^{\ast }.$}} matrix operators $\mathcal{D}$ in (\ref{d}) and the
reciprocal $\mathcal{\tilde{D}}$\ in (\ref{l}) which generally appear in
fermionic theories. This hints at a possible generalization to a
non-perturbative regime, where the non linear (differential) operator $%
\boldsymbol{\hat{D}}$ involves oscillating (trigonometric) functions like
for instance%
\begin{equation}
\boldsymbol{\hat{D}=}\left(
\begin{array}{cc}
0 & \sin q_{L} \\
\sin q_{R} & 0%
\end{array}%
\right) \qquad \rightarrow \qquad \mathcal{\tilde{D}}=\left(
\begin{array}{cc}
0 & q_{L} \\
q_{R} & 0%
\end{array}%
\right)  \label{2}
\end{equation}%
with eigenvalues given by $\delta _{\pm }\sim \pm \sqrt{\left\vert \sin
q_{L}\sin q_{R}\right\vert }$ which tends to $\pm \sqrt{\left\vert
q_{L}q_{R}\right\vert }$ for small values $q_{L/R}\rightarrow 0.$

\subsubsection{Lattice extension}

\begin{figure}[tbph]
\begin{center}
\includegraphics[width=10cm]{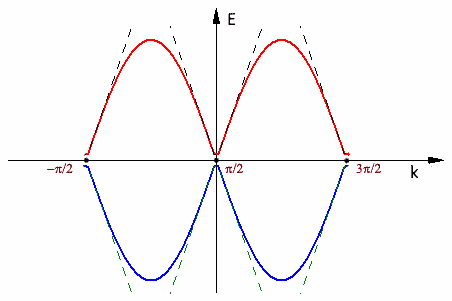}
\end{center}
\par
\vspace{-0.5cm}
\caption{Band structure of the toy model (\protect\ref{hk}). It gives the
variation of $E_{\pm }=\pm \protect\sqrt{tt^{\ast }\cos ^{2}k}$ (solid
line). In dashed line, we draw the behaviour around the Dirac-like points
given by the zeros of $\pm \left\vert t(k-n\frac{\protect\pi }{2}%
)\right\vert .$}
\label{g1}
\end{figure}

In the (euclidian) lattice field description, the continuum fields $%
V_{L}\left( \xi \right) \equiv V^{+}\left( \xi \right) $ and $V_{R}\left(
\xi \right) \equiv V^{-}\left( \xi \right) $ are discretized in terms of the
two following: $\left( \mathbf{i}\right) $ The lattice components $V_{%
\mathbf{\xi }_{\mathbf{n}}}^{+}\equiv V_{\mathbf{n}}^{+}$ and $V_{\mathbf{%
\xi }_{\mathbf{n}}}^{-}\equiv V_{\mathbf{n}}^{-}$ with discrete coordinates $%
\mathbf{\xi }_{\mathbf{n}}$ $=n^{i}\mathbf{e}_{i}$ where $\left( \mathbf{e}%
_{1},\mathbf{e}_{2}\right) $ are generators of the 2D lattice. And $\left(
\mathbf{ii}\right) $ their\emph{\ Fourier transforms} given by the
expansions $\sum_{\mathbf{k}}e^{i\mathbf{k}.\mathbf{\xi }_{\mathbf{n}}}%
\tilde{V}_{\mathbf{\xi }_{\mathbf{n}}}^{+}$ and $\sum_{\mathbf{k}}e^{i%
\mathbf{k}.\mathbf{\xi }_{\mathbf{n}}}\tilde{V}_{\mathbf{\xi }_{\mathbf{n}%
}}^{-}$. Borrowing ideas from tight binding modeling of condensed matter
\textrm{\cite{tm1}-\cite{trs1}}, the quadratic lattice hamiltonian
describing neighboring interactions read as follows%
\begin{equation}
H=\sum_{\mathbf{\xi }_{\mathbf{n}},\mathbf{\xi }_{\mathbf{m}}}\left( V_{%
\mathbf{\xi }_{\mathbf{n}}}^{+\dagger },V_{\mathbf{\xi }_{\mathbf{n}%
}}^{-\dagger }\right) \left(
\begin{array}{cc}
t_{\mathbf{\xi }_{\mathbf{n}},\mathbf{\xi }_{\mathbf{m}}}^{--} & t_{\mathbf{%
\xi }_{\mathbf{n}},\mathbf{\xi }_{\mathbf{m}}}^{-+} \\
t_{\mathbf{\xi }_{\mathbf{n}},\mathbf{\xi }_{\mathbf{m}}}^{+-} & t_{\mathbf{%
\xi }_{\mathbf{n}},\mathbf{\xi }_{\mathbf{m}}}^{++}%
\end{array}%
\right) \left(
\begin{array}{c}
V_{\mathbf{\xi }_{\mathbf{m}}}^{+} \\
V_{\mathbf{\xi }_{\mathbf{m}}}^{-}%
\end{array}%
\right) +hc
\end{equation}%
where the complex $t_{\mathbf{\xi }_{\mathbf{n}},\mathbf{\xi }_{\mathbf{m}%
}}^{\pm \pm }$ are coupling functions that can be thought of as constants
for convenience. By using the Fourier transforms while restricting the
interactions to the first nearest neighbours ($\mathbf{\xi }_{\mathbf{m}}=%
\mathbf{\xi }_{\mathbf{n}}+\mathbf{\theta }_{j}$ with $j=1,...,d$) and
assuming off-diagonal hopping parameters for illustration (eg: $H=t\sum_{%
\mathbf{\xi }_{\mathbf{n}}}\sum_{j}V_{\mathbf{\xi }_{\mathbf{n}}}^{-\dagger
}V_{\mathbf{\xi }_{\mathbf{n}}+\mathbf{\theta }_{j}}^{+}+hc$), it reads in
the Fourier space like $H=\sum_{\mathbf{k}}\mathcal{\tilde{V}}_{\mathbf{k}%
}^{\dagger }H_{\mathbf{k}}\mathcal{\tilde{V}}_{\mathbf{k}}$ with 2$\times $2
matrix $H_{\mathbf{k}}$ given by
\begin{equation}
H_{\mathbf{k}}=\left(
\begin{array}{cc}
0 & f_{\mathbf{k}} \\
f_{\mathbf{k}}^{\ast } & 0%
\end{array}%
\right) \qquad ,\qquad f_{\mathbf{k}}=t\sum_{j}e^{ik_{j}}
\end{equation}%
where we have set $k_{j}=\mathbf{k.\theta }_{j}$ and for which the
eigenvalues $E_{\mathbf{k}}^{\pm }$ are given by $\pm \sqrt{f_{\mathbf{k}%
}^{\ast }f_{\mathbf{k}}};$ see eq(\ref{LH}) for an interesting concrete
model. For the toy example where we only consider the two first closest
neighbors given by $k_{1}=\mathbf{k.\theta }_{1}=k$ and $k_{2}=\mathbf{%
k.\theta }_{2}=-k,$ we get $f_{\mathbf{k}}=t\cos k$ with $t=t_{1}+it_{2}$
and consequently
\begin{equation}
H_{\mathbf{k}}=\left(
\begin{array}{cc}
0 & t\sin \left( \frac{\pi }{2}-k\right) \\
t^{\ast }\sin \left( \frac{\pi }{2}-k\right) & 0%
\end{array}%
\right)  \label{hk}
\end{equation}%
with eigenvalues
\begin{equation}
E_{\pm }=\pm \sqrt{tt^{\ast }\cos ^{2}k}\qquad ,\qquad -\frac{\pi }{2}\leq
k\leq \frac{3\pi }{2}  \label{kh}
\end{equation}%
For this toy example, the zero modes of $H_{\mathbf{k}}$, which follow from
the solving the condition $f_{\mathbf{k}}=0$, are given by $\kappa _{\pm
}=\pm \frac{\pi }{2}$ $\func{mod}2\pi .$ Around these zeros while setting $%
q=\kappa _{\pm }-k<<1$, the non perturbative Hamiltonian $H_{\mathbf{k}}$
reduces to the $h_{\mathbf{q}}$ given by
\begin{equation}
h_{\mathbf{q}}=\left(
\begin{array}{cc}
0 & tq \\
t^{\ast }q & 0%
\end{array}%
\right) +\mathcal{O}\left( q^{3}\right)
\end{equation}%
which is linear in momentum. The band structure of the tight binding model (%
\ref{hk}-\ref{kh}) is depicted in the Figure \textbf{\ref{g1} }where the
\emph{Dirac-like} cones at $k=n\frac{\pi }{2}$ with $n\in \mathbb{Z}^{\ast }$
are drawn in dashed lines.

\section{Extension to SU(3) Lattices}

\label{sec:3} Construction A of code based NCFTs is deeply connected to the
lattice structures of SU(2) and SO(4)\textrm{\ }Lie algebras.\textrm{\ }This
naturally raises the question of whether similar constructions can be
extended to lattices associated with other Lie algebras like for instance
the rank two SU(3), SO(5) and G$_{2}$ \cite{Y1}-\cite{Y3} . In this section,
we focus on the SU(3) case and apply the algorithm introduced in section 2
to develop a new realisation of NCFTs with central charge $c_{\text{\textsc{l%
}}/\text{\textsc{r}}}=2\mathrm{r}$ ($\mathrm{r\geq 1}$) thus generalising
the previous ($\mathbf{\Lambda }_{\mathrm{k}}^{\mathbf{su}_{2}},\mathbf{%
\Lambda }_{\mathrm{k}\mathcal{C}}^{\mathbf{su}_{3}},\mathbf{\Lambda }_{%
\mathrm{k}}^{\mathbf{su}_{2}\ast }$). We denote the new lattice triplet like
\begin{equation}
(\mathbf{\Lambda }_{\mathrm{k}}^{\mathbf{su}_{3}},\mathbf{\Lambda }_{\mathrm{%
k}\mathcal{C}}^{\mathbf{su}_{3}},\mathbf{\Lambda }_{\mathrm{k}}^{\mathbf{su}%
_{3}\ast })  \label{3la}
\end{equation}%
with the embedding sequence $\mathbf{\Lambda }_{\mathrm{k}}^{\mathbf{su}%
_{3}}\subset \mathbf{\Lambda }_{\mathrm{k}\mathcal{C}}^{\mathbf{su}%
_{3}}\subset \mathbf{\Lambda }_{\mathrm{k}}^{\mathbf{su}_{3}\ast }\subset
\mathbb{R}^{2\mathrm{r},2\mathrm{r}}.$ For r=1, the lattices $\mathbf{%
\Lambda }_{\mathrm{k}}^{\mathbf{su}_{3}}$ and $\mathbf{\Lambda }_{\mathrm{k}%
}^{\mathbf{su}_{3}\ast }$ are 4D, factorising respectively as follows%
\begin{equation}
\mathbf{\Lambda }_{\mathrm{k}}^{\mathbf{su}_{3}}=\text{\textsc{R}}_{\mathrm{k%
}}^{\mathbf{su}_{3}}\times \text{\textsc{R}}_{\mathrm{k}}^{\mathbf{su}%
_{3}\prime }\qquad ,\qquad \mathbf{\Lambda }_{\mathrm{k}}^{\mathbf{su}%
_{3}\ast }=\text{\textsc{W}}_{\mathrm{k}}^{\mathbf{su}_{3}}\times \text{%
\textsc{W}}_{\mathrm{k}}^{\mathbf{su}_{3}\prime }  \label{lsu3}
\end{equation}%
where \textsc{R}$_{\mathrm{k}}^{\mathbf{su}_{3}}$ and \textsc{W}$_{\mathrm{k}%
}^{\mathbf{su}_{3}}$ are the 2D root and weight lattices of SU(3). By using
the isomorphism
\begin{equation}
\text{\textsc{W}}_{\mathrm{3}}^{\mathbf{su}_{3}}/\text{\textsc{R}}_{\mathrm{3%
}}^{\mathbf{su}_{3}}\simeq \mathbb{Z}_{\mathrm{3}}
\end{equation}%
it results that the discriminant $\Lambda _{\mathrm{3}}^{\mathbf{su}_{3}\ast
}/\Lambda _{\mathrm{3}}^{\mathbf{su}_{3}}$ is isomorphic to $\mathbb{Z}_{%
\mathrm{3}}\times \mathbb{Z}_{\mathrm{3}}.$ To build the realisation (\ref%
{3la}), we consider the case k=3, supported by the $\mathbb{Z}_{\mathrm{3}}$%
\textrm{\ }centre of SU(3) and report the more general cases k\TEXTsymbol{>}%
3 and k\TEXTsymbol{<}3 to\textrm{\ appendix C}.

\subsection{Realisation at level\emph{\ }k=3}

\begin{figure}[tbph]
\begin{center}
\includegraphics[width=10cm]{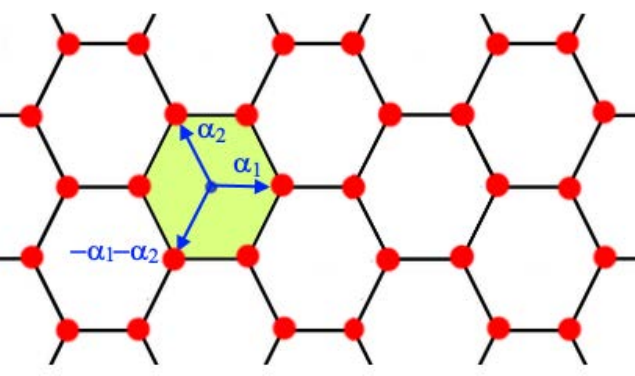}
\end{center}
\par
\vspace{-0.5cm}
\caption{Hexagonal lattice $\mathbb{Z}\mathbf{\protect\alpha }_{1}\oplus
\mathbb{Z}\mathbf{\protect\alpha }_{2}$ generated by the simple roots $%
\mathbf{\protect\alpha }_{1}$ and $\mathbf{\protect\alpha }_{2}$ \textrm{of
SL(3) }with $(\widehat{\mathbf{\protect\alpha }_{1},\mathbf{\protect\alpha }%
_{2}})=2\protect\pi /3.$ It is the root lattice \textsc{R}$_{\mathrm{3}%
}^{sl_{3}}$ of sl(3). The unit cell is in grey color.}
\label{31}
\end{figure}

In light of the factorisation (\ref{lsu3}), our focus narrows to the two
dimensional lattices \textsc{R}$_{\mathrm{3}}^{\mathbf{su}_{3}}$ and \textsc{%
W}$_{\mathrm{3}}^{\mathbf{su}_{3}}$ along with their key properties. The
root lattice \textsc{R}$_{\mathrm{3}}^{\mathbf{su}_{3}}$ is generated by the
two simple roots $\mathbf{\alpha }_{1}$ and $\mathbf{\alpha }_{2}$ \textrm{o}%
f SU(3) and takes the form \textsc{R}$_{\mathrm{3}}^{\mathbf{su}_{3}}=%
\mathbb{Z}\mathbf{\alpha }_{1}\oplus \mathbb{Z}\mathbf{\alpha }_{2}.$ In the
standard Cartesian basis, the simple roots are as follows\textrm{\footnote{%
\ An equivalent basis is given by $\mathbf{\alpha }_{1}=\frac{\sqrt{2}}{2}%
(1,-\sqrt{3})$ and $\mathbf{\alpha }_{2}=\frac{\sqrt{2}}{2}(1,\sqrt{3})$ as
well as $\mathbf{\alpha }_{1}+\mathbf{\alpha }_{2}=\sqrt{2}(0,1).$ We also
have%
\begin{equation*}
A=\frac{\sqrt{2}}{2}\left(
\begin{array}{cc}
1 & 1 \\
-\sqrt{3} & \sqrt{3}%
\end{array}%
\right) ,\qquad B=\frac{\sqrt{2}}{2}\left(
\begin{array}{cc}
1 & -\frac{1}{3}\sqrt{3} \\
1 & \frac{1}{3}\sqrt{3}%
\end{array}%
\right)
\end{equation*}%
}}
\begin{equation}
\mathbf{\alpha }_{1}=\sqrt{2}\left( 1,0\right) ,\qquad \mathbf{\alpha }_{2}=%
\frac{\sqrt{2}}{2}\left( -1,\sqrt{3}\right) ,\qquad \mathbf{\alpha }_{1}+%
\mathbf{\alpha }_{2}=\frac{1}{\sqrt{2}}\left( 1,\sqrt{3}\right)
\label{al12}
\end{equation}%
They satisfy the Euclidean inner product $\mathbf{\alpha }_{i}.\mathbf{%
\alpha }_{j}=K_{ij}$ where $K_{ij}$ is the Cartan matrix. The resulting
\textsc{R}$_{\mathrm{3}}^{\mathbf{su}_{3}}$ is an hexagonal lattice with
angle $(\widehat{\mathbf{\alpha }_{1},\mathbf{\alpha }_{2}})=2\pi /3$ as%
\textrm{\ }depicted by the Figure \textbf{\ref{31} }where the unit cell is
given by the pistachio colored hexagon\textrm{.}
The site vectors $\mathbf{x}\in $\textsc{R}$_{\mathrm{3}}^{\mathbf{su}_{3}}$
have the form $A\mathbf{n}$ with integer $\mathbf{n}=\left(
n_{1},n_{2}\right) ;$ their discrete components $x_{\mathbf{n}}^{i}$ are
given by the linear combinations $\sum A_{j}^{i}n^{j}$ where $A_{j}^{i}$ is
a characteristic matrix whose entries can be identified with the components $%
A_{j}^{i}=\left( \alpha _{j}^{i}\right) $ of the standard vector expansion $%
\mathbf{x}_{\mathbf{n}}=\sum $ $\mathbf{\alpha }_{j}n^{j}.$ Using (\ref{al12}%
), it reads explicitly as follows
\begin{equation}
A_{j}^{i}=\left(
\begin{array}{cc}
\alpha _{1}^{1} & \alpha _{2}^{1} \\
\alpha _{1}^{2} & \alpha _{2}^{2}%
\end{array}%
\right) =\left(
\begin{array}{cc}
\sqrt{2} & -\frac{\sqrt{2}}{2} \\
0 & \frac{\sqrt{6}}{2}%
\end{array}%
\right)   \label{vn}
\end{equation}%
with $\det A=\sqrt{3}>1$. With the help of these relationships, the
euclidian length $\mathbf{x}_{\mathbf{n}}^{2}$ of the site vectors in the
root lattice compute as $\sum n^{i}K_{ij}n^{j}$ with matrix $K=A^{T}A$
coinciding with the Cartan matrix given in (\ref{cart}).

\textbf{The weight lattice} \textsc{W}$_{\mathrm{3}}^{\mathbf{su}_{3}}$

\begin{figure}[tbph]
\begin{center}
\includegraphics[width=10cm]{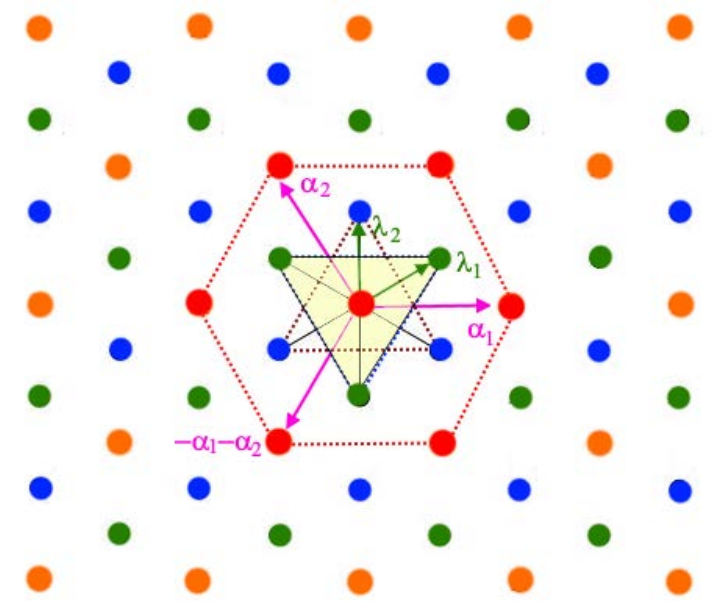}
\end{center}
\par
\vspace{-0.5cm}
\caption{Triangular lattice $\mathbb{Z}\mathbf{\protect\lambda }_{1}\oplus
\mathbb{Z}\mathbf{\protect\lambda }_{2}$ with $(\widehat{\mathbf{\protect%
\lambda }_{1},\mathbf{\protect\lambda }_{2}})=\protect\pi /3.$ It is the
weight lattice \textsc{W}$_{\mathrm{3}}^{\mathbf{su}_{3}}$ of su(3). The
unit cell is a triangle as in pistachio color. Sites in \textsc{W}$_{\mathrm{%
3}}^{\mathbf{su}_{3}}$ are painted in three colors: red, blue and green as
an illustration of the splitting (\protect\ref{416}).}
\label{32}
\end{figure}

Just as with the root lattice \textsc{R}$_{\mathrm{3}}^{\mathbf{su}_{3}}$,
the weight lattice \textsc{W}$_{\mathrm{3}}^{\mathbf{su}_{3}}$ is generated
by the two fundamental weight vectors of SU(3) denoted $\mathbf{\lambda }%
_{1} $ and $\mathbf{\lambda }_{2},$ that is \textsc{W}$_{\mathrm{3}}^{%
\mathbf{su}_{3}}=\mathbb{Z}\mathbf{\lambda }^{1}\oplus \mathbb{Z}\mathbf{%
\lambda }^{2}$ with
\begin{equation}
\mathbf{\lambda }^{1}=\frac{1}{\sqrt{6}}\left( \sqrt{3},1\right) ,\qquad
\mathbf{\lambda }^{2}=\frac{1}{\sqrt{6}}\left( 0,2\right) ,\qquad \mathbf{%
\lambda }^{1}+\mathbf{\lambda }^{2}=\frac{1}{\sqrt{2}}\left( 1,\sqrt{3}%
\right)  \label{W3}
\end{equation}%
and euclidian intersection $\mathbf{\lambda }_{i}.\mathbf{\lambda }_{j}=%
\tilde{K}_{ij}$ where $\tilde{K}_{ij}$ is the inverse of the Cartan matrix.
These weight vectors are related to the simple roots via the duality
relation $\mathbf{\alpha }_{i}.\mathbf{\lambda }^{j}=\delta _{i}^{j}$ which
is solved explicitly as
\begin{equation}
\mathbf{\lambda }_{1}=\frac{1}{3}\left( 2\mathbf{\alpha }_{1}+\mathbf{\alpha
}_{2}\right) ,\qquad \mathbf{\lambda }_{2}=\frac{1}{3}\left( \mathbf{\alpha }%
_{1}+2\mathbf{\alpha }_{2}\right)
\end{equation}%
The lattice \textsc{W}$_{\mathrm{3}}^{\mathbf{su}_{3}}$ has a triangular
structure with angle $(\widehat{\mathbf{\lambda }_{1},\mathbf{\lambda }_{2}}%
)=\pi /3$; it is depicted\textrm{\ }in Figure \textbf{\ref{32} }where the
unit cell is given by the pistachio colored triangle\textrm{.}
The site vectors $\mathbf{k}_{\mathbf{n}}\in $\textsc{W}$_{\mathrm{3}}^{%
\mathbf{su}_{3}}$ can be written as $B\mathbf{m}$ with integer vector $%
\mathbf{m}=\left( m_{1},m_{2}\right) $ and components ($k_{\mathbf{m}})_{i}$
expressed as $\sum m_{j}B_{i}^{j}$ where\textrm{\ }$B_{i}^{j}=(\lambda
_{i}^{j})$\textrm{\ }defines the characteristic matrix of \textsc{W}$_{%
\mathrm{3}}^{\mathbf{su}_{3}}$ reading explicitly as follows%
\begin{equation}
B_{i}^{j}=\left(
\begin{array}{cc}
\lambda _{1}^{1} & \lambda _{2}^{1} \\
\lambda _{1}^{2} & \lambda _{2}^{2}%
\end{array}%
\right) =\left(
\begin{array}{cc}
\frac{\sqrt{2}}{2} & \frac{\sqrt{6}}{6} \\
0 & \frac{\sqrt{6}}{3}%
\end{array}%
\right)  \label{wn}
\end{equation}%
with $\det B=1/\sqrt{3}<1.$ Due to the duality, $B$ is the inverse of the
root matrix\textrm{\ }$A$. The euclidian length $\mathbf{k}_{\mathbf{m}}^{2}$
of a site vector is computed as $\sum m_{i}\tilde{K}^{ij}m_{j}$ with matrix $%
\tilde{K}=BB^{T}$ coinciding with the inverse of the Cartan matrix. By
replacing into (\ref{vn}, \ref{wn})$,$ we get%
\begin{equation}
\begin{tabular}{lll}
$\left( x_{\mathbf{n}}\right) _{1}$ & $=$ & $\sqrt{2}n_{1}-\frac{\sqrt{2}}{2}%
n_{2}$ \\
$\left( x_{\mathbf{n}}\right) _{2}$ & $=$ & $\frac{\sqrt{6}}{2}n_{2}$%
\end{tabular}%
,\qquad
\begin{tabular}{lll}
$\left( k_{\mathbf{m}}\right) _{1}$ & $=$ & $\frac{\sqrt{2}}{2}m_{1}$ \\
$\left( k_{\mathbf{m}}\right) _{2}$ & $=$ & $\frac{\sqrt{6}}{6}m_{1}+\frac{%
\sqrt{6}}{3}m_{2}$%
\end{tabular}%
\end{equation}%
and in\textrm{\ }$K=A^{T}A$\textrm{\ }as well as $\tilde{K}=BB^{T}:$
\begin{equation}
K=\left(
\begin{array}{cc}
2 & -1 \\
-1 & 2%
\end{array}%
\right) \qquad ,\qquad \tilde{K}=\frac{1}{3}\left(
\begin{array}{cc}
2 & 1 \\
1 & 2%
\end{array}%
\right)  \label{cart}
\end{equation}%
with
\begin{equation}
\begin{tabular}{lllll}
$\mathbf{x}_{\mathbf{n}}^{T}.\mathbf{x}_{\mathbf{n}}$ & $=$ & $2\left[
\left( n_{1}\right) ^{2}+\left( n_{2}\right) ^{2}-n_{1}n_{2}\right] $ & $\in
$ & $2\mathbb{Z}$ \\
$\mathbf{k}_{\mathbf{m}}^{T}.\mathbf{k}_{\mathbf{m}}$ & $=$ & $\frac{2}{3}%
\left[ \left( m_{1}\right) ^{2}+\left( m_{2}\right) ^{2}+m_{1}m_{2}\right] $
& $\in $ & $\frac{2}{3}\mathbb{Z}$ \\
$\mathbf{x}_{\mathbf{n}}^{T}.\mathbf{k}_{\mathbf{m}}$ & $=$ & $%
n_{1}m_{1}+n_{2}m_{2}$ & $\in $ & $\mathbb{Z}$%
\end{tabular}%
\end{equation}%
Notice also that substituting $\mathbf{\lambda }_{1}=\left( 2\mathbf{\alpha }%
_{1}+\mathbf{\alpha }_{2}\right) /3$ and $\mathbf{\lambda }_{2}=\left(
\mathbf{\alpha }_{1}+2\mathbf{\alpha }_{2}\right) /3$ into the linear
combination $\mathbf{k}_{\mathbf{m}}=\sum m_{j}\mathbf{\lambda }^{j},$
yields the following
\begin{equation}
\mathbf{k}_{\mathbf{m}}=\frac{1}{3}\left( 2m_{1}+m_{2}\right) \mathbf{\alpha
}_{1}+\frac{1}{3}\left( m_{1}+2m_{2}\right) \mathbf{\alpha }_{2}
\end{equation}%
Introducing new integers p,q and congruence classes\textrm{\ }$\xi ,\zeta
\in \left\{ 0,1,2\right\} $\textrm{\ }$\func{mod}3$\textrm{\ }via $%
2m_{1}+m_{2}=3p+\xi $ and $m_{1}+2m_{2}=3q+\zeta ,$ we rewrite $\mathbf{k}_{%
\mathbf{m}}$ as
\begin{equation}
\begin{tabular}{lll}
$\mathbf{k}_{\mathbf{m}}$ & $=$ & $\mathbf{k}_{\mathbf{m}}^{0}+\frac{1}{3}%
\left( 2\xi \mathbf{-}\zeta \right) \mathbf{\lambda }_{1}+\frac{1}{3}\left(
2\zeta \mathbf{-}\xi \right) \mathbf{\lambda }_{2}$ \\
$\mathbf{k}_{\mathbf{m}}^{0}$ & $=$ & $p\mathbf{\alpha }_{1}+q\mathbf{\alpha
}_{2}$%
\end{tabular}%
\end{equation}%
where $\mathbf{k}_{\mathbf{m}}^{0}\in $\textsc{R}$_{\mathrm{3}}^{\mathbf{su}%
_{3}}.$ The extra term $\frac{1}{3}\left( 2\xi \mathbf{-}\zeta \right)
\mathbf{\lambda }_{1}+\frac{1}{3}\left( 2\zeta \mathbf{-}\xi \right) \mathbf{%
\lambda }_{2}$ introduces global shifts to the sites of \textsc{R}$_{\mathrm{%
3}}^{\mathbf{su}_{3}}$. Setting $\zeta =2\xi $ and $\xi =\varepsilon $ with $%
\varepsilon =0,1,2$ $\func{mod}3,$ $\mathbf{k}_{\mathbf{m}}\left(
\varepsilon \right) $ simplifies to%
\begin{equation}
\mathbf{k}_{\mathbf{m}}\left( \varepsilon \right) =\mathbf{k}_{\mathbf{m}%
}^{0}+\varepsilon \mathbf{\lambda }_{2}\qquad ,\qquad \varepsilon =0,1,2
\end{equation}%
Thus, the weight lattice\textrm{\ }\textsc{W}$_{\mathrm{3}}^{\mathbf{su}%
_{3}} $ decomposes into three superposed sheets (equivalent classes) indexed
by $\varepsilon :$%
\begin{equation}
\begin{tabular}{|c|c||c|c||c|}
\hline
${\small m}_{1}$ & ${\small m}_{2}$ & sites $\mathbf{k}_{\mathbf{m}}^{0}$ &
sites $\mathbf{k}_{\mathbf{m}}$ & weight lattice \textsc{W}$_{\mathrm{3}}^{%
\mathbf{su}_{3}}$ \\ \hline\hline
$2p-q$ & $2q-p+\varepsilon $ & $p\mathbf{\alpha }_{1}+q\mathbf{\alpha }_{2}$
& $\mathbf{k}_{\mathbf{m}}^{0}+\varepsilon \mathbf{\lambda }_{2}$ & \textsc{R%
}$_{\mathrm{3}}^{\mathbf{su}_{3}}\left[ \varepsilon \right] =$\textsc{R}$_{%
\mathrm{3}}^{\mathbf{su}_{3}}+{\small \varepsilon }\mathbf{\lambda }_{2}$ \\
\hline
\end{tabular}
\label{u1}
\end{equation}%
with
\begin{equation}
\text{\textsc{W}}_{\mathrm{3}}^{\mathbf{su}_{3}}=\cup _{\varepsilon =0}^{2}%
\text{\textsc{R}}_{\mathrm{3}}^{\mathbf{su}_{3}}\left[ \varepsilon \right] =%
\text{\textsc{R}}_{\mathrm{3}}^{\mathbf{su}_{3}}\cup \left\{ \text{\textsc{R}%
}_{\mathrm{3}}^{\mathbf{su}_{3}}+\mathbf{\lambda }_{2}\right\} \cup \left\{
\text{\textsc{R}}_{\mathrm{3}}^{\mathbf{su}_{3}}+2\mathbf{\lambda }%
_{2}\right\}  \label{416}
\end{equation}%
This reflects the quotient \textsc{W}$_{\mathrm{3}}^{\mathbf{su}_{3}}/$%
\textsc{R}$_{\mathrm{3}}^{\mathbf{su}_{3}}\simeq \mathbb{Z}_{3}$ where the
cyclic action permutes the three sheets.

\subsection{The triplet $(\mathbf{\Lambda }^{\mathbf{su}_{3}},\mathbf{%
\Lambda }_{\mathcal{C}}^{\mathbf{su}_{3}},\mathbf{\Lambda }^{\ast \mathbf{su}%
_{3}})$}

\begin{figure}[tbph]
\begin{center}
\includegraphics[width=8cm]{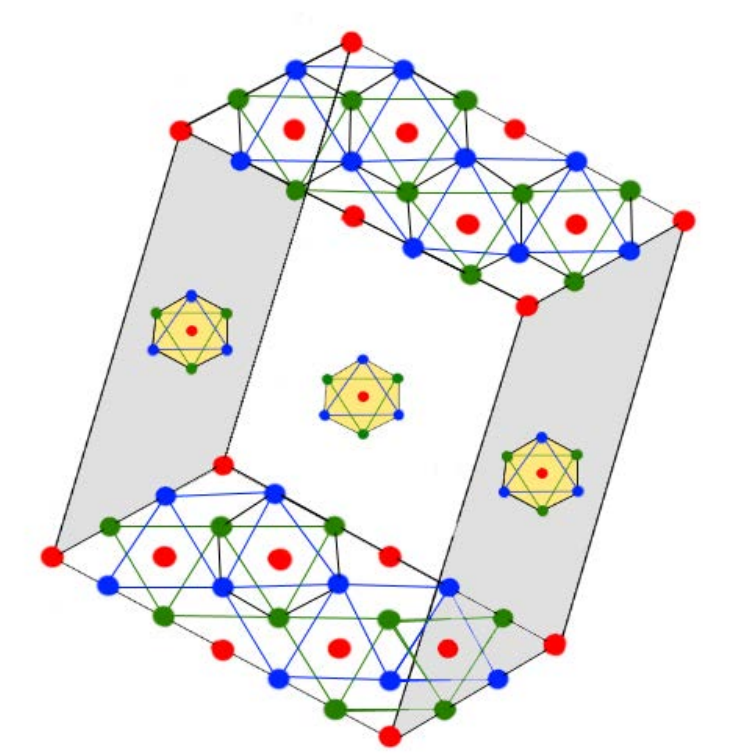}
\end{center}
\par
\vspace{-0.5cm}
\caption{Schematic representation of the 4D lattice $\mathbf{\Lambda }_{%
\mathrm{3}}^{sl_{3}\ast }$ in 3D space by stacking sheets of weight lattices
in the direction normal to the plane of \textsc{W}$_{\mathrm{3}}^{sl_{3}}$.
The 2D base is given by the weight lattice $\mathbb{Z}\mathbf{\protect%
\lambda }_{1}\oplus \mathbb{Z}\mathbf{\protect\lambda }_{2}.$}
\label{33}
\end{figure}

Using the above results and the construction (\ref{lsu3}), we find that in
addition to the root lattice $\mathbf{\Lambda }_{\mathrm{3}}^{\mathbf{su}%
_{3}}=$\textsc{R}$_{\mathrm{3}}^{\mathbf{su}_{3}}\times $\textsc{R}$_{%
\mathrm{3}}^{\mathbf{su}_{3}\prime }$, we have the four dimensional dual
weight given by $\mathbf{\Lambda }_{\mathrm{3}}^{\ast \mathbf{su}_{3}}=$%
\textsc{W}$_{\mathrm{3}}^{\mathbf{su}_{3}}\times $\textsc{W}$_{\mathrm{3}}^{%
\mathbf{su}_{3}\prime }$ which splits into nine sheets as follows
\begin{equation}
\mathbf{\Lambda }_{\mathrm{3}}^{\ast \mathbf{su}_{3}}=\dbigcup\limits_{%
\varepsilon ,\varepsilon ^{\prime }=0}^{3}\text{\textsc{R}}_{\mathrm{3}}^{%
\mathbf{su}_{3}}\left[ \varepsilon \right] \times \text{\textsc{R}}_{\mathrm{%
3}}^{\mathbf{su}_{3}}\left[ \varepsilon ^{\prime }\right]
\end{equation}%
Each sheet is isomorphic to $\mathbf{\Lambda }_{\mathrm{3}}^{\mathbf{su}%
_{3}} $ of $SU_{3}\times SU_{3}$ with metric $\eta $. Explicitly,\textrm{\ }%
we have%
\begin{equation}
\begin{tabular}{lll}
$\mathbf{\Lambda }_{\mathrm{3}}^{\ast \mathbf{su}_{3}}$ & $=$ & $\left\{
\mathbf{\Lambda }_{\mathrm{3}}^{\mathbf{su}_{3}}\right\} \cup \left\{
\mathbf{\Lambda }_{\mathrm{3}}^{\mathbf{su}_{3}}+\mathbf{\lambda }_{2}+%
\mathbf{\lambda }_{2}^{\prime }\right\} \cup \left\{ \mathbf{\Lambda }_{%
\mathrm{3}}^{\mathbf{su}_{3}}+\left( 2\mathbf{\lambda }_{2}+2\mathbf{\lambda
}_{2}^{\prime }\right) \right\} \cup $ \\
&  & $\left\{ \mathbf{\Lambda }_{\mathrm{3}}^{\mathbf{su}_{3}}+\mathbf{%
\lambda }_{2}\right\} \cup \left\{ \mathbf{\Lambda }_{\mathrm{3}}^{\mathbf{su%
}_{3}}+2\mathbf{\lambda }_{2}\right\} \cup $ \\
&  & $\left\{ \mathbf{\Lambda }_{\mathrm{3}}^{\mathbf{su}_{3}}+\mathbf{%
\lambda }_{2}^{\prime }\right\} \cup \left\{ \mathbf{\Lambda }_{\mathrm{3}}^{%
\mathbf{su}_{3}}+2\mathbf{\lambda }_{2}^{\prime }\right\} \cup $ \\
&  & $\left\{ \mathbf{\Lambda }_{\mathrm{3}}^{\mathbf{su}_{3}}+2\mathbf{%
\lambda }_{2}+\mathbf{\lambda }_{2}^{\prime }\right\} \cup \left\{ \mathbf{%
\Lambda }_{\mathrm{3}}^{\mathbf{su}_{3}}+\mathbf{\lambda }_{2}+2\mathbf{%
\lambda }_{2}^{\prime }\right\} $%
\end{tabular}%
\end{equation}%
A schematic illustration of the 4D lattice $\mathbf{\Lambda }_{\mathrm{3}%
}^{\ast \mathbf{su}_{3}}=$\textsc{W}$_{\mathrm{3}}^{\mathbf{su}_{3}}\times $%
\textsc{W}$_{\mathrm{3}}^{\mathbf{su}_{3}}$ embedded in 3D space is shown in
Figure \textbf{\ref{33}} where the base plane is spanned by \textsc{W}$_{%
\mathrm{3}}^{\mathbf{su}_{3}}$ given by $\mathbb{Z}\mathbf{\lambda }%
_{1}\oplus \mathbb{Z}\mathbf{\lambda }_{2}.$
The even self dual lattice $\mathbf{\Lambda }_{\mathrm{3}\mathcal{C}}^{%
\mathbf{su}_{3}}$ is constructed as
\begin{equation}
\mathbf{\Lambda }_{\mathrm{3}\mathcal{C}}^{\mathbf{su}_{3}}=\mathbf{\Lambda }%
_{\mathrm{3}}^{\mathbf{su}_{3}}\cup \left[ \mathbf{\Lambda }_{\mathrm{3}}^{%
\mathbf{su}_{3}}+\mathbf{\lambda }_{2}\right] \cup \left[ \mathbf{\Lambda }_{%
\mathrm{3}}^{\mathbf{su}_{3}}+2\mathbf{\lambda }_{2}\right]
\end{equation}%
corresponding to a three sheet structure, each isomorphic to $\mathbf{%
\Lambda }_{\mathrm{3}}^{\mathbf{su}_{3}}$. The characteristic 4$\times $4
matrices of the three lattices ($\mathbf{\Lambda }_{\mathrm{3}}^{\mathbf{su}%
_{3}},\mathbf{\Lambda }_{\mathrm{3}\mathcal{C}}^{\mathbf{su}_{3}},\mathbf{%
\Lambda }_{\mathrm{3}}^{\ast \mathbf{su}_{3}}$)\ are as follows%
\begin{equation}
\Lambda _{\mathrm{3}}^{\mathbf{su}_{3}}=\left(
\begin{array}{cc}
A & 0 \\
0 & A%
\end{array}%
\right) ,\qquad \Lambda _{\mathrm{3}}^{\ast \mathbf{su}_{3}}=\left(
\begin{array}{cc}
B & 0 \\
0 & B%
\end{array}%
\right) ,\qquad \Lambda _{\mathrm{3}\mathcal{C}}^{\mathbf{su}_{3}}=\left(
\begin{array}{cc}
A & 0 \\
0 & B%
\end{array}%
\right)  \label{su3l}
\end{equation}%
with determinants\textrm{\ }$\det \Lambda _{\mathrm{3}}^{\mathbf{su}_{3}}=3$%
, $\det \Lambda _{\mathrm{3}}^{\ast \mathbf{su}_{3}}=1/3$ and $\det \Lambda
_{\mathrm{3}\mathcal{C}}^{\mathbf{su}_{3}}=1.$ The induced metrics for%
\textrm{\ }the various lattices are defined like $g_{\Lambda _{\mathrm{3}%
}}=(\Lambda _{\mathrm{3}}^{\mathbf{su}_{3}})^{T}\mathbf{\eta }\Lambda _{%
\mathrm{3}}^{\mathbf{su}_{3}}$ and $g_{\Lambda _{\mathrm{3}}^{\ast
}}=(\Lambda _{\mathrm{3}}^{\ast \mathbf{su}_{3}})^{T}\mathbf{\eta }\Lambda _{%
\mathrm{3}}^{\ast \mathbf{su}_{3}}$\ as well as $g_{\Lambda _{\mathrm{3}%
\mathcal{C}}}=(\Lambda _{\mathrm{3}\mathcal{C}}^{\mathbf{su}_{3}})^{T}%
\mathbf{\eta }\Lambda _{\mathrm{3}\mathcal{C}}^{\mathbf{su}_{3}},$\ they are
given by
\begin{equation}
g_{\Lambda _{\mathrm{3}}}=\left(
\begin{array}{cc}
0 & K \\
K & 0%
\end{array}%
\right) ,\qquad g_{\Lambda _{\mathrm{3}}^{\ast }}=\left(
\begin{array}{cc}
0 & K^{-1} \\
K^{-1} & 0%
\end{array}%
\right) ,\qquad g_{\Lambda _{\mathrm{3}\mathcal{C}}}=\left(
\begin{array}{cc}
0 & I_{2} \\
I_{2} & 0%
\end{array}%
\right)
\end{equation}%
with $\det g_{\Lambda _{\mathrm{3}}^{\mathbf{su}_{{\small 3}}}}=3^{2}$, $%
\det g_{\Lambda _{\mathrm{3}}^{\ast \mathbf{su}_{{\small 3}}}}=1/3^{2}$ and $%
\det \left( g_{\Lambda _{\mathrm{3}\mathcal{C}}}\right) =1.$ Based on (\ref%
{su3l}), and using the construction (\ref{csa}) of code-CFTs with $c_{\text{%
\textsc{l}/\textsc{r}}}=2\mathrm{r\geq 2}$, the code $\mathcal{C}$ is
contained into $\left( \mathbb{Z}_{3}\times \mathbb{Z}_{3}\right) ^{\mathrm{r%
}}$ and maps to a 4r dimensional lattice $\mathbf{\Lambda }_{\mathrm{3}%
\mathcal{C}}^{{\small \mathrm{r,r}}}$ satisfying the inclusions
\begin{equation}
\underbrace{\Lambda _{\mathrm{3}}^{\mathbf{su}_{3}}\oplus ...\oplus \Lambda
_{\mathrm{3}}^{\mathbf{su}_{3}}}_{\mathrm{r}\text{ \textrm{times}}}\text{ }%
\subset \text{ }\mathbf{\Lambda }_{\mathrm{3}\mathcal{C}}^{{\small \mathrm{%
r,r}}}\text{ }\subset \text{ }\underbrace{\Lambda _{\mathrm{3}}^{\ast
\mathbf{su}_{3}}\oplus ...\oplus \Lambda _{\mathrm{3}}^{\ast \mathbf{su}_{3}}%
}_{2\mathrm{r}\text{ \textrm{times}}}\text{ }\subset \text{ }\mathbb{R}^{2%
\mathrm{r},2\mathrm{r}}
\end{equation}%
This lattice $\mathbf{\Lambda }_{\mathrm{3}\mathcal{C}}^{{\small \mathrm{r,r}%
}}$ is even and self dual, characterized by the matrices:%
\begin{equation}
\Lambda _{\mathrm{3}\mathcal{C}}^{{\small \mathrm{r,r}}}=\left(
\begin{array}{cc}
\oplus _{i=1}^{\mathrm{r}}A & 0 \\
0 & \oplus _{i=1}^{\mathrm{r}}B%
\end{array}%
\right) ,\qquad g_{\Lambda _{\mathrm{3}\mathcal{C}}}^{{\small [2\mathrm{r}]}%
}=\left(
\begin{array}{cc}
0 & I_{2\mathrm{r}} \\
I_{2\mathrm{r}} & 0%
\end{array}%
\right)
\end{equation}%
with determinants $\det \Lambda _{\mathrm{3}\mathcal{C}}^{{\small \mathrm{r,r%
}}}=1$ and $\det \Lambda _{\mathrm{3}\mathcal{C}}^{\ast {\small \mathrm{r,r}}%
}=1.$ Thus, $\mathbf{\Lambda }_{\mathrm{3}\mathcal{C}}^{{\small \mathrm{r,r}}%
}$ defines a Narain theory with central charge 2$\mathrm{r}$.

\section{Particles in NCFTs and path to topological matter}

\label{sec:4} 
\begin{figure}[tbph]
\begin{center}
\includegraphics[width=12cm]{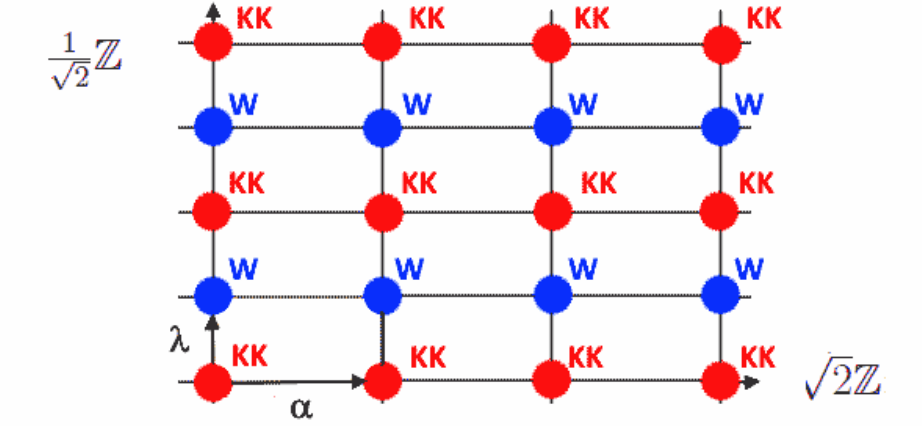}
\end{center}
\caption{The 2D even self dual lattice $\mathbf{\Lambda }_{2\mathcal{C}}^{%
\mathrm{1,1}}\simeq (\protect\sqrt{2}\mathbb{Z})\times (\frac{1}{\protect%
\sqrt{2}}\mathbb{Z})$ with red sites occupied by KKs and blue sites by
windings. A red (blue) site has two first closed neighbors of blue (red)
color; this is because $\protect\alpha =2\protect\lambda $. The area of the
unit cell is equal to 1.}
\label{51}
\end{figure}

\qquad The present and following section lay out a proposal to
extend the holographic ensemble framework to discrete lattice QFT with
applications to topological phases of matter featuring Dirac points with
fermionic zero modes \textrm{\cite{AZ}-\cite{cr1}}. In this context,\textrm{%
\ }Code and Narain CFTs are imagined as sitting at critical points of
quantum fields $\phi \left( \mathbf{r}_{\mathbf{m}}\right) $ populating the
sites of the lattice $\mathbf{\Lambda }_{\mathcal{C}}$ with position vectors
$\mathbf{r}_{\mathbf{m}}=x_{\mathbf{m}}^{a}\epsilon _{a}$ and Fourier
expansions
\begin{equation}
\phi \left( \mathbf{r}_{\mathbf{m}}\right) \simeq \dsum\limits_{\mathbf{k}}%
\mathbf{e}^{i\mathbf{k.r}_{\mathbf{m}}}\tilde{\phi}_{\mathbf{k}}\qquad
,\qquad \tilde{\phi}_{\mathbf{k}}\simeq \dsum\limits_{\mathbf{r}_{\mathbf{m}%
}}\mathbf{e}^{-i\mathbf{k.r}_{\mathbf{m}}}\phi \left( \mathbf{r}_{\mathbf{m}%
}\right)
\end{equation}%
Here the waves $\phi \left( \mathbf{r}_{\mathbf{m}}\right) $ will be
interpreted as KK particles $\left\vert KK_{\mathbf{m}}\right\rangle $ and
winding states $\left\vert W_{\mathbf{m}}\right\rangle $ sitting in $\mathbf{%
\Lambda }_{\mathcal{C}}$ as illustrated in Figure \textbf{\ref{51}}.
Our proposal borrows ideas from tight binding constructions of particles on
lattices to describe critical properties of KKs and winding states. A more
detailed treatment of this tight binding will be developed in \textrm{%
section 5} with explicit computations. Meanwhile, we lay here the groundwork
by studying first the energy and the momentum properties of particle states
in code-CFTs characterised by the triplet ($\mathbf{\Lambda }_{\mathrm{k}},%
\mathbf{\Lambda }_{\mathrm{k}\mathcal{C}},\mathbf{\Lambda }_{\mathrm{k}%
}^{\ast }$) realised in terms of the lattices of SU(2). After that, we
extend this construction to the SU(3) based CFT.

\subsection{Particles in SU(2) based model}

We now investigate the quantum particles in the code CFT description of the
model with central anomaly\textrm{\ }c$_{\text{\textsc{l}/\textsc{r}}}$=1,
corresponding to an SU(2) CS theory in the bulk. It consists of left and
right moving particle states $\left\vert P_{\text{\textsc{l}}},P_{\text{%
\textsc{r}}}\right\rangle $ having quantized left/right moving momenta given
by (\ref{pp}) namely \textrm{\cite{dec}}

\begin{equation}
\begin{tabular}{lll}
$P_{\text{\textsc{l}}}$ & $=$ & $\frac{1}{\sqrt{2\mathrm{k}}}\left( \frac{1}{%
\text{\textsc{r}}\sqrt{2}}\left[ a+\mathrm{k}n\right] +\sqrt{2}\text{\textsc{%
r}}\left[ b+\mathrm{k}m\right] \right) $ \\
$P_{\text{\textsc{r}}}$ & $=$ & $\frac{1}{\sqrt{2\mathrm{k}}}\left( \frac{1}{%
\text{\textsc{r}}\sqrt{2}}\left[ a+\mathrm{k}n\right] -\sqrt{2}\text{\textsc{%
r}}\left[ b+\mathrm{k}m\right] \right) $%
\end{tabular}
\label{2p}
\end{equation}%
These momenta are labeled by two pairs of quantum numbers namely: $\left(
\mathbf{i}\right) $ the pair $(a,b)$ with $a,b=0,...,\mathrm{k}-1$ defining
the ground configurations; and $\left( \mathbf{ii}\right) $ the pair $%
(n,m)\in \mathbb{Z}\times \mathbb{Z}$ encoding the degrees of KK and winding
mode excitations. For convenience, we rewrite these momenta like%
\begin{equation}
\begin{tabular}{lll}
$\left( P_{\text{\textsc{l}}}\right) _{\bar{a}\bar{b}}$ & $=$ & $\frac{1}{%
\sqrt{2\mathrm{k}}}\left( \frac{1}{\text{\textsc{r}}\sqrt{2}}\bar{a}+\sqrt{2}%
\text{\textsc{r}}\bar{b}\right) $ \\
$\left( P_{\text{\textsc{r}}}\right) _{\bar{a}\bar{b}}$ & $=$ & $\frac{1}{%
\sqrt{2\mathrm{k}}}\left( \frac{1}{\text{\textsc{r}}\sqrt{2}}\bar{a}-\sqrt{2}%
\text{\textsc{r}}\bar{b}\right) $%
\end{tabular}
\label{2q}
\end{equation}%
where we have set $\bar{a}=a+\mathrm{k}n$ and $\bar{b}=b+\mathrm{k}m$ (i.e: $%
\bar{a},\bar{b}\in \mathbb{Z}_{\mathrm{k}}$). From the above expressions,
several interesting features emerge. In particular, we highlight the
following four:

\begin{description}
\item[$\left( \mathbf{i}\right) $] \textbf{Hilbert space of zero modes:}
\newline
By setting $n=m=0$ in (\ref{2q}), we find k$^{2}$ left and k$^{2}$ right
moving ground states with momenta $\left( \mathring{p}_{\text{\textsc{l}/%
\textsc{r}}}\right) _{ab}$. They are given by
\begin{equation}
\begin{tabular}{lll}
$\left( \mathring{p}_{\text{\textsc{l}}}\right) _{ab}$ & $=$ & $\frac{1}{%
\sqrt{2\mathrm{k}}}\left( \frac{1}{\text{\textsc{r}}\sqrt{2}}a+\sqrt{2}\text{%
\textsc{r}}b\right) $ \\
$\left( \mathring{p}_{\text{\textsc{r}}}\right) _{ab}$ & $=$ & $\frac{1}{%
\sqrt{2\mathrm{k}}}\left( \frac{1}{\text{\textsc{r}}\sqrt{2}}a-\sqrt{2}\text{%
\textsc{r}}b\right) $%
\end{tabular}%
\end{equation}%
These can be interpreted as classical charges defining distinct vacuum
states.

\item[$\left( \mathbf{ii}\right) $] \textbf{Excitations:} \newline
The full\textrm{\ }momenta in eq(\ref{2p}) split into two sectors: the
ground sector defining ground configurations with $\left( \mathring{p}_{%
\text{\textsc{l}/\textsc{r}}}\right) _{ab}$ and a second sector for the
excitations having momenta $\left( p_{\text{\textsc{l}/\textsc{r}}}\right)
_{nm}$ as follows
\begin{equation}
\begin{tabular}{lll}
$P_{\text{\textsc{l}}}$ & $=$ & $\mathring{p}_{\text{\textsc{l}}}+p_{\text{%
\textsc{l}}}$ \\
$P_{\text{\textsc{r}}}$ & $=$ & $\mathring{p}_{\text{\textsc{r}}}+p_{\text{%
\textsc{r}}}$%
\end{tabular}%
,\qquad
\begin{tabular}{lll}
$\left( p_{\text{\textsc{l}}}\right) _{nm}$ & $=$ & $\frac{\mathrm{k}}{\sqrt{%
2\mathrm{k}}}\left( \frac{1}{\text{\textsc{r}}\sqrt{2}}n+\sqrt{2}\text{%
\textsc{r}}m\right) $ \\
$\left( p_{\text{\textsc{r}}}\right) _{nm}$ & $=$ & $\frac{\mathrm{k}}{\sqrt{%
2\mathrm{k}}}\left( \frac{1}{\text{\textsc{r}}\sqrt{2}}n-\sqrt{2}\text{%
\textsc{r}}m\right) $%
\end{tabular}%
\end{equation}

\item[$\left( \mathbf{iii}\right) $] \textbf{Energy and momentum}: \newline
Because the left and the right $P_{\text{\textsc{l}}/\text{\textsc{r}}}$ are
linear combinations, the KK and the windings can be decoupled via a change
of variables%
\begin{equation}
\begin{tabular}{lll}
$H$ & $=$ & $\frac{1}{\sqrt{2}}\left( P_{\text{\textsc{l}}}+P_{\text{\textsc{%
r}}}\right) $ \\
$P$ & $=$ & $\frac{1}{\sqrt{2}}\left( P_{\text{\textsc{l}}}-P_{\text{\textsc{%
r}}}\right) $%
\end{tabular}%
\end{equation}%
with
\begin{equation}
\begin{tabular}{lll}
$H$ & $=$ & $\mathring{h}+h$ \\
$P$ & $=$ & $\mathring{p}+p$%
\end{tabular}%
\end{equation}%
and%
\begin{equation}
\begin{tabular}{lll}
$\mathring{h}$ & $=$ & $\frac{a}{\sqrt{2\mathrm{k}}}\frac{1}{\text{\textsc{r}%
}\sqrt{2}}$ \\
$\mathring{p}$ & $=$ & $\frac{b}{\sqrt{2\mathrm{k}}}\sqrt{2}$\textsc{r}%
\end{tabular}%
\qquad ,\qquad
\begin{tabular}{lll}
$h$ & $=$ & $\frac{\mathrm{k}n}{\sqrt{2\mathrm{k}}}\frac{1}{\text{\textsc{r}}%
\sqrt{2}}$ \\
$p$ & $=$ & $\frac{\mathrm{k}m}{\sqrt{2\mathrm{k}}}\sqrt{2}$\textsc{r}%
\end{tabular}%
\end{equation}

\item[$\left( \mathbf{iv}\right) $] \textbf{Topological index}: \newline
Using eqs(\ref{2p}-\ref{2q}), we\ introduce a "topological index" $\left( I_{%
\mathrm{k}}\right) _{\text{\textsc{lr}}}$ given by $P_{\text{\textsc{l}}%
}^{2}-P_{\text{\textsc{r}}}^{2}$ which is globally defined on the moduli
space of the NCFT. It factorises like $\left( P_{\text{\textsc{l}}}-P_{\text{%
\textsc{r}}}\right) \left( P_{\text{\textsc{l}}}+P_{\text{\textsc{r}}%
}\right) $ and evaluates to
\begin{equation}
\begin{tabular}{lll}
$\left( I_{\mathrm{k}}\right) _{\text{\textsc{lr}}}$ & $=$ & $\frac{2}{%
\mathrm{k}}\bar{a}\bar{b}$ \\
& $=$ & $\frac{2}{\mathrm{k}}\left( a+\mathrm{k}n\right) \left( b+\mathrm{k}%
m\right) $%
\end{tabular}%
\end{equation}%
Remarkably, this index is independent of the parameter \textsc{r}
coordinating the moduli space of NCFT. By substituting $\bar{a}=a+\mathrm{k}n
$ and $\bar{b}=b+\mathrm{k}m,$ this index expands in terms of the CS level
as follows%
\begin{equation}
\left( I_{\mathrm{k}}\right) _{\text{\textsc{lr}}}=\frac{2}{\mathrm{k}}%
ab+2\left( am+bn\right) +\left( 2\mathrm{k}\right) nm  \label{ind}
\end{equation}
\end{description}

Notice that for $\mathrm{k=1,}$ it reduces to the usual $\left( I_{\mathrm{1}%
}\right) _{\text{\textsc{lr}}}=2nm\in 2\mathbb{Z}$ and for $\mathrm{k=2,}$
we have $\left( I_{\mathrm{2}}\right) _{\text{\textsc{lr}}}=ab+2\left(
am+bn\right) +4nm.$ And because $\left( I_{\mathrm{k}}\right) _{\text{%
\textsc{lr}}}$ is a globally defined feature, we may fix the parameter
\textsc{r} for convenience and set for example \textsc{r}$=\frac{1}{\sqrt{2%
\mathrm{k}}}.$

\subsubsection{Exhibiting the SU(2) symmetry}

The above relations characterising the left/right momenta $P_{\text{\textsc{l%
}/\textsc{r}}}$ admit a natural lattice description in terms of the SU(2)
root and weight systems. They can be imagined as vectors $\boldsymbol{P}_{%
\text{\textsc{l}/\textsc{r}}}$ valued in the root R$_{\mathrm{k}}^{\mathbf{su%
}_{2}}$ and weight W$_{\mathrm{k}}^{\mathbf{su}_{2}}$ lattices of SU(2) as
follows
\begin{equation}
\boldsymbol{P}_{\text{\textsc{l}/\textsc{r}}}\sim n\mathbf{\alpha }\pm m%
\mathbf{\lambda }\quad \sim \quad z_{\pm }\mathbf{\lambda }\qquad ,\qquad
z_{\pm }=2n\pm m\quad \in \quad \mathbb{Z}
\end{equation}%
Recall that \textsc{R}$_{\mathrm{k}}^{\mathbf{su}_{2}}$ and \textsc{W}$_{%
\mathrm{k}}^{\mathbf{su}_{2}}$\ are respectively generated by the simple
root $\mathbf{\alpha }$ and the fundamental $\mathbf{\lambda }$ satisfying
the duality relation $\mathbf{\alpha }.\mathbf{\lambda }=1$ with norms $%
\mathbf{\alpha }^{2}=2$ and $\mathbf{\lambda }^{2}=1/2$. For simplicity, we
may identify these vectors like $\mathbf{\alpha }=\sqrt{2}$ and $\mathbf{%
\lambda }=1/\sqrt{2}$ in which case the momenta (\ref{2p}) take the form:%
\begin{equation}
\begin{tabular}{lll}
$\boldsymbol{P}_{\text{\textsc{l}}}$ & $=$ & $\frac{1}{\sqrt{2\mathrm{k}}}%
\left( \frac{1}{2\text{\textsc{r}}}\left[ a+\mathrm{k}n\right] \mathbf{%
\alpha }+2\text{\textsc{r}}\left[ b+\mathrm{k}m\right] \mathbf{\lambda }%
\right) $ \\
$\boldsymbol{P}_{\text{\textsc{r}}}$ & $=$ & $\frac{1}{\sqrt{2\mathrm{k}}}%
\left( \frac{1}{2\text{\textsc{r}}}\left[ a+\mathrm{k}n\right] \mathbf{%
\alpha }-2\text{\textsc{r}}\left[ b+\mathrm{k}m\right] \mathbf{\lambda }%
\right) $%
\end{tabular}%
\end{equation}%
Moreover, by setting $\mathbf{\tilde{\alpha}}=\frac{1}{2\text{\textsc{r}}}%
\mathbf{\alpha }$ and $\mathbf{\tilde{\lambda}}=2$\textsc{r}$\mathbf{\lambda
},$\textrm{\ }the momenta written as%
\begin{equation}
\begin{tabular}{lll}
$\boldsymbol{P}_{\text{\textsc{l}}}$ & $=$ & $\frac{1}{\sqrt{2\mathrm{k}}}%
\left( \left[ a+\mathrm{k}n\right] \mathbf{\tilde{\alpha}}+\left[ b+\mathrm{k%
}m\right] \mathbf{\tilde{\lambda}}\right) $ \\
$\boldsymbol{P}_{\text{\textsc{r}}}$ & $=$ & $\frac{1}{\sqrt{2\mathrm{k}}}%
\left( \left[ a+\mathrm{k}n\right] \mathbf{\tilde{\alpha}}-\left[ b+\mathrm{k%
}m\right] \mathbf{\tilde{\lambda}}\right) $%
\end{tabular}
\label{PPP}
\end{equation}%
with $\mathbf{\tilde{\alpha}.\tilde{\lambda}}=1$ and $\mathbf{\tilde{\alpha}}%
^{2}=\frac{1}{2\text{\textsc{r}}^{2}}$ as well as $\mathbf{\tilde{\lambda}}%
^{2}=2$\textsc{r}$^{2}.$ Using \textsc{r}$=\frac{1}{\sqrt{2\mathrm{k}}},$ we
get%
\begin{equation}
\mathbf{\tilde{\alpha}}=\sqrt{\frac{\mathrm{k}}{2}}\mathbf{\alpha }\qquad
,\qquad \mathbf{\tilde{\lambda}}=\sqrt{\frac{2}{\mathrm{k}}}\mathbf{\lambda }%
\qquad ,\qquad \mathbf{\tilde{\alpha}.\tilde{\lambda}}=1
\end{equation}%
showing that for k=2, we have $\mathbf{\tilde{\alpha}}=\mathbf{\alpha }$ and
$\mathbf{\tilde{\lambda}}=\mathbf{\lambda }.$ Putting these changes into the
above relationship, we end up with%
\begin{equation}
\begin{tabular}{lll}
$\boldsymbol{\mathring{h}}$ & $=$ & $\frac{a}{\sqrt{2\mathrm{k}}}\mathbf{%
\tilde{\alpha}}$ \\
$\boldsymbol{\mathring{p}}$ & $=$ & $\frac{b}{\sqrt{2\mathrm{k}}}\mathbf{%
\tilde{\lambda}}$%
\end{tabular}%
\qquad ,\qquad
\begin{tabular}{lll}
$\boldsymbol{h}$ & $=$ & $\frac{\mathrm{k}n}{\sqrt{2\mathrm{k}}}\mathbf{%
\tilde{\alpha}}$ \\
$\boldsymbol{p}$ & $=$ & $\frac{\mathrm{k}m}{\sqrt{2\mathrm{k}}}\mathbf{%
\tilde{\lambda}}$%
\end{tabular}%
\end{equation}%
and
\begin{equation}
\begin{tabular}{lllll}
$\boldsymbol{H}$ & $=$ & $\frac{1}{\sqrt{2\mathrm{k}}}\bar{a}\mathbf{\tilde{%
\alpha}}$ & $=$ & $\frac{1}{\sqrt{2\mathrm{k}}}\left( a+\mathrm{k}n\right)
\mathbf{\tilde{\alpha}}$ \\
$\boldsymbol{P}$ & $=$ & $\frac{b}{\sqrt{2\mathrm{k}}}\bar{b}\mathbf{\tilde{%
\lambda}}$ & $=$ & $\frac{b}{\sqrt{2\mathrm{k}}}\left( b+\mathrm{k}m\right)
\mathbf{\tilde{\lambda}}$%
\end{tabular}
\label{HP}
\end{equation}%
and subsequently%
\begin{equation}
\begin{tabular}{lll}
$\boldsymbol{P}_{\text{\textsc{l}}}$ & $=$ & $\frac{1}{\sqrt{2\mathrm{k}}}%
\left( \bar{a}\mathbf{\tilde{\alpha}}+\bar{b}\mathbf{\tilde{\lambda}}\right)
$ \\
& $=$ & $\frac{1}{\sqrt{2\mathrm{k}}}\left( \left[ a+\mathrm{k}n\right]
\mathbf{\tilde{\alpha}}+\left[ b+\mathrm{k}m\right] \mathbf{\tilde{\lambda}}%
\right) $ \\
$\boldsymbol{P}_{\text{\textsc{r}}}$ & $=$ & $\frac{1}{\sqrt{2\mathrm{k}}}%
\left( \bar{a}\mathbf{\tilde{\alpha}}-\bar{b}\mathbf{\tilde{\lambda}}\right)
$ \\
& $=$ & $\frac{1}{\sqrt{2\mathrm{k}}}\left( \left[ a+\mathrm{k}n\right]
\mathbf{\tilde{\alpha}}\left[ b+\mathrm{k}m\right] \mathbf{\tilde{\lambda}}%
\right) $%
\end{tabular}
\label{pq}
\end{equation}%
\begin{equation*}
\end{equation*}%
In this relation, T-duality acts through the change $\left( a,n,\mathbf{%
\tilde{\alpha}}\right) \leftrightarrow (b,m,\mathbf{\tilde{\lambda}});$ it
leaves invariant the spectrum of the left momenta $\boldsymbol{P}_{\text{%
\textsc{l}}}$ while flipping the sign of $\boldsymbol{P}_{\text{\textsc{r}}}$%
. Here are some noteworthy features that arise:

\begin{itemize}
\item \textbf{Ground configurations}: The charge momenta $(\boldsymbol{%
\mathring{h}})_{a}$ and $(\boldsymbol{\mathring{p}})_{b}$ of the ground
configurations span a k- dimensional space. The associated Hilbert space can
be generated by classical ground plane waves $\psi _{a}=e^{i\boldsymbol{%
\mathring{h}}_{a}.\boldsymbol{\mathring{x}}}$ and $\tilde{\psi}_{b}=e^{i%
\boldsymbol{\mathring{p}}_{a}.\boldsymbol{\mathring{y}}}$\ with position%
\textrm{\ }variables $\boldsymbol{\mathring{x}}=\mathring{x}\mathbf{\tilde{%
\lambda}}$ and $\boldsymbol{\mathring{y}}=\mathring{y}\mathbf{\tilde{\alpha}.%
}$ These waves combine into 2j+1 components with label $-j\leq a\leq j$
forming\textrm{\ }an SU(2) representation of isospin
\begin{equation}
j=\frac{\mathrm{k}-1}{2}  \label{js}
\end{equation}%
which for \textrm{k=1} gives an isosinglet with j=0 and for\textrm{\ k=2, }a
doublet with j=1/2.

\item \textbf{Interpolating vacua }$\psi _{a}\pm \tilde{\psi}_{b}$: these
describe\textrm{\ }correlated ground configurations corresponding to the
states with $n=m=0.$ For these values, the eqs(\ref{PPP}) reduce to%
\begin{equation}
\begin{tabular}{lll}
$\boldsymbol{\mathring{p}}_{\text{\textsc{l}}}$ & $=$ & $\frac{1}{\sqrt{2%
\mathrm{k}}}\left( a\mathbf{\tilde{\alpha}}+b\mathbf{\tilde{\lambda}}\right)
$ \\
$\boldsymbol{\mathring{p}}_{\text{\textsc{r}}}$ & $=$ & $\frac{1}{\sqrt{2%
\mathrm{k}}}\left( a\mathbf{\tilde{\alpha}}-b\mathbf{\tilde{\lambda}}\right)
$%
\end{tabular}
\label{ppl}
\end{equation}%
They yield the momenta of the k$^{2}$ ground configurations $\Psi
_{a,b}^{\pm }\sim \psi _{a}\pm \tilde{\psi}_{b}$ which are characterised by
the topological index:
\begin{equation}
\boldsymbol{\mathring{p}}_{\text{\textsc{l}}}^{2}-\boldsymbol{\mathring{p}}_{%
\text{\textsc{r}}}^{2}=2\boldsymbol{\mathring{h}\mathring{p}}\qquad ,\qquad 2%
\boldsymbol{\mathring{h}\mathring{p}}=\frac{2}{\mathrm{k}}ab\in 2\mathbb{Z}_{%
\mathrm{k}}
\end{equation}

\item \textbf{Excitations:} For non vanishing $n$ and $m,$ one obtains
excited states $\Psi _{a,b}^{(n,m)}\sim \psi _{a}^{(n)}\pm \tilde{\psi}%
_{b}^{(m)}$ with relative momenta as follows%
\begin{equation}
\begin{tabular}{lll}
$\boldsymbol{p}_{\text{\textsc{l}}}$ & $=$ & $\sqrt{\frac{\mathrm{k}}{2}}%
\left( n\mathbf{\tilde{\alpha}}+m\mathbf{\tilde{\lambda}}\right) $ \\
$\boldsymbol{p}_{\text{\textsc{r}}}$ & $=$ & $\sqrt{\frac{\mathrm{k}}{2}}%
\left( n\mathbf{\tilde{\alpha}}-m\mathbf{\tilde{\lambda}}\right) $%
\end{tabular}
\label{plpr}
\end{equation}%
and associated topological index
\begin{equation}
\boldsymbol{p}_{\text{\textsc{l}}}^{2}-\boldsymbol{p}_{\text{\textsc{r}}%
}^{2}=\left( 2\mathrm{k}\right) nm\in 2\mathrm{k}\mathbb{Z}
\end{equation}%
For future reference, it is useful to express the difference $\boldsymbol{P}%
_{\text{\textsc{l}}}^{2}-\boldsymbol{P}_{\text{\textsc{r}}}^{2}$ in terms of
the variable $\boldsymbol{H}=\left( \boldsymbol{P}_{\text{\textsc{l}}}+%
\boldsymbol{P}_{\text{\textsc{r}}}\right) /\sqrt{2}$ and $\boldsymbol{P}%
=\left( \boldsymbol{P}_{\text{\textsc{l}}}-\boldsymbol{P}_{\text{\textsc{r}}%
}\right) /\sqrt{2}$ like
\begin{equation}
\boldsymbol{P}_{\text{\textsc{l}}}^{2}-\boldsymbol{P}_{\text{\textsc{r}}%
}^{2}=2\boldsymbol{HP}
\end{equation}%
By substituting $\boldsymbol{H}=\boldsymbol{\mathring{h}}+\boldsymbol{h}$
and $\boldsymbol{P}=\boldsymbol{\mathring{p}}+\boldsymbol{p}$, the
topological index splits into three contributions:%
\begin{equation}
\boldsymbol{P}_{\text{\textsc{l}}}^{2}-\boldsymbol{P}_{\text{\textsc{r}}%
}^{2}=2\boldsymbol{\mathring{h}\mathring{p}}+2(\boldsymbol{\mathring{h}p}+%
\boldsymbol{h\mathring{p})}+2\boldsymbol{hp}  \label{hh}
\end{equation}
\end{itemize}

\subsubsection{About the even integer condition}

\begin{figure}[tbph]
\begin{center}
\includegraphics[width=8cm]{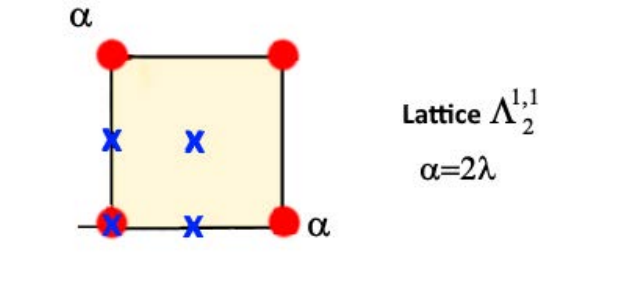}
\end{center}
\par
\vspace{-0.5cm}
\caption{The values of the points ($\mathring{h},\mathring{p}$) in root
lattice of SO(4). These points are given by $(\boldsymbol{\mathring{h}},%
\boldsymbol{\mathring{p}})=(0,0),$ $(\frac{1}{2},0),$ $(0,\frac{1}{2}),$ $(%
\frac{1}{2},\frac{1}{2})$. Here, we have k=2 and r=1. In general, there are k%
$^{2}$ points sitting in the unit cell of $\mathbf{\Lambda }_{\mathrm{k}}^{%
\mathrm{1,1}}$.}
\label{m2}
\end{figure}

Here, we revisit the\ even integer condition defining the lattices $\mathbf{%
\Lambda }_{\mathrm{k}}^{\mathrm{r,r}}$ and $\mathbf{\Lambda }_{\mathrm{k}%
\mathcal{C}}^{\mathrm{r,r}}$ of the discriminant group $\mathbb{Z}_{\mathrm{k%
}}\times \mathbb{Z}_{\mathrm{k}}$. Our formulation involves \textsl{three
constraint relations} stemming from the decomposition (\ref{hh}) which also
reads%
\begin{equation}
\boldsymbol{P}_{\text{\textsc{l}}}^{2}-\boldsymbol{P}_{\text{\textsc{r}}%
}^{2}=(\boldsymbol{\mathring{p}}_{\text{\textsc{l}}}^{2}-\boldsymbol{%
\mathring{p}}_{\text{\textsc{r}}}^{2})+(\boldsymbol{p}_{\text{\textsc{l}}%
}^{2}-\boldsymbol{p}_{\text{\textsc{r}}}^{2})+\left[ (\boldsymbol{\mathring{p%
}}_{\text{\textsc{l}}}^{2}-\boldsymbol{p}_{\text{\textsc{r}}}^{2})+(%
\boldsymbol{p}_{\text{\textsc{l}}}^{2}-\boldsymbol{\mathring{p}}_{\text{%
\textsc{r}}}^{2})\right]
\end{equation}%
It reduces to the familiar even integer condition in the case of a trivial
code with \textrm{k=1} where $\boldsymbol{\mathring{p}}_{\text{\textsc{l}/%
\textsc{r}}}=0.$ The final result we seek to retain\textrm{\ }are given by
the conditions (\ref{cd}). \newline
To that purpose, we start from the standard\textrm{\ }even integer condition
of Narain lattices namely%
\begin{equation}
\boldsymbol{p}_{\text{\textsc{l}}}^{2}-\boldsymbol{p}_{\text{\textsc{r}}%
}^{2}\in 2\mathbb{Z}\qquad ,\qquad \boldsymbol{p}_{\text{\textsc{l}}}\text{ }%
,\text{ }\boldsymbol{p}_{\text{\textsc{r}}}\in \mathbf{\Lambda }_{\mathrm{1}%
\mathcal{C}}^{\mathrm{r,r}}
\end{equation}%
This corresponds to the case where $a=b=0$ for which the momenta $%
\boldsymbol{p}_{\text{\textsc{l}}},\boldsymbol{p}_{\text{\textsc{r}}}$ are
given by (\ref{plpr}) with CS level k=1 in the action (\ref{act}). Using (%
\ref{plpr}), one finds $\boldsymbol{p}_{\text{\textsc{l}}}^{2}-\boldsymbol{p}%
_{\text{\textsc{r}}}^{2}=2\mathrm{k}nm$ which indeed\textrm{\ }belongs to 2$%
\mathbb{Z}$ for k=1. For higher values of the CS level, this generalises to 2%
$\mathrm{k}\mathbb{Z}$; i.e:%
\begin{equation}
\boldsymbol{p}_{\text{\textsc{l}}}^{2}-\boldsymbol{p}_{\text{\textsc{r}}%
}^{2}\in 2\mathrm{k}\mathbb{Z}\qquad for\qquad \mathrm{k\geq 1}
\end{equation}%
To derive the conditions for non trivial values of the labels (a,b), we
compute $\boldsymbol{P}_{\text{\textsc{l}}}^{2}-\boldsymbol{P}_{\text{%
\textsc{r}}}^{2}$ by using (\ref{pq}) which yields $\frac{2}{\mathrm{k}}(a+%
\mathrm{k}n)(b+\mathrm{k}m)$. This expression decomposes as in (\ref{ind})
namely $\frac{2}{\mathrm{k}}ab+2\left( am+bn\right) +2\mathrm{k}nm.$ Given
that $\left( a,b\right) \in \mathbb{Z}_{\mathrm{k}}\times \mathbb{Z}_{%
\mathrm{k}}$ and $\left( n,m\right) \in \mathbb{Z}\times \mathbb{Z}$, it
follows that%
\begin{equation}
\begin{tabular}{lll}
$\frac{2}{\mathrm{k}}ab$ & $\in $ & $\frac{2}{\mathrm{k}}\mathbb{Z}$ \\
$2\mathrm{k}nm$ & $\in $ & $2\mathrm{k}\mathbb{Z}$%
\end{tabular}%
\qquad ,\qquad
\begin{tabular}{lll}
$2am$ & $\in $ & $2\mathbb{Z}$ \\
$2bn$ & $\in $ & $2\mathbb{Z}$%
\end{tabular}
\label{2k}
\end{equation}%
However, equating with (\ref{hh}), indicates that $\boldsymbol{P}_{\text{%
\textsc{l}}}^{2}-\boldsymbol{P}_{\text{\textsc{r}}}^{2}=2\boldsymbol{%
\mathring{h}\mathring{p}}+2(\boldsymbol{\mathring{h}p}+\boldsymbol{h%
\mathring{p})}+2\boldsymbol{hp}$ as in eq(\ref{hh}). Accordingly, we deduce
the three constraint relations that characterise the code CFT%
\begin{equation}
\begin{tabular}{lll}
$2\boldsymbol{\mathring{h}\mathring{p}}$ & $\in $ & $\frac{2}{\mathrm{k}}%
\mathbb{Z}$ \\
$2\boldsymbol{hp}$ & $\in $ & $2\mathrm{k}\mathbb{Z}$%
\end{tabular}%
\qquad ,\qquad
\begin{tabular}{lll}
$2\boldsymbol{\mathring{h}p}$ & $\in $ & $2\mathbb{Z}$ \\
$2\boldsymbol{h\mathring{p}}$ & $\in $ & $2\mathbb{Z}$%
\end{tabular}
\label{cd}
\end{equation}%
Moreover, using the identifications $\boldsymbol{\mathring{h}\mathring{p}}=%
\frac{1}{\mathrm{k}}ab$ and $\boldsymbol{hp}=\mathrm{k}nm$, it also results
the \textrm{two} following: $\left( \mathbf{i}\right) $ the structure of the
lattices hosting the ground momenta ($\boldsymbol{\mathring{h}},\boldsymbol{%
\mathring{p}}$) which are given by%
\begin{equation}
\begin{tabular}{lllll}
$\boldsymbol{\mathring{h}}$ & $=$ & $xa\mathbf{\tilde{\alpha}}$ & $\in $ & $%
x $\textsc{R}$_{\mathrm{k}}^{\mathbf{su}_{2}}$ \\
$\boldsymbol{\mathring{p}}$ & $=$ & $yb\mathbf{\tilde{\lambda}}$ & $\in $ & $%
y$\textsc{W}$_{\mathrm{k}}^{\mathbf{su}_{2}}$%
\end{tabular}%
\qquad \Rightarrow \qquad
\begin{tabular}{lllll}
$\boldsymbol{\mathring{h}}$ & $=$ & $\frac{1}{\mathrm{k}}a\mathbf{\tilde{%
\alpha}}$ & $\in $ & $\frac{1}{\mathrm{k}}$\textsc{R}$_{\mathrm{k}}^{\mathbf{%
su}_{2}}$ \\
$\boldsymbol{\mathring{p}}$ & $=$ & $\frac{1}{\mathrm{k}}b\mathbf{\tilde{%
\alpha}}$ & $\in $ & $\frac{1}{\mathrm{k}}$\textsc{R}$_{\mathrm{k}}^{\mathbf{%
su}_{2}}$%
\end{tabular}
\label{hp}
\end{equation}%
with the property $xy=\frac{1}{\mathrm{k}}.$\ This condition can be
satisfied in multiple ways such as $x=y=\frac{1}{\sqrt{\mathrm{k}}}$ \textrm{%
or} $x=\frac{1}{\mathrm{k}}$ and $y=1$. For the second case, we have $%
\boldsymbol{\mathring{h}}=\frac{1}{\mathrm{k}}a\mathbf{\tilde{\alpha}}\in $%
\textsc{R}$_{\mathrm{k}}^{\mathbf{su}_{2}}$ and $\boldsymbol{\mathring{p}}=b%
\mathbf{\tilde{\lambda}}\in $\textsc{W}$_{\mathrm{k}}^{\mathbf{su}_{2}}$
with $\mathbf{\tilde{\lambda}=}\frac{1}{\mathrm{k}}\mathbf{\tilde{\alpha}}.$
$\left( \mathbf{ii}\right) $ Regarding the excited momenta ($\boldsymbol{h},%
\boldsymbol{p}$), we obtain
\begin{equation}
\begin{tabular}{lllll}
$\boldsymbol{h}$ & $=$ & $zn\mathbf{\tilde{\alpha}}$ & $\in $ & $z$\textsc{R}%
$_{\mathrm{k}}^{\mathbf{su}_{2}}$ \\
$\boldsymbol{p}$ & $=$ & $tm\mathbf{\tilde{\lambda}}$ & $\in $ & $t$\textsc{W%
}$_{\mathrm{k}}^{\mathbf{su}_{2}}$%
\end{tabular}%
\qquad \Rightarrow \qquad
\begin{tabular}{lllll}
$\boldsymbol{h}$ & $=$ & $n\mathbf{\tilde{\alpha}}$ & $\in $ & \textsc{R}$_{%
\mathrm{k}}^{\mathbf{su}_{2}}$ \\
$\boldsymbol{p}$ & $=$ & $\mathrm{k}m\mathbf{\tilde{\lambda}}$ & $\in $ & $%
\mathrm{k}$\textsc{W}$_{\mathrm{k}}^{\mathbf{su}_{2}}$%
\end{tabular}%
\end{equation}%
with $zt=\mathrm{k}$ which can be solved for $z=t=\sqrt{\mathrm{k}}$ or
equivalently $z=1$ and $t=\mathrm{k}$. For the CS level $\mathrm{k=2}$, eq(%
\ref{hp}) reduces to%
\begin{equation}
\begin{tabular}{lllll}
$\boldsymbol{\mathring{h}}$ & $=$ & $\frac{a}{2}\mathbf{\alpha }$ & $=$ & $%
\frac{a}{2}\mathbf{\alpha }$ \\
$\boldsymbol{\mathring{p}}$ & $=$ & $b\mathbf{\lambda }$ & $=$ & $\frac{b}{2}%
\mathbf{\alpha }$%
\end{tabular}%
\qquad ,\qquad
\begin{tabular}{lllll}
$\boldsymbol{h}$ & $=$ & $n\mathbf{\alpha }$ & $=$ & $n\mathbf{\alpha }$ \\
$\boldsymbol{p}$ & $=$ & $\mathrm{2}m\mathbf{\lambda }$ & $=$ & $m\mathbf{%
\alpha }$%
\end{tabular}
\label{ph}
\end{equation}%
The graphic representation of ($\boldsymbol{\mathring{h}},\boldsymbol{%
\mathring{p}}$) within the unit cell of the 2D lattice $\mathbf{\Lambda }%
_{2}^{\mathrm{1,1}}$\ is depicted in Figure \textbf{\ref{m2}}. The ground
states are given by the 4 left moving ($\psi _{a}+\tilde{\psi}_{b})/\sqrt{2}$
and the 4 right moving ($\psi _{a}-\tilde{\psi}_{b})/\sqrt{2}$.

\subsection{Particles in the SU(3) based model}

The properties of particle states in the code CFT based on SU(3), with
discriminant group $\mathbb{Z}_{\mathrm{k}}\times \mathbb{Z}_{\mathrm{k}}$
and lattices ($\mathbf{\Lambda }_{\mathrm{k}}^{\mathbf{su}_{3}},\mathbf{%
\Lambda }_{\mathrm{k}\mathcal{C}}^{\mathbf{su}_{3}},\mathbf{\Lambda }_{%
\mathrm{k}}^{\ast \mathbf{su}_{3}}$) realised in terms of the root \textsc{R}%
$_{\mathrm{k}}^{\mathbf{su}_{3}}$ and weight \textsc{W}$_{\mathrm{k}}^{%
\mathbf{su}_{3}}$ systems, follow an analogous structure to that of the
SU(2) theory. Here, the momenta of the particle states
\begin{equation}
P_{\text{\textsc{l}}/\text{\textsc{r}}}:=\left( P_{\text{\textsc{l}}/\text{%
\textsc{r}}}\right) _{\mathbf{a},\mathbf{b}}^{\mathbf{n},\mathbf{m}}
\end{equation}%
are labeled by two pairs of integer vectors: $\left( \mathbf{i}\right) $ a
pair $(\mathbf{a},\mathbf{b})$ with $\mathbf{a}=(a_{1},a_{2})$ and $\mathbf{b%
}=(b_{1},b_{2})$ ranging in $a_{i},b_{i}=0,...,\mathrm{k}-1$; they label the
ground configurations. $\left( \mathbf{ii}\right) $ A pair $(\mathbf{n,m}%
)\in \mathbb{Z}^{2}\times \mathbb{Z}^{2}$ with $\mathbf{n}=(n_{1},n_{2})$
and $\mathbf{m}=(m_{1},m_{2})$ determining the degrees of excitations of the
KK and the windings modes.

\subsubsection{Momenta $P_{L/R}$ and SU(3) symmetry}

To derive the left/right momenta $P_{\text{\textsc{l}}/\text{\textsc{r}}}$
for the SU(3) based model, we extend the relation (\ref{pq}) from the SU(2)
theory. For that, we start by $\boldsymbol{P}_{\text{\textsc{l}/\textsc{r}}}=%
\frac{1}{\sqrt{2}}(\boldsymbol{H}\pm \boldsymbol{P)}$\textrm{\ }where%
\begin{equation}
\boldsymbol{H}=\frac{1}{\sqrt{2}}\left( \boldsymbol{P}_{\text{\textsc{l}}}+%
\boldsymbol{P}_{\text{\textsc{r}}}\right) \qquad ,\qquad \boldsymbol{P}=%
\frac{1}{\sqrt{2}}\left( \boldsymbol{P}_{\text{\textsc{l}}}-\boldsymbol{P}_{%
\text{\textsc{r}}}\right)  \label{PP}
\end{equation}%
and
\begin{equation}
\boldsymbol{H}=\boldsymbol{\mathring{h}}+\boldsymbol{h\qquad ,\qquad P}=%
\boldsymbol{\mathring{p}}+\boldsymbol{p}
\end{equation}%
here $(\boldsymbol{\mathring{h}},\boldsymbol{\mathring{p}})$ encode the
ground configurations while\textrm{\ }$(\boldsymbol{h},\boldsymbol{p})$ are
used to describe the KK and winding excitations. Then, we require the SU(2)
conditions (\ref{cd}) to remain valid for the SU(3) model. These constraints
generalise as follows%
\begin{equation}
\begin{tabular}{lllcl}
$\boldsymbol{\mathring{h}}.\boldsymbol{\mathring{p}}$ & $=$ & $\frac{1}{%
\mathrm{k}}\mathbf{a}.\mathbf{b}$ & $\in $ & $\frac{1}{\mathrm{k}}\mathbb{Z}$
\\
$\boldsymbol{h}.\boldsymbol{p}$ & $=$ & $\mathrm{k}\mathbf{n.m}$ & $\in $ & $%
\mathrm{k}\mathbb{Z}$%
\end{tabular}%
\qquad ,\qquad
\begin{tabular}{lllcl}
$\boldsymbol{\mathring{h}}.\boldsymbol{p}$ & $=$ & $\mathbf{a.m}$ & $\in $ &
$\mathbb{Z}$ \\
$\boldsymbol{h}.\boldsymbol{\mathring{p}}$ & $=$ & $\mathbf{b.n}$ & $\in $ &
$\mathbb{Z}$%
\end{tabular}
\label{QP}
\end{equation}%
where each pairing, like $\mathbf{a.b,}$ stands for the scalar products, $%
a_{1}b_{1}+a_{2}b_{2}$. A solution to these constraints (\ref{QP}) is given
as follows:

\begin{description}
\item[$\left( \mathbf{i}\right) $] \textbf{values of }$(\boldsymbol{%
\mathring{h}},\boldsymbol{\mathring{p}}):$
\begin{equation}
\begin{tabular}{lllll}
$\boldsymbol{\mathring{h}}$ & $=$ & $\frac{1}{\mathrm{k}}\left( a_{1}\mathbf{%
\tilde{\alpha}}_{1}+a_{2}\mathbf{\tilde{\alpha}}_{2}\right) $ & $\in $ & $%
\frac{1}{\mathrm{k}}$\textsc{R}$_{\mathrm{k}}^{\mathbf{su}_{3}}$ \\
$\boldsymbol{\mathring{p}}$ & $=$ & $b\mathbf{\tilde{\lambda}}_{1}+b_{2}%
\mathbf{\tilde{\lambda}}_{2}$ & $\in $ & \textsc{W}$_{\mathrm{k}}^{\mathbf{su%
}_{3}}$%
\end{tabular}%
\end{equation}%
with%
\begin{equation}
\mathbf{\tilde{\alpha}}_{i}=\sqrt{\frac{\mathrm{k}}{3}}\mathbf{\alpha }%
_{i}\qquad ,\qquad \mathbf{\tilde{\lambda}}^{i}=\sqrt{\frac{3}{\mathrm{k}}}%
\mathbf{\lambda }^{i}\qquad ,\qquad \mathbf{\tilde{\alpha}}_{i}.\mathbf{%
\tilde{\lambda}}^{j}=\delta _{i}^{j}  \label{al}
\end{equation}%
Using the relations $\mathbf{\tilde{\lambda}}_{1}=\frac{1}{\mathrm{k}}\left(
2\mathbf{\tilde{\alpha}}_{1}+\mathbf{\tilde{\alpha}}_{2}\right) $ and $%
\mathbf{\tilde{\lambda}}_{2}=\frac{1}{\mathrm{k}}\left( \mathbf{\tilde{\alpha%
}}_{1}+2\mathbf{\tilde{\alpha}}_{2}\right) ,$ we also have
\begin{equation}
\begin{tabular}{lllll}
$\boldsymbol{\mathring{h}}$ & $=$ & $\frac{1}{\mathrm{k}}\left( a_{1}\mathbf{%
\tilde{\alpha}}_{1}+a_{2}\mathbf{\tilde{\alpha}}_{2}\right) $ & $\in $ & $%
\frac{1}{\mathrm{k}}$\textsc{R}$_{\mathrm{k}}^{\mathbf{su}_{3}}$ \\
$\boldsymbol{\mathring{p}}$ & $=$ & $\frac{1}{\mathrm{k}}\left(
b_{1}^{\prime }\mathbf{\tilde{\alpha}}_{1}+b_{2}^{\prime }\mathbf{\tilde{%
\alpha}}_{2}\right) $ & $\in $ & $\frac{1}{\mathrm{k}}$\textsc{R}$_{\mathrm{k%
}}^{\mathbf{su}_{3}}$%
\end{tabular}
\label{hhp}
\end{equation}%
where we have set $b_{1}^{\prime }=2b_{1}+b_{2}$ and $b_{2}^{\prime
}=b_{1}+2b_{2}.$

\item[$\left( \mathbf{ii}\right) $] \textbf{values of }$(\boldsymbol{h},%
\boldsymbol{p}):$%
\begin{equation}
\begin{tabular}{lllll}
$\boldsymbol{h}$ & $=$ & $n_{1}\mathbf{\tilde{\alpha}}_{1}+n_{2}\mathbf{%
\tilde{\alpha}}_{2}$ & $\in $ & \textsc{R}$_{\mathrm{k}}^{\mathbf{su}_{3}}$
\\
$\boldsymbol{p}$ & $=$ & $\mathrm{k}\left( m_{1}\mathbf{\tilde{\lambda}}%
_{1}+m_{2}\mathbf{\tilde{\lambda}}_{2}\right) $ & $\in $ & $\mathrm{k}$%
\textsc{W}$_{\mathrm{k}}^{\mathbf{su}_{3}}$%
\end{tabular}%
\end{equation}%
reading also like%
\begin{equation}
\begin{tabular}{lllll}
$\boldsymbol{h}$ & $=$ & $n_{1}\mathbf{\tilde{\alpha}}_{1}+n_{2}\mathbf{%
\tilde{\alpha}}_{2}$ & $\in $ & \textsc{R}$_{\mathrm{k}}^{\mathbf{su}_{3}}$
\\
$\boldsymbol{p}$ & $=$ & $m_{1}^{\prime }\mathbf{\tilde{\alpha}}%
_{1}+m_{2}^{\prime }\mathbf{\tilde{\alpha}}_{2}$ & $\in $ & \textsc{R}$_{%
\mathrm{k}}^{\mathbf{su}_{3}}$%
\end{tabular}%
\end{equation}%
with $m_{1}^{\prime }=2m_{1}+m_{2}$ and $m_{2}^{\prime }=m_{1}+2m_{2}$.
\end{description}

\begin{figure}[tbph]
\begin{center}
\includegraphics[width=14cm]{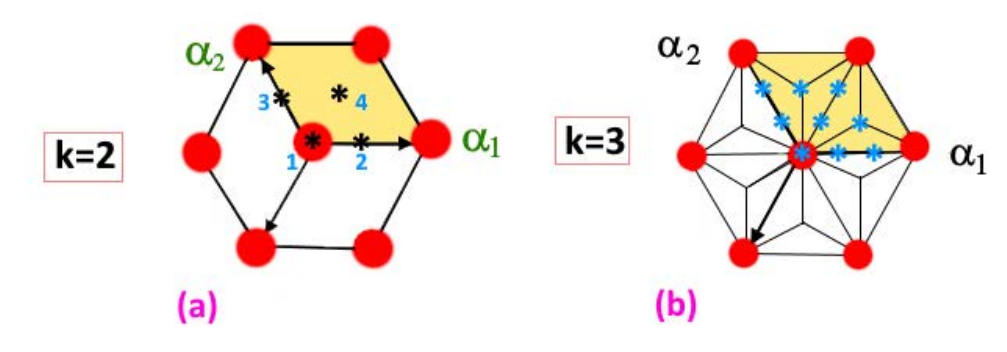}
\end{center}
\par
\vspace{-0.5cm}
\caption{The values of the points $\mathring{h}$ in root lattices \textsc{R}$%
_{\mathrm{k}}^{\mathbf{su}_{3}}$. On left we have 4 ground points. On right
there are 9 points in unit cell.}
\label{m3}
\end{figure}

By substituting these expressions into (\ref{PP}), we get%
\begin{equation}
\begin{tabular}{lll}
$\boldsymbol{H}$ & $=$ & $\frac{1}{\sqrt{2\mathrm{k}}}\left( a^{i}+\mathrm{k}%
n^{i}\right) \mathbf{\tilde{\alpha}}_{i}$ \\
$\boldsymbol{P}$ & $=$ & $\frac{1}{\sqrt{2\mathrm{k}}}\left( b_{i}+\mathrm{k}%
m_{i}\right) \mathbf{\tilde{\lambda}}^{i}$%
\end{tabular}%
\qquad \Leftrightarrow \qquad
\begin{tabular}{lll}
$\boldsymbol{H}$ & $=$ & $\frac{1}{\sqrt{2\mathrm{k}}}\bar{a}^{i}\mathbf{%
\tilde{\alpha}}_{i}$ \\
$\boldsymbol{P}$ & $=$ & $\frac{1}{\sqrt{2\mathrm{k}}}\bar{b}_{i}\mathbf{%
\tilde{\lambda}}^{i}$%
\end{tabular}%
\end{equation}%
and subsequently
\begin{equation}
\begin{tabular}{lll}
$\boldsymbol{P}_{\text{\textsc{l}}}$ & $=$ & $\frac{1}{\sqrt{2\mathrm{k}}}%
\left( \bar{a}^{i}\mathbf{\tilde{\alpha}}_{i}+\bar{b}_{i}\mathbf{\tilde{%
\lambda}}^{i}\right) $ \\
& $=$ & $\frac{1}{\sqrt{2\mathrm{k}}}\left[ \left( a^{i}+\mathrm{k}%
n^{i}\right) \mathbf{\tilde{\alpha}}_{i}+\left( b_{i}+\mathrm{k}m_{i}\right)
\mathbf{\tilde{\lambda}}^{i}\right] $ \\
$\boldsymbol{P}_{\text{\textsc{r}}}$ & $=$ & $\frac{1}{\sqrt{2\mathrm{k}}}%
\left( \bar{a}^{i}\mathbf{\tilde{\alpha}}_{i}-\bar{b}_{i}\mathbf{\tilde{%
\lambda}}^{i}\right) $ \\
& $=$ & $\frac{1}{\sqrt{2\mathrm{k}}}\left[ \left( a^{i}+\mathrm{k}%
n^{i}\right) \mathbf{\tilde{\alpha}}_{i}-\left( b_{i}+\mathrm{k}m_{i}\right)
\mathbf{\tilde{\lambda}}^{i}\right] $%
\end{tabular}%
\end{equation}%
We conclude by giving the locii of the ground states characterised by the
pair ($\boldsymbol{\mathring{h}},\boldsymbol{\mathring{p}}$) written in
terms of the $\mathbf{\tilde{\alpha}}_{i}$ as in eq(\ref{al}). Focussing on $%
\boldsymbol{\mathring{h}},$ the loci for the cases k=2 and k=3 are depicted
in the Figure \textbf{\ref{m3}}.

\subsubsection{Ground configurations and SU(3) representations}

We begin by recalling that, for SU(2) based model, the ground state
configurations for a CS level k transform in the representation with isospin
$j=(k-1)/2$ as in (\ref{js}). In the case of code CFTs based on SU(3), a
similar structure emerges\textrm{.} By using $\mathbf{\tilde{\alpha}}_{1}=%
\frac{\mathrm{k}}{3}(2\mathbf{\tilde{\lambda}}_{1}-\mathbf{\tilde{\lambda}}%
_{2})$ and $\mathbf{\tilde{\alpha}}_{2}=\frac{\mathrm{k}}{3}(2\mathbf{\tilde{%
\lambda}}_{2}-\mathbf{\tilde{\lambda}}_{1}),$ the scaled quantum numbers $(%
\boldsymbol{\mathring{h}}^{\prime },\boldsymbol{\mathring{p}}^{\prime })=(%
\sqrt{3\mathrm{k}}\boldsymbol{\mathring{h}},\sqrt{\mathrm{k}/3}\boldsymbol{%
\mathring{p}})$ characterising the ground configurations are given by the
weight vectors%
\begin{equation}
\begin{tabular}{lll}
$\boldsymbol{\mathring{h}}^{\prime }$ & $=$ & $r_{1}\mathbf{\lambda }%
_{1}+r_{2}\mathbf{\lambda }_{2}$ \\
$\boldsymbol{\mathring{p}}^{\prime }$ & $=$ & $b_{1}\mathbf{\lambda }%
_{1}+b_{2}\mathbf{\lambda }_{2}$%
\end{tabular}%
\end{equation}%
where the integers $r_{1}=2a_{1}-a_{2}$ and $r_{2}=2a_{2}-a_{1}$ (modulo k)%
\textrm{\ }take values in the range $\left[ 0,k-1\right] .$ These quantum
numbers correspond to the\textrm{\ }highest weight vectors of SU(3). Recall
that highest weight representations $\mathcal{D}^{(s_{1},s_{2})}$ of the
symmetry SU(3) are defined by the highest weight vector $\mathbf{\omega }%
=s_{1}\mathbf{\tilde{\lambda}}_{1}+s_{2}\mathbf{\tilde{\lambda}}_{2}$\
parameterised by two positive integers $\left( s_{1},s_{2}\right) .$ Their
dimensions are given by%
\begin{equation}
d_{s_{1},s_{2}}=\frac{1}{2}\left( s_{1}+1\right) \left( s_{2}+1\right)
\left( s_{1}+s_{2}+2\right)
\end{equation}%
For the particular situation $s_{2}=0$ where the $SU(3)$ is broken down to $%
SU(2)\times U(1),$ the dimension reduces to $d_{s,0}=\frac{1}{2}\left(
s+1\right) \left( s+2\right) .$ As for the case $s=2j,$ we have $%
d_{2j,0}=\left( 2j+1\right) \left( j+1\right) $ that splits into $j+1$
multiplets like%
\begin{equation}
d_{2j,0}=\underbrace{\left( 2j+1\right) +...+\left( 2j+1\right) }_{j+1\text{
iso-multiplets}}
\end{equation}

\section{Lattice fermions and topological phases}

\label{sec:5} This section\textrm{\ }builds on the previous and further
develops the connection between Narain CFT and topological matter with Dirac
cones formulated in terms of fermionic lattices. To establish this link
explicitly, we present three \emph{guiding arguments} (\underline{\textbf{A1}%
}, \underline{\textbf{A2}} and \underline{\textbf{A3}}) while focussing on
the CS level k=2 to illustrate the basic ideas using the associated 2D
lattices \textsc{R}$_{\mathrm{2}}^{\mathbf{su}_{3}}$ and \textsc{W}$_{%
\mathrm{2}}^{\mathbf{su}_{3}}$:

\begin{itemize}
\item \underline{\textbf{A1}}: \emph{lattice} \textsc{W}$_{\mathrm{2}}^{%
\mathbf{su}_{3}}$ = \emph{honeycomb} $\mathbb{A}_{\mathrm{2}}+\mathbb{B}_{%
\mathrm{2}}$. \newline
The weight lattice \textsc{W}$_{\mathrm{2}}^{\mathbf{su}_{3}}$ contains the
root lattice \textsc{R}$_{\mathrm{2}}^{\mathbf{su}_{3}}\subset $\textsc{W}$_{%
\mathrm{2}}^{\mathbf{su}_{3}}$ leading to the discriminant \textsc{W}$_{%
\mathrm{2}}^{\mathbf{su}_{3}}/\text{\textsc{R}}_{\mathrm{2}}^{\mathbf{su}%
_{3}}\simeq \mathbb{Z}_{2},$ this allows to split the weight lattice as
follows%
\begin{equation}
\text{\textsc{W}}_{\mathrm{2}}^{\mathbf{su}_{3}}=\ \ \mathbb{A}_{\mathrm{2}%
}\ \ \dbigcup \text{ \ }\mathbb{B}_{\mathrm{2}}\quad ,\qquad
\begin{tabular}{lll}
$\mathbb{A}_{\mathrm{2}}$ & $=$ & $\text{\textsc{R}}_{\mathrm{2}}^{\mathbf{su%
}_{3}}$ \\
$\mathbb{B}_{\mathrm{2}}$ & $=$ & $\text{\textsc{W}}_{\mathrm{2}}^{\mathbf{su%
}_{3}}\backslash \text{\textsc{R}}_{\mathrm{2}}^{\mathbf{su}_{3}}$%
\end{tabular}
\label{hc}
\end{equation}%
To distinguish the two sublattices of \textsc{W}$_{\mathrm{2}}^{\mathbf{su}%
_{3}}$, we introduce a physical hypothesis: we\textrm{\ }interpret the
lattices $\mathbb{A}_{\mathrm{2}}$ and $\mathbb{B}_{\mathrm{2}}$ as being
occupied by different (\emph{colored}) particles, each localized on one
sublattice as described in \underline{\textbf{A2}}.

\item \underline{\textbf{A2}}: \emph{the KK/W correspondence}\newline
Leveraging the root/weight duality and the KK{\small \ }$\leftrightarrow $
Winding correspondence, we propose\textrm{\ }that quantum excitations $%
\left\vert {\small KK}_{\mathbf{m}}\right\rangle $ sit in $\mathbb{A}_{%
\mathrm{2}}$ while the winding modes $\left\vert {\small W}_{\mathbf{m}%
}\right\rangle $ live in $\mathbb{B}_{\mathrm{2}}$; see Figure \textbf{\ref%
{KK} }for illustration.\newline
\begin{figure}[tbph]
\begin{center}
\includegraphics[width=8cm]{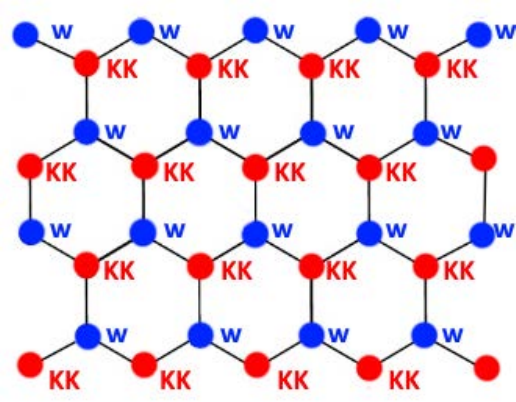}
\end{center}
\par
\vspace{-0.5cm}
\caption{Lattice of KK and winding particles. Each unit cell contains a pair
KK/W in agreement with $\mathbb{Z}_{2}$ symmetry of code CFT at CS level
k=2. }
\label{KK}
\end{figure}

This assumption implies that both type of states $\left\vert {\small KK}_{%
\mathbf{m}}\right\rangle $ and $\left\vert {\small W}_{\mathbf{m}%
}\right\rangle $ are hosted by the honeycomb \textsc{W}$_{\mathrm{2}}^{%
\mathbf{su}_{3}}$ where each unit cell is occupied by a $\mathbb{Z}_{2}$
doublet as also depicted in Figure \textbf{\ref{i3}}),
\begin{equation}
\left(
\begin{array}{c}
\left\vert {\small KK}\right\rangle \\
\left\vert {\small W}\right\rangle%
\end{array}%
\right)
\end{equation}%
Observe that the NCFT based on \textsc{W}$_{\mathrm{2}}^{\mathbf{su}_{3}}$
has a central anomaly $c_{\text{\textsc{l/r}}}=2$ which suggests that the
theory has two bosonic conformal fields $X^{1}(\tau ,\sigma )$ and $%
X^{2}(\tau ,\sigma )$ splitting into left $X_{L}^{i}\left( \tau +\sigma
\right) $ and right $X_{R}^{i}\left( \tau -\sigma \right) $ moving fields.
In complex coordinates $z=\tau +i\sigma ,$ these become the holomorphic $%
X_{L}^{i}=X_{L}^{i}\left( z\right) $ and the anti-holomorphic $%
X_{R}^{i}=X_{R}^{i}\left( \bar{z}\right) .$

\item \underline{\textbf{A3}}: \emph{fermionisation}\newline
We apply fermionisation techniques\footnote{%
\ Free conformal fermions $\psi _{L/R}$ can be expressed in terms of bosons $%
\phi _{L/R}$ by using vertex operators $e^{\pm i\phi _{L/R}}$. For
constructions using bosonisation and tight binding models, see \textrm{\cite%
{sen1,sen2,sen3}}} in two dimensions \textrm{\cite{ferm1,ferm2}} as in the
table Eq(\ref{tab1}) to reinterpret the bosonic theory as a fermionic one
with a tight binding model having Dirac cones, enabling thus the modeling of
critical points in topological phases of matter\textrm{.} The main lines of
this proposal can be formally schematised in the following algorithm%
\begin{equation}
\begin{tabular}{ccc}
\textbf{Code-CFT} &  &  \\
${\small \downarrow }$ &  &  \\
\textbf{Narain CFT} &  &  \\
${\small \downarrow }$ &  &  \\
\textbf{bosonic CFT} &  &  \\
$\downarrow \uparrow $ &  &  \\ \hline
\multicolumn{1}{|c}{\textbf{fermionic CFT}} & ${\small \rightarrow }$ &
\multicolumn{1}{c|}{\textbf{Dirac cones}} \\
\multicolumn{1}{|c}{${\small \uparrow }$} & $\circlearrowright $ &
\multicolumn{1}{c|}{${\small \downarrow }$} \\
\multicolumn{1}{|c}{\textbf{gapless states}} & ${\small \leftarrow }$ &
\multicolumn{1}{c|}{\textbf{lattice model}} \\ \hline
\end{tabular}%
\end{equation}%
\begin{equation*}
\end{equation*}
\end{itemize}

Using the arguments (\underline{\textbf{A1}}, \underline{\textbf{A2}},
\underline{\textbf{A3}}), we show throughout this section how known results
from 2D Lattice matter such as 2D graphene \textrm{\cite{graf1,graf2}} can
be recast in terms of a NCFT. These particular insulators feature gapless
states protected by discrete symmetry and carry non trivial Chern invariant
\textrm{\cite{tm1,tm2}} similarly to topological Haldane theory of the
anomalous quantum Hall effect.

\subsection{KK/W system on the lattice}

\qquad We begin by recalling that the Narain CFT with c$_{\text{\textsc{l}/%
\textsc{r}}}$=2 can be realised using two free real bosonic fields $%
X^{i}(\tau ,\sigma )$ splitting like $X_{L}^{i}(\tau +\sigma
)+X_{R}^{i}(\tau -\sigma )$. The Narain compactification imposes the
periodicity condition $X_{L/R}(\sigma +2\pi )=X_{L/R}(\sigma )+2\pi Rm$ and
the KK\ constraint $e^{i2\pi RP_{L/R}}=1$ which captures the single
valuedness of $X_{L/R}$ and $X_{L/R}+2\pi R.$ The allowed values of the
momenta $P_{L/R}$ are quantized and described in detail in section 4.

\subsubsection{Fermionisation}

In the fermionic representation, we reinterpret the central charge value c$_{%
\text{\textsc{l}/\textsc{r}}}=2$ by expressing the two real free conformal
bosons ($X^{1},X^{2}$) in terms of four free real Fermi fields as shown in
the table below%
\begin{equation}
\begin{tabular}{|c|c|c|}
\hline
{\small \ \ 2 real bosons \ \ } & {\small \ \ 4 real fermions }$\rightarrow $%
{\small \ 2 complex\ } & \ \ c$_{\text{\textsc{l}/\textsc{r}}}$ \ \  \\
\hline
$X^{1}$ & $\left(
\begin{array}{c}
\psi ^{1} \\
\psi ^{2}%
\end{array}%
\right) \rightarrow \psi =\psi ^{1}+i\psi ^{2}$ & 1 \\ \hline
$X^{2}$ & $\left(
\begin{array}{c}
\chi ^{1} \\
\chi ^{2}%
\end{array}%
\right) \rightarrow \chi =\chi ^{1}+i\chi ^{2}$ & 1 \\ \hline
\end{tabular}
\label{tab1}
\end{equation}%
\begin{equation*}
\end{equation*}%
By using the complex field $X=X_{1}+iX_{2}$, the left $J_{z}^{L}$ and the
right $J_{\bar{z}}^{R}$ conformal currents are respectively given by $%
\partial _{z}X_{L}$ and $\partial _{\bar{z}}X_{R}$. In terms of the two
complex Fermi fields $\psi =\psi _{1}+i\psi _{2}$ and $\chi =\chi _{1}+i\chi
_{2}$, these conformal currents read as $J_{z}^{L}\sim $ $:\psi _{L}\psi
_{L}^{\dagger }+\chi _{L}\chi _{L}^{\dagger }:$ and $J_{\bar{z}}^{R}\sim $ $%
:\psi _{R}\psi _{R}^{\dagger }+\chi _{R}\chi _{R}^{\dagger }:.$ Notice that
the two Fermi fields may be also thought of as the two Weyls of a Dirac
spinor: $f_{\uparrow }=\psi $ and $f_{\downarrow }=\chi $. With the compact
notation$\ (\psi ,\chi ):=\mathbf{\varphi }^{a}$, the left and right
currents can be written more concisely as$:\mathbf{\varphi }_{L}^{a}\mathbf{%
\varphi }_{La}^{\dagger }\mathbf{:}$ and $:\mathbf{\varphi }_{R}^{a}\mathbf{%
\varphi }_{Ra}^{\dagger }\mathbf{:}$

Moreover, the classical dynamics of these free conformal fermions is
governed by the massless 2D Dirac equation $i\gamma ^{\mu }\partial _{\mu }%
\mathbf{\varphi }=\mathbf{0}$. After performing a Fourier transform, it
reads as follows $\left( \gamma ^{\mu }q_{\mu }\right) \mathbf{\tilde{\varphi%
}}=\mathbf{0}$ with $\mathbf{\tilde{\varphi}}=\mathbf{\tilde{\varphi}}\left(
q\right) $ and $\gamma ^{\mu }$ defining Dirac matrices in 2D. The above
Dirac equation describes gapless states in continuum; however it also
appears in lattice field theory in the vicinity of the so-called Dirac
points where the Dirac Hamiltonian satisfies \textrm{\cite{tm}}%
\begin{equation}
H_{Dirac}\left\vert \mathbf{\tilde{\varphi}}\right\rangle =0\qquad ,\qquad
E_{g}=E_{+}-E_{-}=0  \label{eg}
\end{equation}%
In lattice modeling, the fermionic fields $\mathbf{\varphi }_{\mathbf{r}%
_{n}}=(\mathbf{\psi }_{\mathbf{r}_{n}},\mathbf{\chi }_{\mathbf{r}_{n}})$ are
defined on a honeycomb (\ref{hc}) with position sites $\mathbf{r}_{n}^{A}\in
\mathbb{A}_{\mathrm{2}}$ and $\mathbf{r}_{n}^{B}\in \mathbb{B}_{\mathrm{2}}.$
The local fermionic fields $\mathbf{\psi }_{\mathbf{r}_{n}}$ and $\mathbf{%
\chi }_{\mathbf{r}_{n}}$ sit respectively at the two superposed lattices as%
\begin{equation}
\begin{tabular}{lll}
$\mathbf{\psi }_{\mathbf{r}_{n}}$ & $=$ & $\mathbf{\varphi }\left( \mathbf{r}%
_{n}^{A}\right) $ \\
$\mathbf{\chi }_{\mathbf{r}_{n}}$ & $=$ & $\mathbf{\varphi }\left( \mathbf{r}%
_{n}^{B}\right) $%
\end{tabular}%
\qquad ,\qquad
\begin{tabular}{lll}
$\mathbf{r}_{\mathbf{n}}^{A}$ & $=$ & $n_{1}\mathbf{\alpha }_{1}+n_{2}%
\mathbf{\alpha }_{2}$ \\
$\mathbf{r}_{\mathbf{n}}^{B}$ & $=$ & $\mathbf{r}_{\mathbf{n}}^{A}+\mathbf{%
\omega }$%
\end{tabular}%
\end{equation}%
where the highest weight vector $\mathbf{\omega }\ $is a fundamental weight
of SU(3); say $\mathbf{\omega =\lambda }_{1}$. They expand like
\begin{equation}
\mathbf{\psi }_{\mathbf{r}_{n}}=\dsum_{\mathbf{k}}e^{i\mathbf{k.r}_{n}}%
\mathbf{\tilde{\psi}}_{\mathbf{k}}\qquad ,\qquad \mathbf{\chi }_{\mathbf{r}%
_{n}}=\dsum_{\mathbf{k}}\sum_{l=1}^{3}e^{i\mathbf{k.}\left( \mathbf{r}_{%
\mathbf{n}}+\mathbf{\omega }_{l}\right) }\mathbf{\tilde{\chi}}_{\mathbf{k}}
\label{ft}
\end{equation}%
with the three $\mathbf{\omega }_{1},\mathbf{\omega }_{2},\mathbf{\omega }%
_{3}$ as in eq(\ref{omi}). In this description, the lattice Hamiltonian
\textsc{H}$_{lat}$ of the KK/W system is given by the sum over the quadratic
interactions $\mathbf{\varphi }_{\mathbf{r}_{n}}^{\dagger }$\textrm{t}$_{nm}%
\mathbf{\varphi }_{\mathbf{r}_{m}}+hc$ expanding as%
\begin{equation}
\begin{tabular}{lll}
$\mathbf{\varphi }_{\mathbf{r}_{n}}^{\dagger }\mathrm{t}_{nm}\mathbf{\varphi
}_{\mathbf{r}_{m}}$ & $=$ & $\mathbf{\psi }_{\mathbf{r}_{n}^{A}}^{\dagger
}t_{\mathbf{r}_{n}^{A}\mathbf{r}_{m}^{A}}\mathbf{\psi }_{\mathbf{r}_{m}^{A}}+%
\mathbf{\chi }_{\mathbf{r}_{n}^{B}}^{\dagger }t_{\mathbf{r}_{n}^{B}\mathbf{r}%
_{m}^{B}}\mathbf{\chi }_{\mathbf{r}_{m}^{B}}+$ \\
&  & $\mathbf{\psi }_{\mathbf{r}_{n}^{A}}^{\dagger }t_{\mathbf{r}_{n}^{A}%
\mathbf{r}_{m}^{B}}\mathbf{\chi }_{\mathbf{r}_{m}^{B}}+\mathbf{\chi }_{%
\mathbf{r}_{n}^{B}}^{\dagger }t_{\mathbf{r}_{n}^{B}\mathbf{r}_{m}^{A}}%
\mathbf{\psi }_{\mathbf{r}_{m}^{A}}$%
\end{tabular}%
\end{equation}%
with coupling tensors \textrm{t}$_{nm}$ function of the positions ($\mathbf{r%
}_{n}^{A},\mathbf{r}_{m}^{B}).$ In models such as the Haldane theory\textrm{%
, }these couplings are subject to the Peierls substitution \textrm{t}$%
_{nm}\rightarrow \mathrm{t}_{nm}\exp (-i\int_{\Upsilon _{nm}}\vec{A}.%
\overrightarrow{dl})$ generating Aharonov-Bohm phases during hoppings. In
practice, one typically includes only the first and the second\textrm{\ }%
coupling orders when calculating the spectrum of lattice hamiltonians
\textsc{H}$_{lat}$ involving the first and the second closest neighbouring
as given below
\begin{equation}
\text{\textsc{H}}_{lat}\simeq \text{\textsc{H}}_{lat}^{(0)}+\dsum\limits_{%
\left\langle \mathbf{r}_{n},\mathbf{r}_{m}\right\rangle }\mathbf{\psi }_{%
\mathbf{r}_{n}}^{\dagger }t_{nm}^{\left( 1\right) }\mathbf{\chi }_{\mathbf{r}%
_{m}}+\dsum\limits_{\left\langle \left\langle \mathbf{r}_{n},\mathbf{r}%
_{m}\right\rangle \right\rangle }t_{nm}^{\left( 2\right) }\left( \mathbf{%
\psi }_{\mathbf{r}_{n}}^{\dagger }\mathbf{\psi }_{\mathbf{r}_{m}}+\mathbf{%
\chi }_{\mathbf{r}_{n}}^{\dagger }\mathbf{\chi }_{\mathbf{r}_{m}}\right) +hc
\label{hlat}
\end{equation}%
which we refer to as \textsc{H}$_{lat}^{(0)}+$\textsc{H}$_{lat}^{(1)}+$%
\textsc{H}$_{lat}^{(2)}$ where we have added the on site energy \textsc{H}$%
_{lat}^{(0)}$ having the form $M(\mathbf{\psi }_{\mathbf{r}_{n}}^{\dagger }%
\mathbf{\psi }_{\mathbf{r}_{n}}-\mathbf{\chi }_{\mathbf{r}_{n}}^{\dagger }%
\mathbf{\chi }_{\mathbf{r}_{n}})$ breaking the inversion symmetry. This
tight binding model mirrors the structure of the Haldane theory for the
anomalous quantum Hall effect \cite{tm,tm1}. In this regard, we recall that
the Haldane theory has non trivial topological phases for the filled band in
2D graphene. Consequently, the Haldane results apply to the KK/W system
imagined in terms of the $\psi /\chi $ fermions. As such, the eq(\ref{hlat})
has a non vanishing first Chern number ($C_{1}\neq 0$) given by the integral
of the curvature of the Berry connection $A=\left\langle \mathbf{k}|\frac{1}{%
i}\nabla _{\mathbf{k}}|\mathbf{k}\right\rangle $ \textrm{\cite{tm,tm1,trs}},
\begin{equation}
C_{1}=\dint\nolimits_{BZ}F
\end{equation}%
where the Berry curvature $F=dA$ (intrinsic magnetic field) is induced by
the second closest neighbors ($M,t_{nm}^{\left( 2\right) }\neq 0$) which
open the gap $E_{g}$ in (\ref{eg}) therefore breaking inversion and time
reversal symmetries due to the non vanishing Berry curvature.

\subsubsection{Hamiltonian Eq(\protect\ref{hlat}) in reciprocal space}

Here, we consider the lattice hamiltonian (\ref{hlat}) and draw the path
leading to Dirac cones, which are essential for the gapless states (massless
Dirac modes) and 2D topological matter. Starting from the lattice \textsc{H}$%
_{lat},$ the analysis ultimately yields the 2-dimensional massless Dirac
equation%
\begin{equation}
\mathcal{D}_{\mathbf{q}}\mathbf{\tilde{\varphi}}=\left( \sigma
^{x}q_{x}+\sigma ^{y}q_{y}\right) \mathbf{\tilde{\varphi}}=0
\end{equation}

\paragraph{A) Expression of \textsc{H}$_{lat}^{(1)}:$}

We first consider the bloc hamiltonian
\begin{equation}
\text{\textsc{H}}_{lat}^{(1)}=\dsum_{\mathbf{r}_{n}\in \mathbb{A}%
}\dsum_{l=1}^{3}t_{l}\mathbf{\psi }_{\mathbf{r}_{n}}^{\dagger }\mathbf{\chi }%
_{\mathbf{r}_{n}+\mathbf{\omega }_{l}}+hc  \label{lh}
\end{equation}%
with coupling parameters $t_{l}$ (in general complex) and where the vectors%
\textrm{\ }($\mathbf{\omega }_{1},\mathbf{\omega }_{2},\mathbf{\omega }_{3})$%
\textrm{\ }represent the displacements to the three nearest neighbors of the
site $\mathbf{r}_{\mathbf{n}}^{A}$ given by $\mathbf{r}_{\mathbf{n}}^{B}=%
\mathbf{r}_{\mathbf{n}}^{A}+\mathbf{\omega }_{l}$; these neighbours generate
the fundamental dimensional representation of SU(3) and can be expressed in
terms of the highest weight vector $\mathbf{\lambda }_{1}$ as follows
\begin{equation}
\left(
\begin{array}{c}
\mathbf{\omega }_{1} \\
\mathbf{\omega }_{2} \\
\mathbf{\omega }_{3}%
\end{array}%
\right) =\left(
\begin{array}{c}
\mathbf{\lambda }_{1} \\
\mathbf{\lambda }_{1}-\mathbf{\alpha }_{1} \\
\mathbf{\lambda }_{1}-\mathbf{\alpha }_{1}-\mathbf{\alpha }_{2}%
\end{array}%
\right) =\frac{1}{3}\left(
\begin{array}{c}
2\mathbf{\alpha }_{1}+\mathbf{\alpha }_{2} \\
-\mathbf{\alpha }_{1}+\mathbf{\alpha }_{2} \\
-\mathbf{\alpha }_{1}-2\mathbf{\alpha }_{2}%
\end{array}%
\right)  \label{omi}
\end{equation}%
Notice that these weight vectors satisfy $\mathbf{\omega }_{1}+\mathbf{%
\omega }_{2}+\mathbf{\omega }_{3}=0$ reflecting the traceleness condition of
SU(3) matrices. Furthermore, the weight vectors satisfy some useful
relations such as $\mathbf{\omega }_{1}-\mathbf{\omega }_{2}=\mathbf{\alpha }%
_{1}$\ and $\mathbf{\omega }_{1}-\mathbf{\omega }_{3}=\mathbf{\alpha }_{3}$
where we set $\mathbf{\alpha }_{3}=\mathbf{\alpha }_{1}+\mathbf{\alpha }%
_{2}. $ Substituting the Fourier transform of the lattice fields ( $\mathbf{%
\varphi }_{\mathbf{r}_{n}}=\sum_{\mathbf{k}}e^{i\mathbf{k.r}_{n}}\mathbf{%
\tilde{\varphi}}_{\mathbf{k}}$) in the hamiltonian (\ref{lh}), we can bring
the \textsc{H}$_{lat}^{(1)}$ into the quadratic form%
\begin{equation}
\text{\textsc{H}}_{lat}^{(1)}=\dsum_{\mathbf{k}\in \mathbb{R}^{2}}\left(
\mathbf{\tilde{\psi}}_{\mathbf{k}}^{\dagger },\mathbf{\tilde{\chi}}_{\mathbf{%
k}}^{\dagger }\right) \mathbb{H}_{lat}^{(1)}\left(
\begin{array}{c}
\mathbf{\tilde{\psi}}_{\mathbf{k}} \\
\mathbf{\tilde{\chi}}_{\mathbf{k}}%
\end{array}%
\right)  \label{1k}
\end{equation}%
with the 2$\times $2 matrix $\mathbb{H}_{lat}^{(1)}$ given by%
\begin{equation}
\mathbb{H}_{lat}^{(1)}=\left(
\begin{array}{cc}
0 & f_{\mathbf{k}} \\
f_{\mathbf{k}}^{\ast } & 0%
\end{array}%
\right) ,\qquad
\begin{tabular}{lll}
$f_{\mathbf{k}}$ & $=$ & $t_{1}e^{i\mathbf{k.\omega }_{1}}+t_{2}e^{i\mathbf{%
k.\omega }_{2}}+t_{3}e^{i\mathbf{k.\omega }_{3}}$ \\
& $=$ & $te^{i\mathbf{k.\omega }_{1}}\left( 1+e^{-i\mathbf{k.\alpha }%
_{1}}+e^{-i\mathbf{k.\alpha }_{3}}\right) $%
\end{tabular}
\label{LH}
\end{equation}%
where we have assumed real $t_{i}=t$. By decomposing the function $f_{%
\mathbf{k}}$ like $h_{\mathbf{k}}^{(x)}+ih_{\mathbf{k}}^{(y)}$ with $h_{%
\mathbf{k}}^{(x)}=t\sum_{l=1}^{3}\cos \mathbf{k.\omega }_{l}$ and $h_{%
\mathbf{k}}^{(y)}=-t\sum_{l=1}^{3}\sin \mathbf{k.\omega }_{l},$ we finally
obtain
\begin{equation}
\mathbb{H}_{lat}^{(1)}=h_{\mathbf{k}}^{x}\sigma ^{x}+h_{\mathbf{k}%
}^{y}\sigma ^{y}
\end{equation}

\paragraph{B) Expression of \textsc{H}$_{lat}^{(2)}:$}

We now consider the block hamiltonian \textsc{H}$_{lat}^{(2)}$ describing
the couplings between second closest neighbours on the lattice%
\begin{equation}
\text{\textsc{H}}_{lat}^{(2)}=\dsum_{\mathbf{r}_{n}}\dsum_{a=1}^{6}\left(
\mathbf{\psi }_{\mathbf{r}_{n}}^{\dagger }t_{\mathbf{r}_{n},\mathbf{r}_{n}+%
\mathbf{\varpi }_{a}}^{\psi \psi }\mathbf{\psi }_{\mathbf{r}_{n}+\mathbf{%
\varpi }_{a}}+\mathbf{\chi }_{\mathbf{r}_{n}}^{\dagger }t_{\mathbf{r}_{n},%
\mathbf{r}_{n}+\mathbf{\varpi }_{a}}^{\chi \chi }\mathbf{\chi }_{\mathbf{r}%
_{n}+\mathbf{\varpi }_{a}}\right) +hc
\end{equation}%
here, $t_{\mathbf{r}_{n},\mathbf{r}_{n}+\mathbf{\varpi }_{a}}^{\psi \psi }$
and $t_{\mathbf{r}_{n},\mathbf{r}_{n}+\mathbf{\varpi }_{a}}^{\chi \chi }$
are complex coupling functions of ($\mathbf{r}_{n},\mathbf{r}_{m}$). The six
$\mathbf{\varpi }_{a}$s label the second nearest neighbours of a given site $%
\mathbf{r}_{n}$ with $\mathbf{r}_{n}-\mathbf{r}_{m}=\mathbf{\varpi }_{a}.$
They read in terms of the simple roots of SU(3) as follows,
\begin{equation}
\begin{tabular}{lll}
$\mathbf{\varpi }_{1}$ & $=$ & $+\mathbf{\alpha }_{1}$ \\
$\mathbf{\varpi }_{4}$ & $=$ & $-\mathbf{\alpha }_{1}$%
\end{tabular}%
\quad ,\quad
\begin{tabular}{lll}
$\mathbf{\varpi }_{2}$ & $=$ & $+\mathbf{\alpha }_{2}$ \\
$\mathbf{\varpi }_{5}$ & $=$ & $-\mathbf{\alpha }_{2}$%
\end{tabular}%
\quad ,\quad
\begin{tabular}{lll}
$\mathbf{\varpi }_{3}$ & $=$ & $+\mathbf{\alpha }_{3}$ \\
$\mathbf{\varpi }_{6}$ & $=$ & $-\mathbf{\alpha }_{3}$%
\end{tabular}%
\end{equation}%
where we have set $\mathbf{\alpha }_{3}=\mathbf{\alpha }_{1}+\mathbf{\alpha }%
_{2}$. Due to the homogeneity of the honeycomb lattice, the coupling
parameters can be assumed to depend only on the difference $\mathbf{r}_{n}-%
\mathbf{r}_{m}$ leading to constant $t_{a}^{\psi \psi }$ and $t_{a}^{\chi
\chi }$. Being complex, these couplings can be moreover parameterised like $%
t_{a}^{\psi \psi }=t^{\psi \psi }e^{i\Phi _{a}}$ and $t_{a}^{\chi \chi
}=t^{\chi \chi }e^{i\Phi _{a}}.$ And by using the Fourier transforms, we can
put \textsc{H}$_{lat}^{(2)}$ in the form%
\begin{equation}
\text{\textsc{H}}_{lat}^{(2)}=\dsum_{\mathbf{k}\in \mathbb{R}^{2}}\left(
\mathbf{\tilde{\psi}}_{\mathbf{k}}^{\dagger },\mathbf{\tilde{\chi}}_{\mathbf{%
k}}^{\dagger }\right) \mathbb{H}_{lat}^{(2)}\left(
\begin{array}{c}
\mathbf{\tilde{\psi}}_{\mathbf{k}} \\
\mathbf{\tilde{\chi}}_{\mathbf{k}}%
\end{array}%
\right)
\end{equation}%
with%
\begin{equation}
\mathbb{H}_{lat}^{(2)}=\left(
\begin{array}{cc}
\func{Re}g_{\mathbf{k}}^{\psi \psi }+M & 0 \\
0 & \func{Re}g_{\mathbf{k}}^{\chi \chi }-M%
\end{array}%
\right)
\end{equation}%
and%
\begin{equation}
g_{\mathbf{k}}^{\psi \psi }=\dsum_{a=1}^{6}t_{a}^{\psi \psi }e^{i\mathbf{%
k.\omega }_{a}}\qquad ,\qquad g_{\mathbf{k}}^{\chi \chi
}=\dsum_{a=1}^{6}t_{a}^{\chi \chi }e^{i\mathbf{k.\omega }_{a}}  \label{k2}
\end{equation}%
Adding (\ref{1k},\ref{k2})$,$ we get%
\begin{equation}
\mathbb{H}_{lat}=\left(
\begin{array}{cc}
\func{Re}g_{\mathbf{k}}^{\psi \psi }+M & f_{\mathbf{k}} \\
f_{\mathbf{k}}^{\ast } & \func{Re}g_{\mathbf{k}}^{\chi \chi }-M%
\end{array}%
\right)
\end{equation}%
with eigenvalues%
\begin{equation}
E_{\pm }=\func{Re}[g_{\mathbf{k}}^{\psi \psi }+g_{\mathbf{k}}^{\chi \chi
}]\pm \sqrt{\left( \func{Re}[g_{\mathbf{k}}^{\psi \psi }-g_{\mathbf{k}%
}^{\chi \chi }]-M\right) ^{2}+f_{\mathbf{k}}f_{\mathbf{k}}^{\ast }}
\end{equation}%
and gap energy $E_{g}=2[(\func{Re}[g_{\mathbf{k}}^{\psi \psi }-g_{\mathbf{k}%
}^{\chi \chi }]-M)^{2}+f_{\mathbf{k}}f_{\mathbf{k}}^{\ast }]^{1/2}.$

\subsection{Dirac cones and gapless states}

\begin{figure}[tbph]
\begin{center}
\includegraphics[width=8cm]{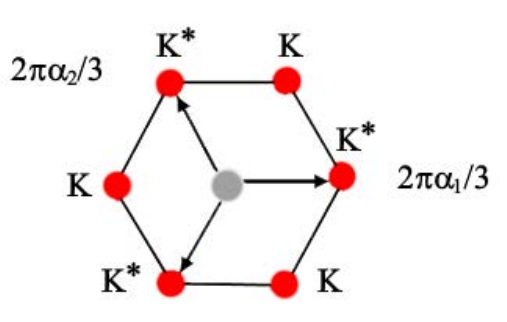}
\end{center}
\par
\vspace{-0.5cm}
\caption{Dirac points $\mathbf{\protect\kappa }_{0}$ and $\mathbf{\protect%
\kappa }_{0}^{\ast }$ as well as the action do the $\mathbb{Z}_{3}$
transformations.}
\label{KE}
\end{figure}

In this subsection, we consider the gapless regime by setting the mass and
diagonal couplings to $M=t_{a}^{\psi \psi }=t_{a}^{\chi \chi }=0$ in order
to derive the location of the Dirac cones by computing the zero modes of $%
\mathbb{H}_{lat}^{(1)}$. To that purpose, observe first that $f_{\mathbf{k}}$
is invariant under cyclic permutation ($\mathbf{\omega }_{1},\mathbf{\omega }%
_{2},\mathbf{\omega }_{3}$) generating $\mathbb{Z}_{3};$ and as such, $f_{%
\mathbf{k}}$ can be also factorised in two equivalent ways as follows:
\begin{equation}
f_{\mathbf{k}}=te^{i\mathbf{k.\omega }_{2}}\left( 1+e^{i\mathbf{k.\alpha }%
_{1}}+e^{-i\mathbf{k.\alpha }_{2}}\right) ,\qquad f_{\mathbf{k}}=te^{i%
\mathbf{k.\omega }_{3}}\left( 1+e^{-i\mathbf{k.\alpha }_{2}}+e^{-i\mathbf{%
k.\alpha }_{3}}\right)
\end{equation}%
The two energy eigenvalues E$_{\pm }$ of the matrix $\mathbb{H}_{lat}$ are
given by $\pm \sqrt{f_{\mathbf{k}}f_{\mathbf{k}}^{\ast }}$ from which we
deduce the expression of the gap energy which is equal to $E_{g}=2\sqrt{f_{%
\mathbf{k}}f_{\mathbf{k}}^{\ast }}.$ Hence, the condition for a vanishing
gap is given by $f_{\mathbf{k}}=0$; its solution can be inspired from the
well known identity $1+e^{i\frac{2\pi }{3}}+e^{i\frac{4\pi }{3}}=0$
concerning the three roots of $z^{3}=1.$ It leads to two different solutions
(mod $\mathbb{Z}_{3}$ transformations), one at $\mathbf{k}=\mathbf{\kappa }%
_{0}$ and the other at $\mathbf{k}=\mathbf{\kappa }_{0}^{\ast }$. They are
often termed as the Dirac points and are given by the particular momenta%
\begin{equation}
\begin{tabular}{|c|c|c|}
\hline
\multicolumn{3}{|c|}{Dirac point $\mathbf{\kappa }_{0}$} \\ \hline
$\ \ \mathbf{\kappa }_{0}\mathbf{.\alpha }_{1}$ & $=$ & $\ \ \frac{2\pi }{3}$
$\func{mod}2\pi $ \ \ \  \\ \hline
$\ \ \mathbf{\kappa _{0}.\alpha }_{3}$ & $=$ & $\ \ \frac{4\pi }{3}$ $\func{%
mod}2\pi $ \ \ \  \\ \hline
\end{tabular}%
\qquad \Rightarrow \qquad
\begin{tabular}{lll}
$\mathbf{\kappa }_{0}$ & $=$ & $\frac{2\pi }{3}\left( \mathbf{\lambda }_{1}+%
\mathbf{\lambda }_{2}\right) $ \\
& $=$ & $\frac{2\pi }{3}\left( \mathbf{\alpha }_{1}+\mathbf{\alpha }%
_{2}\right) $%
\end{tabular}%
\end{equation}%
or equivalently $\mathbf{\kappa }_{\mathbf{m}}=\mathbf{\kappa }_{0}+2\pi
\left( m_{1}\mathbf{\lambda }_{1}+m_{2}\mathbf{\lambda }_{2}\right) $ with
integers $m_{i}.$ Additionally, we have%
\begin{equation}
\begin{tabular}{|c|c|c|}
\hline
\multicolumn{3}{|c|}{Dirac point $\mathbf{\kappa }_{0}^{\ast }$} \\ \hline
$\ \ \mathbf{\kappa }_{0}^{\ast }\mathbf{.\alpha }_{1}$ & $=$ & $\ \ \frac{%
4\pi }{3}$ $\func{mod}2\pi $ \ \ \  \\ \hline
$\ \ \mathbf{\kappa }_{0}^{\ast }\mathbf{.\alpha }_{3}$ & $=$ & $\ \ \frac{%
2\pi }{3}$ $\func{mod}2\pi $ \ \ \  \\ \hline
\end{tabular}%
\qquad \Rightarrow \qquad
\begin{tabular}{lll}
$\mathbf{\kappa }_{0}^{\ast }$ & $=$ & $\frac{2\pi }{3}\left( 2\mathbf{%
\lambda }_{1}-\mathbf{\lambda }_{2}\right) $ \\
& $=$ & $\frac{2\pi }{3}\mathbf{\alpha }_{1}$%
\end{tabular}%
\end{equation}%
or equivalently $\mathbf{\kappa }_{\mathbf{m}}^{\ast }=\mathbf{\kappa }%
_{0}^{\ast }+2\pi \left( m_{1}\mathbf{\alpha }_{1}+m_{2}\mathbf{\alpha }%
_{2}\right) $ giving the Brillouin zone. The Dirac points together with $%
\mathbb{Z}_{3}$ symmetry are depicted by the Figure \textbf{\ref{KE}}.
Near these Dirac points; say $\mathbf{k=\kappa _{0}+q}$ with small $\mathbf{q%
},$ the function $f$ approximates at leading order like $3te^{i\frac{\pi }{6}%
}\mathbf{q.\lambda }_{1}$. Putting into $f_{\mathbf{k}}=te^{i\mathbf{%
k.\omega }_{2}}\left( 1+e^{i\mathbf{k.\alpha }_{1}}+e^{-i\mathbf{k.\alpha }%
_{2}}\right) $ with $\mathbf{\kappa }_{0}=\frac{2\pi }{3}\left( \mathbf{%
\lambda }_{1}+\mathbf{\lambda }_{2}\right) $ and $\mathbf{\kappa }_{0}%
\mathbf{.\omega }_{2}=2\pi ,$ then using $\mathbf{\kappa }_{0}\mathbf{%
.\alpha }_{1}=\frac{2\pi }{3}$ and $\mathbf{\kappa }_{0}\mathbf{.\alpha }%
_{2}=\frac{2\pi }{3},$ we get%
\begin{equation}
\begin{tabular}{lll}
$f_{\mathbf{q}}$ & $\simeq $ & $te^{i\mathbf{\kappa }_{0}\mathbf{.\alpha }%
_{1}}\left( \mathbf{q.\alpha }_{1}\right) +te^{-i\mathbf{\kappa }_{0}\mathbf{%
.\alpha }_{2}}\left( \mathbf{q.\alpha }_{2}\right) +O\left( \mathbf{q}%
^{2}\right) $ \\
& $\simeq $ & $-\frac{t}{2}\mathbf{q.}\left( \mathbf{\alpha }_{1}+\mathbf{%
\alpha }_{2}\right) +i\frac{t\sqrt{3}}{2}\mathbf{q.}\left( \mathbf{\alpha }%
_{1}-\mathbf{\alpha }_{2}\right) +O\left( \mathbf{q}^{2}\right) $%
\end{tabular}%
\end{equation}%
Substituting in the hamiltonian (\ref{LH}) while setting $f_{\mathbf{q}%
}:=q_{x}-iq_{y}+O\left( q^{2}\right) $ with $q_{x}=-\frac{t}{2}\mathbf{q.}%
\left( \mathbf{\alpha }_{2}+\mathbf{\alpha }_{1}\right) $ and $q_{y}=\frac{t%
\sqrt{3}}{2}\mathbf{q.}\left( \mathbf{\alpha }_{2}-\mathbf{\alpha }%
_{1}\right) $, we end up with%
\begin{equation}
\mathbb{H}_{loc}=\left(
\begin{array}{cc}
0 & q_{x}-iq_{y} \\
q_{x}+iq_{y} & 0%
\end{array}%
\right) +O\left( \mathbf{q}^{2}\right)
\end{equation}%
where the leading linear term is the Dirac operator $\mathcal{D}_{\mathbf{q}%
} $.

\section{Conclusion and discussion}

\label{sec:6} In this study, we began by revisiting the so-called
construction A of code CFTs realising Narain CFTs with an initial emphasis%
\textrm{\ }on the lattices ($\mathbf{\Lambda }_{\mathrm{k}}^{\mathrm{1,1}},%
\mathbf{\Lambda }_{\mathrm{k}\mathcal{C}}^{\mathrm{1,1}},\mathbf{\Lambda }_{%
\mathrm{k}}^{\ast \mathrm{1,1}}$) sitting in $\mathbb{R}^{1,1}$ and then%
\textrm{\ }extended the analysis to their higher dimensional generalisations
($\mathbf{\Lambda }_{\mathrm{k}}^{\mathrm{r,r}},\mathbf{\Lambda }_{\mathrm{k}%
\mathcal{C}}^{\mathrm{r,r}},\mathbf{\Lambda }_{\mathrm{k}}^{\ast \mathrm{r,r}%
})$. These lattices are\textrm{\ }characterised by the central charges of
the 2D NCFT ($c_{\text{\textsc{l}}}=c_{\text{\textsc{r}}}=\mathrm{r}$) and
the CS level $\mathrm{k}$ of the 3D holographic dual. Then, in \textrm{%
section 2}, we showed that the lattice $\mathbf{\Lambda }_{\mathrm{k}}^{%
\mathrm{1,1}}$ and its dual $\mathbf{\Lambda }_{\mathrm{k}}^{\ast \mathrm{1,1%
}}$ as well as the even self dual $\mathbf{\Lambda }_{\mathrm{k}\mathcal{C}%
}^{\mathrm{1,1}}$ exhibit remarkable properties; in particular the following:

$\left( \mathbf{1}\right) $ They can be imagined via the nested structure:%
\begin{equation}
\mathbf{\Lambda }_{\mathrm{k}}^{\mathbf{su}_{2}}\subset \mathbf{\Lambda }_{%
\mathrm{k}\mathcal{C}}^{\mathbf{su}_{2}}\subset \mathbf{\Lambda }_{\mathrm{k}%
}^{\ast \mathbf{su}_{2}}\subset \mathbb{R}^{1,1}  \label{su1}
\end{equation}%
with components provided by fibrations of root-like \textsc{R}$_{\mathrm{k}%
}^{\mathbf{su}_{2}}$ and weight-like \textsc{W}$_{\mathrm{k}}^{\mathbf{su}%
_{2}}$ lattices of the su(2) algebra, given explicitly by:
\begin{equation}
\mathbf{\Lambda }_{\mathrm{k}}^{\mathbf{su}_{2}}=\text{\textsc{R}}_{\mathrm{k%
}}^{\mathbf{su}_{2}}\times \text{\textsc{R}}_{\mathrm{k}}^{\mathbf{su}%
_{2}},\qquad \mathbf{\Lambda }_{\mathrm{k}\mathcal{C}}=\text{\textsc{R}}_{%
\mathrm{k}}^{\mathbf{su}_{2}}\times \text{\textsc{W}}_{\mathrm{k}}^{\mathbf{%
su}_{2}},\qquad \mathbf{\Lambda }_{\mathrm{k}}^{\ast }=\text{\textsc{W}}_{%
\mathrm{k}}^{\mathbf{su}_{2}}\times \text{\textsc{W}}_{\mathrm{k}}^{\mathbf{%
su}_{2}}
\end{equation}%
We refer to them as root-like and weight-like because only for k=2, the
lattices coincide exactly with the standard root and weight lattices of
su(2). Generally speaking, \textsc{R}$_{\mathrm{k}}^{\mathbf{su}_{2}}$
consists of points of the form$\{n\mathbf{\tilde{\alpha}}$ $\mathbf{,}$%
\textbf{\ }$n\in \mathbb{Z\}}$ while \textsc{W}$_{\mathrm{k}}^{\mathbf{su}%
_{2}}$ is defined by $\{m\mathbf{\tilde{\lambda},}$ $m$ $\in \mathbb{Z}\}$
with dilated root $\mathbf{\tilde{\alpha}}$ and compressed fundamental
weight $\mathbf{\tilde{\lambda}}$ related to their standard counterparts $%
\mathbf{\alpha }$ and $\mathbf{\lambda }$ as
\begin{equation*}
\mathbf{\tilde{\alpha}=}\sqrt{\frac{\mathrm{k}}{2}}\mathbf{\alpha \qquad
,\qquad \tilde{\lambda}}=\sqrt{\frac{2}{\mathrm{k}}}\mathbf{\lambda }
\end{equation*}%
For physical applications, we interpreted the sites \textbf{x}$_{n}\in $%
\textsc{R}$_{\mathrm{k}}^{\mathbf{su}_{2}}$ as hosts of Kaluza-Klein
excitations $\left\vert KK_{n}\right\rangle $ and the sites \textbf{k}$%
_{m}\in $\textsc{W}$_{\mathrm{k}}^{\mathbf{su}_{2}}\backslash $\textsc{R}$_{%
\mathrm{k}}^{\mathbf{su}_{2}}$ as carriers of winding modes $\left\vert
W_{n}\right\rangle $. A detailed discussion of this description was given
\textrm{in section 5}.

$\left( \mathbf{2}\right) $ The one dimensional lattices \textsc{R}$_{%
\mathrm{k}}=\sqrt{\mathrm{k}}\mathbb{Z}$ and \textsc{W}$_{\mathrm{k}}=\frac{1%
}{\sqrt{\mathrm{k}}}\mathbb{Z}$ defined for a generic CS level\textrm{\ k, }%
exhibit several notable properties: First, for\textrm{\ k=1}, they coincide
\textsc{R}$_{\mathrm{1}}^{\mathbf{su}_{\mathbf{2}}}=$\textsc{W}$_{\mathrm{1}%
}^{\mathbf{su}_{\mathbf{2}}}.$ In this case, the lattice is integral and
self dual with generators obeying $\mathbf{\tilde{\alpha}}^{2}=\mathbf{%
\tilde{\lambda}}^{2}=1$\textbf{.} Second, for\textrm{\ k=2,} \textsc{R}$_{%
\mathrm{2}}^{\mathbf{su}_{\mathbf{2}}}=\sqrt{\mathrm{2}}\mathbb{Z}$
correspond to the root lattice of SU(2)\textrm{\ }and\textrm{\ }\textsc{W}$_{%
\mathrm{2}}^{\mathbf{su}_{\mathbf{2}}}=\frac{1}{\sqrt{\mathrm{2}}}\mathbb{Z}$
is its weight lattice. The former is even integer valued whereas the latter
is half integer.\textrm{\ }Third, these discrete lattices \textsc{R}$_{%
\mathrm{2}}^{\mathbf{su}_{\mathbf{2}}}$ and \textsc{W}$_{\mathrm{2}}^{%
\mathbf{su}_{\mathbf{2}}}$ can be identified geometrically as\textrm{\ }the
(anti)-diagonal 1D sublattices of the ambient $\mathbb{Z}\times \mathbb{Z}$
and the $\frac{1}{\mathrm{2}}\mathbb{Z}\times \frac{1}{\mathrm{2}}\mathbb{Z}$
square lattices respectively.

$(\mathbf{3})$ A detailed inspection of these properties leads to the key
identification:\textrm{\ }\textsc{R}$_{\mathrm{k}}^{\mathbf{su}_{\mathbf{2}%
}}=\sqrt{\frac{\mathrm{k}}{2}}\mathbb{Z}\mathbf{\alpha }$ and \textsc{W}$_{%
\mathrm{k}}^{\mathbf{su}_{\mathbf{2}}}=\sqrt{\frac{2}{\mathrm{k}}}\mathbb{Z}%
\mathbf{\lambda }$ where $\mathbf{\alpha }$\ and $\mathbf{\lambda }$\ are
the usual simple root and the fundamental weight $\mathbf{\lambda }$ vectors
of SU(2) which satisfy $\mathbf{\alpha }^{2}=2,$ $\mathbf{\lambda }^{2}=1/2$
as well as $\mathbf{\alpha .\lambda }=1.$ For generic CS levels\textrm{\ k%
\TEXTsymbol{>}2}, the inner products scale like%
\begin{equation*}
\mathbf{\tilde{\alpha}}^{2}=\mathrm{k},\qquad \mathbf{\tilde{\lambda}}^{2}=%
\frac{1}{\mathrm{k}},\qquad \mathbf{\tilde{\alpha}.\tilde{\lambda}}=1
\end{equation*}%
From the above description,\ several important consequences follow:

\begin{description}
\item[$\left( \mathbf{i}\right) $] there is an embedding hierarchy \textsc{R}%
$_{\mathrm{k}}^{\mathbf{su}_{\mathbf{2}}}\subseteq $\textsc{W}$_{\mathrm{k}%
}^{\mathbf{su}_{\mathbf{2}}}$ for all $\mathrm{k\geq 1}$. The discriminant
group \textsc{W}$_{\mathrm{k}}^{\mathbf{su}_{\mathbf{2}}}/$\textsc{R}$_{%
\mathrm{k}}^{\mathbf{su}_{\mathbf{2}}}$ is isomorphic to $\mathbb{Z}_{%
\mathrm{k}}=\mathbb{Z}/(\mathrm{k}\mathbb{Z})$, in agreement with the well
known property regarding the group centre of SU(2)\textrm{\ }implying
\textsc{W}$_{\mathrm{2}}^{\mathbf{su}_{\mathbf{2}}}/$\textsc{R}$_{\mathrm{2}%
}^{\mathbf{su}_{\mathbf{2}}}\simeq \mathbb{Z}_{\mathrm{2}}$.

\item[$\left( \mathbf{ii}\right) $] For $\mathrm{k>2,}$ \textsc{W}$_{\mathrm{%
k}}^{\mathbf{su}_{\mathbf{2}}}$ can be decomposed as a union of\textrm{\ }k
isomorphic sublattices \{\textsc{R}$_{\mathrm{k}}^{\mathbf{su}_{\mathbf{2}%
}}+\varepsilon \mathbf{\lambda \}}$ labeled as
\begin{equation}
\text{\textsc{W}}_{\mathrm{k}}^{\mathbf{su}_{\mathbf{2}}}=\text{\textsc{R}}_{%
\mathrm{k}}^{\mathbf{su}_{\mathbf{2}}}+\varepsilon \mathbf{\lambda }\quad ,%
\mathbf{\qquad }\varepsilon =0,...,k-1\func{mod}\mathrm{k}
\end{equation}

\item[$\left( \mathbf{iii}\right) $] Given these \textsc{R}$_{\mathrm{k}}^{%
\mathbf{su}_{\mathbf{2}}}$ and \textsc{W}$_{\mathrm{k}}^{\mathbf{su}_{%
\mathbf{2}}}$, one can then build the triplet $\mathbf{\Lambda }_{\mathrm{k}%
}^{\mathbf{su}_{\mathbf{2}}},\mathbf{\Lambda }_{\mathrm{k}\mathcal{C}}^{%
\mathbf{su}_{\mathbf{2}}},\mathbf{\Lambda }_{\mathrm{k}}^{\ast \mathbf{su}_{%
\mathbf{2}}}$ following the fibration prescription. For example, the even
self dual lattice $\mathbf{\Lambda }_{\mathrm{k}\mathcal{C}}^{\mathbf{su}_{%
\mathbf{2}}}$ is realised as the product \textsc{R}$_{\mathrm{k}}^{\mathbf{su%
}_{\mathbf{2}}}\times $\textsc{W}$_{\mathrm{k}}^{\mathbf{su}_{\mathbf{2}}}$.
Vector sites $\mathbf{u}_{n,m}\in \mathbf{\Lambda }_{\mathrm{k}\mathcal{C}}^{%
\mathbf{su}_{\mathbf{2}}}$ are labeled by two integers with components $n%
\mathbf{\tilde{\alpha}}\oplus m\mathbf{\tilde{\lambda}}$; its Euclidian
pairing $\mathbf{u}_{n,m}^{T}\mathbf{u}_{n,m}$ is given by $\mathrm{k}n^{2}+%
\frac{1}{\mathrm{k}}m^{2}$\ while its Lorentzian product reads as follows%
\begin{equation}
\mathbf{u}_{n,m}^{T}\mathbf{\eta u}_{n,m}=2nm\in 2\mathbb{Z}
\end{equation}%
The self duality feature of $\mathbf{\Lambda }_{\mathrm{k}\mathcal{C}}$ is
captured by the unimodular property of the matrix $\tilde{A}_{i}^{j}\oplus
\tilde{B}_{i}^{j}$ which is manifestly ensured by the duality relation $%
\mathbf{\tilde{\alpha}.\tilde{\lambda}}=1$. For graphic illustrations of the
discrete structures of the various lattices, see the Figures \textbf{\ref{j1}%
}, \textbf{\ref{j2}}, \textbf{\ref{21a}}, \textbf{\ref{22} }and \textbf{\ref%
{23}}.
\end{description}

\ \ \newline
Moreover, building on the SU(2) results, we constructed new realisations of
Narain CFTs based on the root-like \textsc{R}$_{\mathrm{k}}^{\mathbf{su}%
_{3}} $ and the weight-like \textsc{W}$_{\mathrm{k}}^{\mathbf{su}_{3}}$
lattices. Several structural features established for su(2) carry over
naturally to the SU(3)- based CFT. Particularly, for generic CS levels%
\textrm{\ k}, the inclusion \textsc{R}$_{\mathrm{k}}^{\mathbf{su}%
_{3}}\subset $\textsc{W}$_{\mathrm{k}}^{\mathbf{su}_{3}}$ holds and the
discriminant \textsc{W}$_{\mathrm{k}}^{\mathbf{su}_{3}}/$\textsc{R}$_{%
\mathrm{k}}^{\mathbf{su}_{3}}$ group is isomorphic to $\mathbb{Z}_{\mathrm{k}%
}$. As with SU(2), the case of \textrm{k=3}, the symmetry $\mathbb{Z}_{%
\mathrm{3}}$ emerging from the discriminant corresponds to the centre of
SU(3) generated by the 3-cycle $(1,2,3)$ acting on the fundamental 3D
representations of SU(3) as given by (\ref{omi}).

This SU(3) based construction was initially developed for central charges
\textrm{c}$_{L/R}=2$ but it generalizes directly to \textrm{c}$_{L/R}=2%
\mathrm{d}$ by using higher dimensional self dual lattices $\mathbf{\Lambda }%
_{\mathrm{k}\mathcal{C}}^{\mathbf{su}_{3}(\mathrm{d,d})}\subset \mathbb{R}^{2%
\mathrm{d},2\mathrm{d}}$ with nested structure as:%
\begin{equation}
\mathbf{\Lambda }_{\mathrm{k}}^{\mathbf{su}_{3}(\mathrm{d,d})}\subset
\mathbf{\Lambda }_{\mathrm{k}\mathcal{C}}^{\mathbf{su}_{3(\mathrm{d,d}%
)}}\subset \mathbf{\Lambda }_{\mathrm{k}}^{\ast \mathbf{su}_{3}(\mathrm{d,d}%
)}\subset \mathbb{R}^{2\mathrm{d},2\mathrm{d}}  \label{seq}
\end{equation}%
For d=1, the three lattices in the above sequence (\ref{seq}) correspond
respectively to the fibrations \textsc{R}$_{\mathrm{k}}^{\mathbf{su}%
_{3}}\times $\textsc{R}$_{\mathrm{k}}^{\mathbf{su}_{3}}$ and \textsc{R}$_{%
\mathrm{k}}^{\mathbf{su}_{3}}\times $\textsc{W}$_{\mathrm{k}}^{\mathbf{su}%
_{3}}$ as well as \textsc{W}$_{\mathrm{k}}^{\mathbf{su}_{3}}\times $\textsc{W%
}$_{\mathrm{k}}^{\mathbf{su}_{3}}$. To illustrate this construction, we gave
several explicit realisations as depicted by the Figures \textbf{\ref{31}},
\textbf{\ref{32}}, \textbf{\ref{33} }and \textbf{\ref{34}. }The
generalisation to d\TEXTsymbol{>}1 follows straightforwardly.

After establishing this new framework in terms of algebraic data of code
based Narain CFT, we exploited the new setting to develop a pathway towards
topological phases of matter by linking Narain CFT to critical lattice QFT.%
\textrm{\ }We set out by describing the particle content of the model by
studying the lattice fields excitations that manifest as KK and winding
modes.\textrm{\ }We then proposed three guiding arguments based on: (\textbf{%
i}) the structure of the weight lattice, (\textbf{ii}) the duality with the
root grid and the correspondence with the KK/winding system in addition to (%
\textbf{iii}) the possibility of fermionising the bosonic excitations to
introduce a tight binding model with Dirac cones.\textrm{\ }By deriving the
particle states populating the critical lattice points, we were thus able to
capture key features of gapless topological systems akin to that of the
Haldane theory of quantum anomalous Hall effect.\textrm{\ }This unlocked
several promising research directions among which is the embedding of these
lattice dualities into holographic ones with a one to one correspondence
between the bulk duals.\textrm{\ }It will be interesting to investigate
whether one can construct a bulk theory to the 2D lattice QFT describing the
topological matter as a realisation of the AB Chern-Simons theory dual to
the code based Narain CFT.

\section*{Appendices:}

The following three appendices provide technical details omitted from the
main text in order to alleviate and streamline the presentation. In appendix
A, we give a review on the link between code CFTs and Narain ensembles. In
appendix B, we describe construction A of code Narain CFT with explicit
realisations. In appendix C, we complete the analysis of section 3 by
discussing the cases of k\TEXTsymbol{>}3 and k\TEXTsymbol{<}3. \appendix

\section{From code-CFT to Narain ensemble}

\label{sec:A} \renewcommand{\theequation}{\Alph{section}.\arabic{equation}}
In this appendix, we review the construction of two-dimensional Narain
theories from the perspective of code-based CFTs.\textrm{\ }We begin by
outlining the key elements of code-CFT constructions and establishing their
correspondence with Narain CFTs through the associated partition functions%
\textrm{. }We then examine the sequence of lattices these constructions
generate and revisit their interpretation.

\subsection*{A.1 Overview on code-CFT}

Over the past decade, an increasing interest has been devoted to the
connection between additive codes and a class of Narain conformal field
theories (NCFT) with U(1)$^{\mathrm{c}_{\text{\textsc{l}}}}\times $U(1)$^{%
\mathrm{c}_{\text{\textsc{r}}}}$ where $\mathrm{c}_{\text{\textsc{l}}}=%
\mathrm{c}_{\text{\textsc{r}}}=\mathrm{r}$ \textrm{\cite{1}-\cite{1kA}}.
Starting from an even self dual code $\mathcal{C}$ defined as a discrete set
of codewords $\left\{ \text{\textsc{c}}=(\mathbf{a},\mathbf{b})\in G_{%
\mathrm{k}}^{\mathrm{r},\mathrm{r}}\right\} $ subject to additional
conditions to be specified later, one can build realisations of both the
NCFT and its holographic dual given by the AB Chern-Simons (CS) theory with
coupling constant $\mathrm{k}\in \mathbb{N}^{\ast }$ (CS level) \textrm{\cite%
{1B}}. These constructions also allow to perform explicit calculations of
generalised Narain partition functions using code based algorithms. The
structure of the even self dual codes $\mathcal{C}$ enables to build
discrete sets $\mathbf{\Lambda }_{\mathrm{k}\mathcal{C}}^{\mathrm{r},\mathrm{%
r}}$ sitting in $\mathbb{R}^{\mathrm{r},\mathrm{r}}$ which in turn\textrm{\ }%
realise 2r dimensional Narain lattices $\mathbf{\Lambda }_{\text{\textsc{%
narain}}}$. For a given code $\mathcal{C}$ over the abelian group $G_{%
\mathrm{k}}^{\mathrm{r},\mathrm{r}}$, with codewords satisfying the
algebraic relations \textsc{c}$^{T}\eta $\textsc{c}$\in 2\mathbb{Z}$ and
\textsc{c}$^{T}\eta $\textsc{c}$^{\prime }\in \mathbb{Z}$ \textrm{\cite%
{1aB,1bB}}, \textrm{the} associated even self dual lattice $\mathbf{\Lambda }%
_{\mathrm{k}\mathcal{C}}^{\mathrm{r},\mathrm{r}}$ is imagined as embedded
between a pair of lattices like $\mathbf{\Lambda }_{\mathrm{k}}^{\mathrm{r},%
\mathrm{r}}\subset \mathbf{\Lambda }_{\mathrm{k}\mathcal{C}}^{\mathrm{r},%
\mathrm{r}}\subset (\mathbf{\Lambda }_{\mathrm{k}}^{\mathrm{r},\mathrm{r}%
})^{\ast }$. In this\textrm{\ }setting, the discrete group $G_{\mathrm{k}}^{%
\mathrm{r},\mathrm{r}}$ is identified with the discriminant $\mathbf{\Lambda
}_{\mathrm{k}}^{\mathrm{r},\mathrm{r}\ast }/\mathbf{\Lambda }_{\mathrm{k}}^{%
\mathrm{r},\mathrm{r}}$ taken in this study \textrm{to be} $\mathbb{Z}_{%
\mathrm{k}}^{\mathrm{r}}\times \mathbb{Z}_{\mathrm{k}}^{\mathrm{r}}$.
Moreover, one can represent the usual genus- one partition function of the
NCFT namely \textrm{\cite{1cB, 1dB}-\cite{1iB}}
\begin{equation}
Z_{\text{\textsc{ncft}}}\left[ \tau ,\xi ,\zeta \right] =Tr[q^{L_{0}-\mathrm{%
c}_{\text{\textsc{l}}}/24}q^{\bar{L}_{0}-\mathrm{c}_{\text{\textsc{r}}%
}/24}e^{2\pi i\left( \xi Q-\tilde{\zeta}\tilde{Q}\right) }],\qquad q=e^{2\pi
i\tau }  \label{Z}
\end{equation}%
in term of Mac-Williams functionals $W_{\mathcal{C}}$ that are
characteristic functions of coding theory with particular realisations as in
the eq(\ref{WX}); see also \textrm{\cite{1dB} }for other examples. In this
representation of $Z_{\text{\textsc{ncft}}}$, the above partition function
has been found to exhibit intrinsic features amongst which we cite:

The $Z_{\text{\textsc{ncft}}}$ is generally labeled by a pair of integer
vectors $(\mathbf{a,b})\in $ $\mathbb{Z}_{\mathrm{k}}^{\mathrm{r}}\times
\mathbb{Z}_{\mathrm{k}}^{\mathrm{r}}$ versus the familiar situation where it
behaves as a scalar under the discrete group $\mathbb{Z}_{\mathrm{k}}$ as it
corresponds to the special value k=1 ( for which \{$\mathbb{Z}_{\mathrm{1}%
}=I_{id}\}$) and roughly speaking to the neutral element I$_{id}\in \mathbb{Z%
}_{\mathrm{k}}.$ And as such, the above partition (\ref{Z}) should be
imagined in general as a \emph{matrix }function ($Z_{\text{\textsc{ncft}}}$)$%
_{\mathbf{a},\mathbf{b}}$ often denoted like $\Psi _{\mathbf{a},\mathbf{b}}$%
. This tensorial quantity has been found to describe wave functions in the
holographic dual \textrm{\cite{1eB,1fB} }that capture interesting
information. For\ instance, in the case \textrm{r=1}, the $\Psi _{a,b}$
indexed by $a,b=0,...,\mathrm{k}-1$ generate a Hilbert space with k$^{2}$
dimensions. For k=1, it reduces to the one dimensional $\Psi _{0,0}$ which
is associated with the trivial code word \textsc{c}$=(0,0)$ and which
describes the familiar torus partition function of Narain CFTs. The $\Psi _{%
\mathbf{a},\mathbf{b}}$ has been also found to describe multi-holomorphic
wave functions $\Psi _{\mathbf{a},\mathbf{b}}(\tau ,\xi _{i},\tilde{\zeta}%
_{i})$ where the complex variables $\xi _{i},\tilde{\zeta}_{i}$ can be
imagined in terms of boundary values of \textrm{r} gauge potential pairs ($%
\mathcal{A}_{i},\mathcal{B}_{i}$)$_{1\leq i\leq \mathrm{r}}$ with action as
in eq(\ref{act}) describing the 3D holographic dual of the Narain theory
\textrm{\cite{1B,1iB}}.

From the view of code theory, the $Z_{\text{\textsc{ncft}}}$ has been
represented by the code enumerator polynomials W$_{\mathcal{C}}\left(
X\right) $ shown to play an important role in code CFTs. These polynomials
have the typical structure \textrm{\cite{1jB}}%
\begin{equation}
W_{\mathcal{C}}\left( X\right) =\sum_{(\mathbf{a},\mathbf{b})\in \mathcal{C}%
}\dprod\limits_{i=1}^{\mathrm{r}}X_{a_{i},b_{i}}\sim \sum_{a_{i},b_{i}\in
\mathbb{Z}_{\mathrm{k}}}X_{a_{1},b_{1}}...X_{a_{\mathrm{r}},b_{\mathrm{r}}}
\label{WX}
\end{equation}%
where $\mathbf{a}=\left( a_{1},...,a_{\mathrm{r}}\right) $ and $\mathbf{b}%
=\left( b_{1},...,b_{\mathrm{r}}\right) $ are integer vectors with
components $a_{i},b_{i}\in \mathbb{Z}_{\mathrm{k}}$. The complex $%
X_{a,b}:=X_{a,b}\left( \tau ,\xi \right) $ are formal variables; they are
functions of the complex parameter $\tau $ of the 2-torus and the fugacities
$\xi _{i},\tilde{\zeta}_{i}$. In NCFT, they are realised in terms of the
generalized Siegel theta $\Theta _{a,b}^{\mathrm{r,r}}$ and the Dedekin $%
\eta \left( \tau \right) $ functions like $\frac{1}{\left\vert \eta \left(
\tau \right) \right\vert ^{2\mathrm{r}}}\Theta _{a,b}^{\mathrm{r,r}}\left(
\tau \right) .$ For the conformal model with $\mathrm{c}_{\text{\textsc{l}/%
\textsc{r}}}=\mathrm{r=1,}$ these thetas have the structure $e^{i\pi \tau p_{%
\text{\textsc{l}}}^{2}-i\pi \bar{\tau}p_{\text{\textsc{r}}}^{2}}$ with
quantized left/right momenta p$_{\text{\textsc{l}}/\text{\textsc{r}}}$
reading in terms of the codewords \textsc{c=}($a,b$), the Kaluza-Klein (KK)
mode n and the winding integer m as follows
\begin{equation}
\left( p_{\text{\textsc{l}}/\text{\textsc{r}}}\right) _{\left( a,n\right)
}^{\left( b,m\right) }=\frac{1}{\sqrt{2\mathrm{k}}}\left[ \frac{1}{\text{%
\textsc{r}}\sqrt{2}}(a+\mathrm{k}n)\pm \text{\textsc{r}}\sqrt{2}(b+\mathrm{k}%
m)\right]  \label{pp}
\end{equation}%
with $a,b\in \mathbb{Z}_{\mathrm{k}}$ and $p_{\text{\textsc{l}}}^{2}-p_{%
\text{\textsc{r}}}^{2}\in 2\mathbb{Z}$. Here, the real number \textsc{r} is
the radius of the Narain circle\textrm{;} it plays an important role in
Ensemble averaging \textrm{\cite{1iB,Raja1,1C}}; but in the present study
\textrm{it is fixed to (eg }\textsc{r}$=1$/$\sqrt{2}$\textrm{)}. For $n=m=0$
and CS level k=2, there are several \emph{ground} (un-excited)
configurations with left/right momenta ($p_{\text{\textsc{l}}/\text{\textsc{r%
}}})_{a,b}$ given by $\frac{a}{2}\epsilon _{x}\pm \frac{b}{2}\epsilon _{y}.$
As we will see throughout this study, these \emph{fundamental} momenta
define 2k$^{2}$ ground states [k$^{2}$ for $p_{\text{\textsc{l}}a,b}$ and k$%
^{2}$ for $p_{\text{\textsc{r}}a,b}$ ] within the unit cell $\epsilon
_{x}\times \epsilon _{y}$ of the rectangular lattice%
\begin{equation}
\left( \epsilon _{x}\mathbb{Z}\right) \times \left( \epsilon _{y}\mathbb{Z}%
\right)
\end{equation}%
with parameters $\epsilon _{x}=\frac{1}{\text{\textsc{r}}\sqrt{2}}$ and $%
\epsilon _{y}=$\textsc{r}$\sqrt{2}.$ For \textsc{r}$=1$/$\sqrt{2}$, this
lattice reduces to the square lattice $\mathbb{Z}\times \mathbb{Z}.$
Moreover, the usual modular symmetry of the Narain CFT translates into the
invariance $W_{\mathcal{C}}\left( X\right) =W_{\mathcal{C}}\left( X^{\prime
}\right) $ with $X^{\prime }$ referring to the two following transformations
(translation and inversion of SL(2,$\mathbb{Z}$) \textrm{\cite{1B,1C}}
\begin{equation}
\begin{tabular}{lllll}
$\left( i\right) $ & : & $X_{a,b}\left( \tau +1\right) $ & $=$ & $e^{2\pi
iab/\mathrm{k}}X_{a,b}\left( \tau \right) $ \\
$\left( ii\right) $ & : & $X_{a,b}\left( -1/\tau \right) $ & $=$ & $%
X_{a,b}^{\prime }=U_{ab}^{a^{\prime }b^{\prime }}X_{a^{\prime },b^{\prime
}}\left( \tau \right) $%
\end{tabular}%
\end{equation}%
where the phase $U_{ab}^{a^{\prime }b^{\prime }}$ is given by $e^{2\pi
i\left( ab^{\prime }+ba^{\prime }\right) /\mathrm{k}}$.

\subsection*{A.2 Lattice sequence: $\mathbf{\Lambda }_{\mathrm{k}}^{\mathrm{r%
},\mathrm{r}}\subset \mathbf{\Lambda }_{\mathrm{k}\mathcal{C}}^{\mathrm{r},%
\mathrm{r}}\subset \mathbf{\Lambda }_{\mathrm{k}}^{\ast \mathrm{r},\mathrm{r}%
}$}

\begin{figure}[tbph]
\begin{center}
\includegraphics[width=14cm]{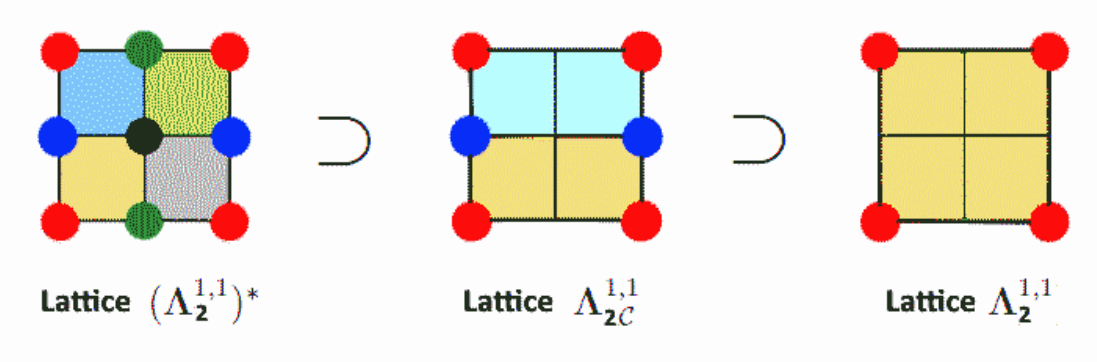}
\end{center}
\par
\vspace{-0.5cm}
\caption{The unit cells of the 2D lattices of construction A of the $%
c_{L/R}=1$ theory. Sites in these lattices are colored, the number of colors
is variable: k$^{2}$ colors for $\mathbf{\Lambda }_{\mathrm{k}}^{\mathrm{1},%
\mathrm{1}\ast }$, k colors for $\mathbf{\Lambda }_{\mathrm{k}\mathcal{C}}^{%
\mathrm{1},\mathrm{1}}$ and one color for $\mathbf{\Lambda }_{\mathrm{k}}^{%
\mathrm{1},\mathrm{1}}.$ The areas of the unit cells are exhibited in
increasing order. On left, $\mathbf{\Lambda }_{\mathrm{2}}^{\mathrm{1},%
\mathrm{1}\ast }$ with 4 unit cells. In middle, the $\mathbf{\Lambda }_{%
\mathrm{2}\mathcal{C}}^{\mathrm{1},\mathrm{1}}$ with 2 unit cells. On right,
the unit cell of $\mathbf{\Lambda }_{\mathrm{2}}^{\mathrm{1},\mathrm{1}}$. }
\label{i1}
\end{figure}

As we revisit the so-called \emph{construction A} of code-CFT (see also
appendix B), focusing on the lattice suite $\mathbf{\Lambda }_{\mathrm{k}}^{%
\mathrm{r},\mathrm{r}}\subset \mathbf{\Lambda }_{\mathrm{k}\mathcal{C}}^{%
\mathrm{r},\mathrm{r}}\subset \mathbf{\Lambda }_{\mathrm{k}}^{\ast \mathrm{r}%
,\mathrm{r}},$ in terms of the root and weight lattices of Lie algebras
\textbf{g}, we then present new realisations of this construction, in which
the discrete groups $G_{\mathrm{k}}^{\mathrm{r},\mathrm{r}}$ \textrm{are }%
interpreted in terms of the symmetries of unit cells of the real lattice $%
\mathbf{\Lambda }_{\mathrm{k}}^{\mathrm{r},\mathrm{r}},$ its dual $\mathbf{%
\Lambda }_{\mathrm{k}}^{\ast \mathrm{r},\mathrm{r}}$ and the even self dual $%
\mathbf{\Lambda }_{\mathrm{k}\mathcal{C}}^{\mathrm{r},\mathrm{r}}$. We begin
by showing that the construction A of the $\mathrm{c}_{\text{\textsc{l/r}}}=%
\mathrm{1}$ Narain theory has a natural interpretation in terms of the root
\textsc{R}$^{\mathbf{su}_{2}}:=\left. \text{\textsc{R}}_{\mathrm{k}}^{%
\mathbf{su}_{2}}\right\vert _{\mathrm{k=2}}$ and the weight \textsc{W}$^{%
\mathbf{su}_{2}}=\left. \text{\textsc{W}}_{\mathrm{k}}^{\mathbf{su}%
_{2}}\right\vert _{\mathrm{k=2}}$ lattices of SU(2) and their fibrations as
given by eq(\ref{lrw}) valid for generic values of k (including the special
value k=2). Recall that the group SU(2) has rank r=1, and for k=2 the
associated discriminant \textsc{W}$^{\mathbf{su}_{2}}/$\textsc{R}$^{\mathbf{%
su}_{2}}$ is isomorphic to $\mathbb{Z}_{2}$. This discriminant shows that
the unit cell of \textsc{W}$^{\mathbf{su}_{2}}$ has two different types of
sites (two colored sites) versus one site-type for \textsc{R}$^{\mathbf{su}%
_{2}}$ as shown on the Figure \textbf{\ref{i1}}. In this illustrative image,
we used 4 types of colored sites (red, blue, green, black\textbf{)} to draw
graphic representations of the particular \textrm{k=2} suite $\mathbf{%
\Lambda }_{\mathrm{2}}^{\mathrm{1},\mathrm{1}}\subset $ $\mathbf{\Lambda }_{%
\mathrm{2}\mathcal{C}}^{\mathrm{1},\mathrm{1}}\subset \mathbf{\Lambda }_{%
\mathrm{2}}^{\ast \mathrm{1},\mathrm{1}}$ where we used: $\left( \mathbf{i}%
\right) $ four colors for the bigger $\mathbf{\Lambda }_{\mathrm{2}}^{\ast
\mathrm{1},\mathrm{1}}=\text{\textsc{W}}_{\mathrm{2}}^{\mathbf{su}%
_{2}}\times \text{\textsc{W}}_{\mathrm{2}}^{\mathbf{su}_{2}},$ $\left(
\mathbf{ii}\right) $ two colors for $\mathbf{\Lambda }_{\mathrm{2}\mathcal{C}%
}^{\mathrm{1},\mathrm{1}}=$\textsc{R}$_{\mathrm{2}}^{\mathbf{su}_{2}}\times $%
\textsc{W}$_{\mathrm{2}}^{\mathbf{su}_{2}}$; and $\left( \mathbf{iii}\right)
$ one color for $\mathbf{\Lambda }_{\mathrm{2}}^{\mathrm{1},\mathrm{1}}=$%
\textsc{R}$_{\mathrm{2}}^{\mathbf{su}_{2}}\times $\textsc{W}$_{\mathrm{2}}^{%
\mathbf{su}_{2}}.$
For generic values of the CS level k, we need\textrm{\ k}$^{2}$ colored
sites to draw graphs for the triplet in the suite $\mathbf{\Lambda }_{%
\mathrm{k}}^{\mathrm{1},\mathrm{1}}\subset \mathbf{\Lambda }_{\mathrm{k}%
\mathcal{C}}^{\mathrm{1},\mathrm{1}}\subset \mathbf{\Lambda }_{\mathrm{k}}^{%
\mathrm{1},\mathrm{1}\ast }$ sitting in $\mathbb{R}^{\mathrm{1},\mathrm{1}},$
they are remarkably realized by the fibrations
\begin{equation}
\begin{tabular}{lll}
$\mathbf{\Lambda }_{\mathrm{k}}^{\mathrm{1},\mathrm{1}}$ & $=$ & $\text{%
\textsc{R}}_{\mathrm{k}}^{\mathbf{su}_{2}}\times \text{\textsc{R}}_{\mathrm{k%
}}^{\mathbf{su}_{2}}$ \\
$\mathbf{\Lambda }_{\mathrm{k}\mathcal{C}}^{\mathrm{1},\mathrm{1}}$ & $=$ & $%
\text{\textsc{R}}_{\mathrm{k}}^{\mathbf{su}_{2}}\times $\textsc{W}$_{\mathrm{%
k}}^{\mathbf{su}_{2}}$ \\
$\mathbf{\Lambda }_{\mathrm{k}}^{\ast \mathrm{1},\mathrm{1}}$ & $=$ &
\textsc{W}$_{\mathrm{k}}^{\mathbf{su}_{2}}\times $\textsc{W}$_{\mathrm{k}}^{%
\mathbf{su}_{2}}$%
\end{tabular}%
\qquad \Leftrightarrow \qquad
\begin{tabular}{lll}
$\mathbf{\Lambda }_{\mathrm{k}}^{\mathbf{su}_{2}}$ & $=$ & $\text{\textsc{R}}%
_{\mathrm{k}}^{\mathbf{su}_{2}}\times \text{\textsc{R}}_{\mathrm{k}}^{%
\mathbf{su}_{2}}$ \\
$\mathbf{\Lambda }_{\mathrm{k}\mathcal{C}}^{\mathbf{su}_{2}}$ & $=$ & $\text{%
\textsc{R}}_{\mathrm{k}}^{\mathbf{su}_{2}}\times $\textsc{W}$_{\mathrm{k}}^{%
\mathbf{su}_{2}}$ \\
$\mathbf{\Lambda }_{\mathrm{k}}^{\ast \mathbf{su}_{2}}$ & $=$ & \textsc{W}$_{%
\mathrm{k}}^{\mathbf{su}_{2}}\times $\textsc{W}$_{\mathrm{k}}^{\mathbf{su}%
_{2}}$%
\end{tabular}
\label{lrw}
\end{equation}%
In these fibrations, the \textsc{R}$_{\mathrm{k}}^{\mathbf{su}_{2}}$ is
isomorphic to the 1D lattice $\sqrt{\mathrm{k}}\mathbb{Z}$ while the \textsc{%
W}$_{\mathrm{k}}^{\mathbf{su}_{2}}$ is isomorphic to its dual $\frac{1}{%
\sqrt{\mathrm{k}}}\mathbb{Z}$ with coset \textsc{W}$_{\mathrm{k}}^{\mathbf{su%
}_{2}}/$\textsc{R}$_{\mathrm{k}}^{\mathbf{su}_{2}}\simeq \mathbb{Z}_{\mathrm{%
k}}$. From this discriminant, one can imagine \textsc{W}$_{\mathrm{k}}^{%
\mathbf{su}_{2}}$ like the product $\mathbb{Z}_{\mathrm{k}}\times $\textsc{R}%
$_{\mathrm{k}}^{\mathbf{su}_{2}}$ letting understand that \textsc{W}$_{%
\mathrm{k}}^{\mathbf{su}_{2}}$ is made of the superposition of k sublattices
\textsc{R}$_{\mathrm{k}}^{\mathbf{su}_{2}}$ distant by some weight vector as
explicitly shown in the core of paper (\emph{sections 3 and 4}). Sites in
the above lattices $\mathbf{\Lambda }$ describe quantum particle states $%
\left\vert \psi _{a,n}^{b,m}\right\rangle $ labeled, in addition to the
indices (a,b) of the ground configurations, by Kaluza modes $\left\vert
n\right\rangle $ and windings $\left\vert m\right\rangle $.

\subsubsection*{a) Constraints using unit cells}

Notice that in the language of colored sites and unit cells, which for the
1D lattices $a_{\rho }\mathbb{Z}$ are given by the parameter $a_{\rho }$
[below $uc(a_{\rho }\mathbb{Z}):=a_{\rho }$], a set of interesting features
follow: First, we have the condition $uc$\textsc{W}$_{\mathrm{k}}^{\mathbf{su%
}_{2}}<uc$\textsc{R}$_{\mathrm{k}}^{\mathbf{su}_{2}}$ because the number of
sites \textbf{k}$_{m}$ in \textsc{W}$_{\mathrm{k}}^{\mathbf{su}_{2}}$ is
bigger than that those \textbf{x}$_{n}$ in \textsc{R}$_{\mathrm{k}}^{\mathbf{%
su}_{2}}$. Consequently the suite of the construction A gets mapped into the
inequalities%
\begin{equation}
uc\mathbf{\Lambda }_{\mathrm{k}}^{\ast \mathbf{su}_{2}}<uc\mathbf{\Lambda }_{%
\mathrm{k}\mathcal{C}}^{\mathbf{su}_{2}}<uc\mathbf{\Lambda }_{\mathrm{k}}^{%
\mathbf{su}_{2}}
\end{equation}%
Second, by using the relations $uc$\textsc{W}$_{\mathrm{k}}^{\mathbf{su}%
_{2}}=\frac{1}{\sqrt{\mathrm{k}}}$ and $uc$\textsc{R}$_{\mathrm{k}}^{\mathbf{%
su}_{2}}=\sqrt{\mathrm{k}}$ and subsequently $uc$\textsc{R}$_{\mathrm{k}}^{%
\mathbf{su}_{2}}=\mathrm{k(}uc$\textsc{W}$_{\mathrm{k}}^{\mathbf{su}_{2}}),$
we end up with the characteristic properties $uc\mathbf{\Lambda }_{\mathrm{k}%
}^{\ast \mathbf{su}_{2}}=\frac{1}{\mathrm{k}}$ and $uc\mathbf{\Lambda }_{%
\mathrm{k}\mathcal{C}}^{\mathbf{su}_{2}}=1$ as well as $uc\mathbf{\Lambda }_{%
\mathrm{k}}^{\mathbf{su}_{2}}=\mathrm{k}$. We also have the following
result:
\begin{equation}
uc\mathbf{\Lambda }_{\mathrm{k}}^{\mathbf{su}_{2}}=\mathrm{k(}uc\mathbf{%
\Lambda }_{\mathrm{k}\mathcal{C}}^{\mathbf{su}_{2}})=\mathrm{k}^{2}(uc%
\mathbf{\Lambda }_{\mathrm{k}}^{\ast \mathbf{su}_{2}})
\end{equation}%
Regarding new realisations of construction A of code CFT going beyond the
SU(2) buildings briefly presented above, we extend the SU(2) realisation (%
\ref{lrw}) to higher dimensional lattices of Lie algebras especially those
symmetries with rank 2 such as SU(3), SO(5) and G2. Focussing on the
interesting case of SU(3), we get for the 4D lattices involved in the basis
sequence $\mathbf{\Lambda }_{\mathrm{k}}^{\mathbf{su}_{3}}\subset \mathbf{%
\Lambda }_{\mathrm{k}\mathcal{C}}^{\mathbf{su}_{3}}\subset \mathbf{\Lambda }%
_{\mathrm{k}}^{\ast \mathbf{su}_{3}}$ hosted by $\mathbb{R}^{\mathrm{2},%
\mathrm{2}}$ the following representation%
\begin{equation}
\begin{tabular}{lll}
$\mathbf{\Lambda }_{\mathrm{k}}^{\mathbf{su}_{3}}$ & $=$ & $\text{\textsc{R}}%
_{\mathrm{k}}^{\mathbf{su}_{3}}\times \text{\textsc{R}}_{\mathrm{k}}^{%
\mathbf{su}_{3}}$ \\
$\mathbf{\Lambda }_{\mathrm{k}\mathcal{C}}^{\mathbf{su}_{3}}$ & $=$ & $\text{%
\textsc{R}}_{\mathrm{k}}^{\mathbf{su}_{3}}\times $\textsc{W}$_{\mathrm{k}}^{%
\mathbf{su}_{3}}$ \\
$\mathbf{\Lambda }_{\mathrm{k}}^{\ast \mathbf{su}_{3}}$ & $=$ & \textsc{W}$_{%
\mathrm{k}}^{\mathbf{su}_{3}}\times $\textsc{W}$_{\mathrm{k}}^{\mathbf{su}%
_{3}}$%
\end{tabular}%
\qquad ,\qquad uc\mathbf{\Lambda }_{\mathrm{k}}^{\ast \mathbf{su}_{3}}<uc%
\mathbf{\Lambda }_{\mathrm{k}\mathcal{C}}^{\mathbf{su}_{3}}<uc\mathbf{%
\Lambda }_{\mathrm{k}}^{\mathbf{su}_{3}}
\end{equation}%
where for k=3, the 2D discrete sets \textsc{R}$_{\mathrm{k}}^{\mathbf{su}%
_{3}}$ and \textsc{W}$_{\mathrm{k}}^{\mathbf{su}_{3}}$ sitting in $\mathbb{R}%
^{\mathrm{2}}$ are nothing but the root and the weight lattices of SU(3).
For this generalisation, we distinguish three cases according to the values
of the Chern-Simons level \textrm{k} of the holographic dual of the Narain
CFT.

\subsubsection*{ b) Three sectors}

These cases concern the following intervals:

\begin{description}
\item[$\left( \mathbf{i}\right) $] \textbf{Case k=3}: this is the canonical
CS level for which the discriminant group is equal to $\mathbb{Z}_{\mathrm{3}%
}\times \mathbb{Z}_{\mathrm{3}}$. Here, the abelian set $\mathbb{Z}_{\mathrm{%
3}}$ is just the group centre of SU(3). For this value, we find\textrm{\ }%
that the 4D lattice $\mathbf{\Lambda }_{\mathrm{3}}^{\mathbf{su}_{3}}$ is
made of the fibration of two hexagonal lattices like \textsc{R}$_{\mathrm{3}%
}^{\mathbf{su}_{3}}\times $\textsc{R}$_{\mathrm{3}}^{\mathbf{su}_{3}}$ while
the dual $\mathbf{\Lambda }_{\mathrm{3}}^{\ast \mathbf{su}_{3}}$ is
constructed out of the fibration of two triangular lattices as \textsc{W}$_{%
\mathrm{3}}^{\mathbf{su}_{3}}\times $\textsc{W}$_{\mathrm{3}}^{\mathbf{su}%
_{3}}$. The areas of the unit cells of these lattices are depicted by the
Figure \textbf{\ref{i2} }in yellow color where for convenience we have also
exhibited the triangulation of the underlying surfaces.
\begin{figure}[tbph]
\begin{center}
\includegraphics[width=16.5cm]{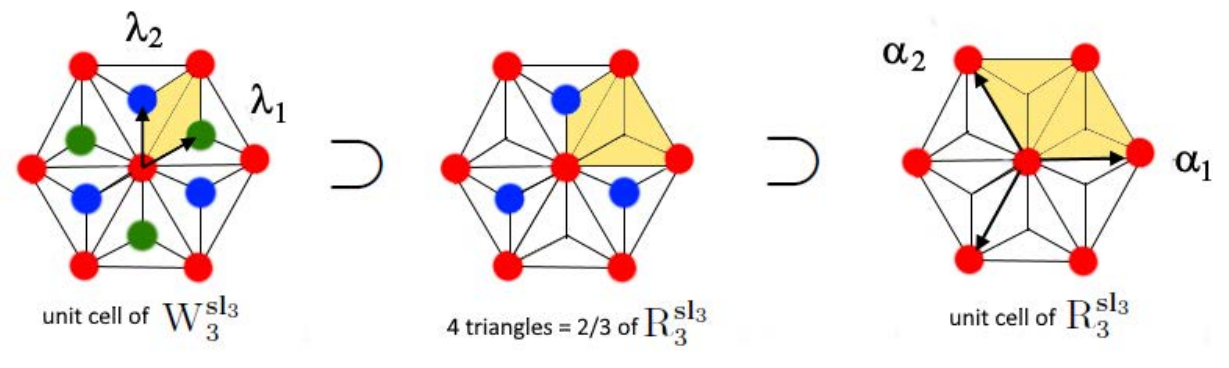}
\end{center}
\par
\vspace{-0.5cm}
\caption{The areas of unit cells of the weight \textsc{W}$_{\mathrm{3}}^{%
\mathbf{sl}_{3}}$ and the root $\text{\textsc{R}}_{\mathrm{3}}^{\mathbf{sl}%
_{3}}$ lattices of SU(3). The unit cell of \textsc{R}$_{\mathrm{3}}^{\mathbf{%
sl}_{3}}$ is three times the unit cell of \textsc{W}$_{\mathrm{3}}^{\mathbf{%
sl}_{3}}$. In the middle we also give the configuration with 2 times uc%
\textsc{W}$_{\mathrm{3}}^{\mathbf{sl}_{3}}$.}
\label{i2}
\end{figure}
In this Figure, we have used three different colors (red, blue, green) to
distinguish the three superposed sublattices $\mathbb{A}+\mathbb{B}+\mathbb{C%
}$ making \textsc{W}$_{\mathrm{3}}^{\mathbf{su}_{3}}$ with red color for
sublattice $\mathbb{A}$; blue color for $\mathbb{B}$ and green color for $%
\mathbb{C}$. The unit cells of \textsc{W}$_{\mathrm{3}}^{\mathbf{su}_{3}}$
and \textsc{R}$_{\mathrm{3}}^{\mathbf{su}_{3}}$ in the Figure \textbf{\ref%
{i2}} are respectively given by%
\begin{equation}
\begin{tabular}{lll}
$uc\text{\textsc{R}}_{\mathrm{3}}^{\mathbf{su}_{3}}$ & $=$ & $\sqrt{3}$ \\
$uc$\textsc{W}$_{\mathrm{3}}^{\mathbf{su}_{3}}$ & $=$ & $\frac{1}{\sqrt{3}}$%
\end{tabular}%
\qquad \Rightarrow \qquad
\begin{tabular}{lll}
$uc\mathbf{\Lambda }_{\mathrm{3}}^{\mathbf{su}_{3}}$ & $=$ & $3$ \\
$uc\mathbf{\Lambda }_{\mathrm{3}}^{\ast \mathbf{su}_{3}}$ & $=$ & $\frac{1}{3%
}$%
\end{tabular}%
\end{equation}%
showing that $uc$\textsc{R}$_{\mathrm{3}}^{\mathbf{su}_{3}}=3\times uc$%
\textsc{W}$_{\mathrm{3}}^{\mathbf{su}_{3}}$ and then $uc\mathbf{\Lambda }_{%
\mathrm{3}}^{\mathbf{su}_{3}}=9\times uc\mathbf{\Lambda }_{\mathrm{3}}^{\ast
\mathbf{su}_{3}}.$ In these relations the $\mathbf{\lambda }_{i}$'s are the
two fundamental weight vectors of SU(3) with angle $(\widehat{\mathbf{%
\lambda }_{1},\mathbf{\lambda }_{2}})=\frac{2\pi }{6}$ and the $\mathbf{%
\alpha }_{i}$'s are the associated simple roots with $(\widehat{\mathbf{%
\alpha }_{1},\mathbf{\alpha }_{2}})=\frac{2\pi }{3}.$ In the Figure \textbf{%
\ref{i2}}, the representative graphs have 18 fundamental elementary
triangles forming 9 unit cells of \textsc{W}$_{\mathrm{3}}^{\mathbf{su}_{3}}$
($18=2\times 9$) and 3 unit cells of \textsc{R}$_{\mathrm{3}}^{\mathbf{su}%
_{3}}$ ($18=6\times 3$).

\item[$\left( \mathbf{ii}\right) $] \textbf{Case k\TEXTsymbol{>}3}: the
results obtained for k=3 extends also to k\TEXTsymbol{>}3 for which we have
the discriminants \textsc{W}$_{\mathrm{k}}^{\mathbf{su}_{3}}/$\textsc{R}$_{%
\mathrm{k}}^{\mathbf{su}_{3}}\simeq \mathbb{Z}_{\mathrm{k}}$ and $\mathbf{%
\Lambda }_{\mathrm{k}}^{\ast \mathbf{su}_{3}}/\mathbf{\Lambda }_{\mathrm{k}%
}^{\mathbf{su}_{3}}\simeq \mathbb{Z}_{\mathrm{k}}\times \mathbb{Z}_{\mathrm{k%
}}.$ For instance, the group $\mathbb{Z}_{\mathrm{k}}$ indicates that
\textsc{W}$_{\mathrm{k}}^{\mathbf{su}_{3}}$ is made of the superposition of
k sublattices%
\begin{equation}
\mathbb{A}_{0}\dbigcup \mathbb{A}_{1}\dbigcup \cdots \dbigcup \mathbb{A}_{%
\mathrm{k-2}}\dbigcup \mathbb{A}_{\mathrm{k-1}}
\end{equation}%
each one of them is isomorphic to \textsc{R}$_{\mathrm{k}}^{\mathbf{su}_{3}}$
but distant by the weight vectors $\varepsilon \mathbf{\lambda }$ with $%
\varepsilon =0,...,\mathrm{k}-1\mathrm{.}$ We also have the following
relationships between the values of the areas of the unit cells of the real
\textsc{R}$_{\mathrm{k}}^{\mathbf{su}_{3}}$ and the dual \textsc{W}$_{%
\mathrm{k}}^{\mathbf{su}_{3}}$ lattices namely
\begin{equation}
uc\text{\textsc{R}}_{\mathrm{k}}^{\mathbf{su}_{3}}=\frac{\mathrm{k}^{2}}{3}%
\times uc\text{\textsc{W}}_{\mathrm{k}}^{\mathbf{su}_{3}}\qquad \Rightarrow
\qquad uc\mathbf{\Lambda }_{\mathrm{k}}^{\mathbf{su}_{3}}=\frac{\mathrm{k}%
^{4}}{9}\times uc\mathbf{\Lambda }_{\mathrm{k}}^{\ast \mathbf{su}_{3}}
\end{equation}%
They can be checked by using the results
\begin{equation}
\begin{tabular}{lll}
$uc\text{\textsc{R}}_{\mathrm{k}}^{\mathbf{su}_{3}}$ & $=$ & $\frac{\mathrm{k%
}}{\sqrt{3}}$ \\
$uc$\textsc{W}$_{\mathrm{k}}^{\mathbf{su}_{3}}$ & $=$ & $\frac{\sqrt{3}}{%
\mathrm{k}}$%
\end{tabular}%
\qquad \Rightarrow \qquad
\begin{tabular}{lll}
$uc\mathbf{\Lambda }_{\mathrm{3}}^{\mathbf{su}_{3}}$ & $=$ & $\frac{\mathrm{k%
}^{2}}{3}$ \\
$uc\mathbf{\Lambda }_{\mathrm{3}}^{\ast \mathbf{su}_{3}}$ & $=$ & $\frac{3}{%
\mathrm{k}^{2}}$%
\end{tabular}%
\end{equation}%
from which we also learn $uc\mathbf{\Lambda }_{\mathrm{3}\mathcal{C}}^{%
\mathbf{su}_{3}}=1$ independently of the value of the CS level.

\item[$\left( \mathbf{iii}\right) $] \textbf{Case k\TEXTsymbol{<}3}: for
those integer values of the CS level k less than 3, we distinguish two
possibilities namely $\mathrm{k}=1,2$. While the CS level k=1 is somehow
"exotic" because the condition $uc$\textsc{W}$_{\mathrm{1}}^{\mathbf{su}%
_{3}}<uc$\textsc{R}$_{\mathrm{1}}^{\mathbf{su}_{3}}$ is violated since $uc$%
\textsc{W}$_{\mathrm{1}}^{\mathbf{su}_{3}}=\sqrt{3}$ while $uc$\textsc{R}$_{%
\mathrm{1}}^{\mathbf{su}_{3}}=1/\sqrt{3}$. This condition holds only for k$%
\geq \sqrt{3}$ coming from solving $\frac{\sqrt{3}}{\mathrm{k}}<\frac{%
\mathrm{k}}{\sqrt{3}}$. This bound rules out k=1 but includes k=2 for which
the weight lattice \textsc{W}$_{\mathrm{2}}^{\mathbf{su}_{3}}$ is given by
the 2D honeycomb while the \textsc{R}$_{\mathrm{2}}^{\mathbf{su}_{3}}$ is an
hexagonal lattice. For illustration, see the pictures depicted by the Figure
\textbf{\ref{i3}}.
\begin{figure}[tbph]
\begin{center}
\includegraphics[width=12cm]{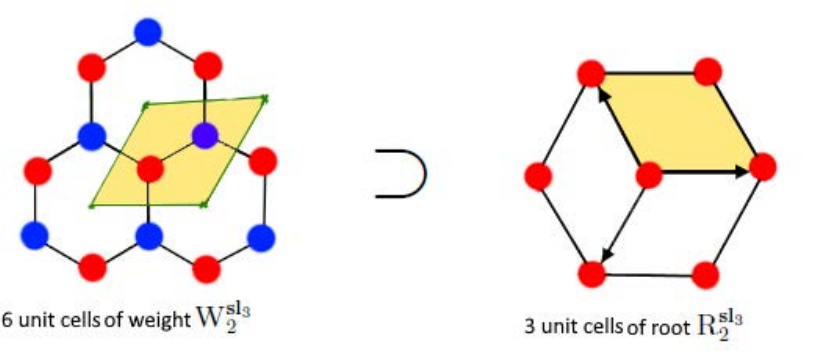}
\end{center}
\par
\vspace{-0.5cm}
\caption{Unit cells of weight \textsc{W}$_{\mathrm{2}}^{\mathbf{sl}_{3}}$
(honeycomb) and root $\text{\textsc{R}}_{\mathrm{2}}^{\mathbf{sl}_{3}}$
(hexagonal) lattices. \textsc{W}$_{\mathrm{2}}^{\mathbf{sl}_{3}}$ is given
by the superposition of two isomorphic sublattices (red and blue)}
\label{i3}
\end{figure}
For this CS level, the discriminant \textsc{W}$_{\mathrm{2}}^{\mathbf{su}%
_{3}}/$\textsc{R}$_{\mathrm{2}}^{\mathbf{su}_{3}}$ is equal to $\mathbb{Z}_{%
\mathrm{2}}$ indicating in turn that \textsc{W}$_{\mathrm{2}}^{\mathbf{su}%
_{3}}$ is made of the superposition $\mathbb{A}+\mathbb{B}$ of two
isomorphic sublattices (red and blue in the Figure \textbf{\ref{i3}})
distant by a weight vector. The areas of the unit cell $uc$\textsc{R}$_{%
\mathrm{2}}^{\mathbf{su}_{3}}$ is equal to $\frac{\mathrm{4}}{3}$ times $uc$%
\textsc{W}$_{\mathrm{k}}^{\mathbf{su}_{3}}.$
\end{description}

\section{Construction A of code Narain CFT}

\label{sec:B} \setcounter{equation}{0} Following \textrm{\cite{1B,1C}}, the
construction A of code based NCFT establishes a correspondence between
algebraic coding theory on the finite abelian group $G_{\mathrm{k}}^{\mathrm{%
r,r}}\sim \mathbb{Z}_{\mathrm{k}}^{\mathrm{r}}\times \mathbb{Z}_{\mathrm{k}%
}^{\mathrm{r}}$ and CFTs defined on toroidal backgrounds. This\ framework
provides a systematic method to build Narain CFTs, describing strings on
torii, by using code theory based on the Mac-Williams functional W$_{%
\mathcal{C}}\left( X\right) $ \textrm{\cite{21,22}}. The structure of linear
codes $\mathcal{C}\subset G_{\mathrm{k}}^{\mathrm{r,r}}$ determine the
fundamental features of the associated NCFTs. In the holographic dual of
these NCFTs \textrm{\cite{23}}, the positive integer k corresponds to the
Chern-Simons (CS) level of the so-called AB theory which consists of two
gauge blocks of abelian potentials $\mathcal{A}:=\sum_{i=1}^{\mathrm{r}}%
\mathcal{A}^{i}Q_{i}$ and $\mathcal{B}:=\sum_{i=1}^{\mathrm{r}}\mathcal{B}%
_{i}\tilde{Q}^{i}$ valued in [U(1)$_{A}\times $U(1)$_{B}]^{\mathrm{r}}$ and
coupled as follows%
\begin{equation}
\mathcal{S}_{\text{\textsc{cs}}}+\mathcal{S}_{\text{\textsc{bnd}}}=\frac{i%
\mathrm{k}}{4\pi }\dint\nolimits_{\mathcal{M}}Tr\left( \mathcal{A}\wedge d%
\mathcal{B}+\mathcal{B}\wedge d\mathcal{A}\right) +\dint\nolimits_{\partial
\mathcal{M}}Tr\left( \mathcal{A}_{z}\mathcal{A}_{\bar{z}}+\mathcal{B}_{z}%
\mathcal{B}_{\bar{z}}\right)  \label{act}
\end{equation}%
with trace relations as $Tr(Q_{i}\tilde{Q}^{j})=\delta _{i}^{j}$ and $%
Tr(Q_{i}Q_{j})=K_{ij}$ as well as $Tr(\tilde{Q}^{i}\tilde{Q}^{j})=\tilde{K}%
^{ij}.$ In the eq(\ref{act}), $\mathcal{M}$ is generally a 3D handlebody
\textrm{(e.g.} a solid torus) with boundary $\partial \mathcal{M}$ given by
a Riemann surface $\Sigma _{g}$ such as the 2-torus for genus $g=1$. For the
leading value of the integral central charge $c_{\text{\textsc{l}/\textsc{r}}%
}=1$, corresponding to the model\textrm{\ r=1}, the code-CFT constructed
over $G_{\mathrm{k}}^{\mathrm{1,1}}:=G_{\mathrm{k}}$ maps codes $\mathcal{C}$
into even self dual\textrm{\ }lattices $\Lambda _{\mathrm{k}\mathcal{C}}^{%
\mathrm{1,1}}$\ constrained like
\begin{equation}
\mathbf{\Lambda }_{\mathrm{k}}^{\mathrm{1,1}}\subset \mathbf{\Lambda }_{%
\mathrm{k}\mathcal{C}}^{\mathrm{1,1}}\subset \mathbf{\Lambda }_{\mathrm{k}%
}^{\ast \mathrm{1,1}}\subset \mathbb{R}^{1,1}
\end{equation}%
In this\ hierarchy,\textrm{\ }the discrete set $\mathbf{\mathbf{\Lambda }_{%
\mathrm{k}}^{\mathrm{1,1}}:=\Lambda }$ is a 2D even integral lattice
embedded in the Lorentzian plane $\mathbb{R}^{1,1}$ endowed with the metric:
\begin{equation}
\eta =\left(
\begin{array}{cc}
0 & 1 \\
1 & 0%
\end{array}%
\right) \qquad \leftrightarrow \qquad \eta _{ij}\equiv \left(
\begin{array}{cc}
0_{AA} & 1_{AB} \\
1_{BA} & 0_{BB}%
\end{array}%
\right)
\end{equation}%
and scalar product $<\mathbf{v,v}>_{\eta }=\mathbf{v}^{T}\eta \mathbf{v}$
evaluating to $2v_{1}v_{2}.$ The metric $\eta $ couples the two gauge
sectors $\mathcal{A}$ and $\mathcal{B}$ in eq(\ref{act}). The dual lattice $%
\mathbf{\Lambda }^{\ast }$ is defined as the set%
\begin{equation}
\mathbf{\Lambda }^{\ast }=\left\{ \mathbf{w}\text{ }|\text{ }\mathbf{v}%
^{T}\eta \mathbf{w}\in \mathbb{Z},\text{ }\forall \mathbf{v\in }\Lambda
\right\}
\end{equation}%
where $\mathbf{\Lambda }=\left\{ \mathbf{v}\text{ }|\text{ }\mathbf{v}%
^{T}\eta \mathbf{v}\in 2\mathbb{Z}\right\} $ is an even integer lattice
satisfying the usual inclusion property $\mathbf{\Lambda }\subset \mathbf{%
\Lambda }^{\ast }.$ The intermediate lattice $\mathbf{\Lambda }_{\mathcal{C}%
} $ is both even and self dual,\emph{\ }that is $\mathbf{\Lambda }_{\mathcal{%
C}}\sim \mathbf{\Lambda }_{\mathcal{C}}^{\ast },$ and is nested between $%
\mathbf{\Lambda }$\ and $\mathbf{\Lambda }^{\ast };$ see Figure \textbf{\ref%
{i1}} for illustration.

\subsection*{B.1 An explicit realisation}

Here, we give an explicit realisation of the triplet ($\mathbf{\Lambda },%
\mathbf{\Lambda }_{\mathcal{C}},\mathbf{\Lambda }^{\ast }$) which has been
exploited in our analysis, sections 2 \& 3, where it has been given a
natural interpretation in terms of the root and the weight lattices of SU(2)
and SU(3). To gain further insight into the embedding chain $\mathbf{\Lambda
}\subset \mathbf{\Lambda }_{\mathcal{C}}\subset \mathbf{\Lambda }^{\ast },$
we start by describing the sites $\mathbf{v}$ populating the abelian lattice
$\mathbf{\Lambda }.$ They are represented by discrete 2D vectors $\mathbf{v}%
_{\mathbf{n}}=v_{\mathbf{n}}^{i}\mathbf{e}_{i}$ labeled by: $\left( \mathbf{i%
}\right) $ the integer vector $\mathbf{n}=(n_{1},n_{2})$ parameterising the
square lattice $\mathbb{Z}\times \mathbb{Z}$ with unit spacing \texttt{a}=1;
in addition to $\left( \mathbf{ii}\right) $ the usual canonical basis
vectors $\mathbf{e}_{1}=(\mathtt{a},0)$ and $\mathbf{e}_{2}=(0,\mathtt{a})$
with Euclidian $\mathbf{e}_{1}.\mathbf{e}_{2}=0$ but Lorentzian $\mathbf{e}%
_{1}\eta \mathbf{e}_{2}=\mathtt{a}^{2}=1.$ For convenience, we express the
vector $\mathbf{v}_{\mathbf{n}}$ as a linear combination of the integer
vector $\mathbf{n}$ like%
\begin{equation}
v_{\mathbf{n}}^{i}=\sum_{j=1}^{2}\Lambda _{j}^{i}n^{j}\qquad ,\qquad \Lambda
_{j}^{i}=\frac{\partial v_{\mathbf{n}}^{i}}{\partial n^{j}}\qquad ,\qquad
\Lambda :=\left(
\begin{array}{cc}
a & b \\
c & d%
\end{array}%
\right)  \label{vl}
\end{equation}%
where $\Lambda _{j}^{i}$ is a characteristic matrix of the real lattice $%
\mathbf{\Lambda },$ acting as a linear map between $\mathbb{Z}^{2}$\ and $%
\mathbf{\Lambda }$ with $\det \Lambda _{ij}\neq 0$ (i.e: $\Lambda _{ij}:%
\mathbf{n}\in \mathbb{Z}^{2}\rightarrow \mathbf{v}_{\mathbf{n}}\in \mathbf{%
\Lambda }$). For any lattice vector $\mathbf{v}_{\mathbf{n}}\in \mathbf{%
\Lambda },$ the Lorentzian scalar product $\mathbf{v}_{\mathbf{n}}^{T}\eta
\mathbf{v}_{\mathbf{n}}$ must be an even integer (i.e. belongining to 2$%
\mathbb{Z}$) thus imposing a strong constraint on the form of the generating
matrix $\Lambda _{j}^{i}$. Indeed, substituting $\mathbf{v}_{\mathbf{n}%
}=\Lambda .\mathbf{n}$ into the Lorentzian scalar yields the conditions
\begin{equation}
\mathbf{n}^{T}\left( g_{\Lambda }\right) \mathbf{n}\in 2\mathbb{Z}\text{ }%
\qquad with\qquad g_{\Lambda }=\Lambda ^{T}\eta \Lambda
\end{equation}%
where (\textbf{i}) the induced metric $g_{\Lambda }$ is a quadratic function
of $\Lambda $, with (\textbf{ii}) diagonal entries taking even integer values%
\begin{equation}
g_{\Lambda }=\left(
\begin{array}{cc}
2m & l \\
l & 2n%
\end{array}%
\right) ,\qquad m,n,l\in \mathbb{Z}  \label{gl}
\end{equation}%
this implies that $\det g_{\Lambda }=-\det \Lambda ^{2}.$ Additionally, the
characteristic matrix $\Lambda $ is not unique; it is defined up to $%
\mathcal{O}$(1,1,$\mathbb{R}$) orthogonal transformations namely $\Lambda
^{\prime }=\mathcal{O}\Lambda $ with\textrm{\ }$\mathcal{O}^{T}\eta \mathcal{%
O}=\eta .$ A simple example of\emph{\ }such transformation is given by $%
\mathcal{O}=\eta $ leading to $\Lambda ^{\prime }=\eta \Lambda .$ Moreover%
\textrm{, }using (\ref{vl}), we obtain the relations $ac=m,$ $bd=n$ and $%
ad+bc=l$ along with the identity $4mn-l^{2}=-\left( ad-bc\right) ^{2}$ that
factorises like
\begin{equation}
\det g_{\Lambda }=\left( m+n+\sqrt{l^{2}+\left( m-n\right) ^{2}}\right)
\left( m+n-\sqrt{l^{2}+\left( m-n\right) ^{2}}\right)
\end{equation}%
In what follows, we exhibit the dependence on the CS level $\mathrm{k}$ by
labeling the above 2D lattice triplet like $\left( \mathbf{\Lambda }_{%
\mathrm{k}},\mathbf{\Lambda }_{\mathrm{k}\mathcal{C}},\mathbf{\Lambda }_{%
\mathrm{k}}^{\ast }\right) $ with (prime) integer $\mathrm{k}\geq 2$. These
lattices can be realised in terms of scaled\ cubic lattices with spacings $a=%
\sqrt{\mathrm{k}}$ and $a_{\ast }=\frac{1}{\sqrt{\mathrm{k}}}$ as follows
\begin{equation}
\begin{tabular}{lllll}
$\mathbf{\Lambda }_{\mathrm{k}}$ & $\sim $ & $\sqrt{\mathrm{k}}$ $\mathbb{Z}$
& $\mathbb{\times }$ & $\sqrt{\mathrm{k}}$ $\mathbb{Z}$ \\
$\mathbf{\Lambda }_{\mathrm{k}}^{\ast }$ & $\sim $ & $\frac{1}{\sqrt{\mathrm{%
k}}}$ $\mathbb{Z}$ & $\times $ & $\frac{1}{\sqrt{\mathrm{k}}}$ $\mathbb{Z}$
\\
$\mathbf{\Lambda }_{\mathrm{k}\mathcal{C}}$ & $\sim $ & $\sqrt{\mathrm{k}}$ $%
\mathbb{Z}$ & $\mathbb{\times }$ & $\frac{1}{\sqrt{\mathrm{k}}}$ $\mathbb{Z}$%
\end{tabular}
\label{3l}
\end{equation}%
Formally, the intermediate lattice\textrm{\ }$\mathbf{\Lambda }_{\mathrm{k}%
\mathcal{C}}$ may be imagined as the product $\mathbf{R}_{\mathrm{k}}\times
\mathbf{W}_{\mathrm{k}}^{\ast }$ where $\mathbf{R}_{\mathrm{k}}$ and $%
\mathbf{W}_{\mathrm{k}}$ correspond, respectively, to the root and weight
lattices as defined in section 2. The realisation (\ref{31}) satisfies the
following features: $\left( \mathbf{1}\right) $ each site $v_{\mathbf{n}%
}^{i} $ generating the real lattice $\mathbf{\Lambda }_{\mathrm{k}}$ can be
written as $\sqrt{\mathrm{k}}X_{j}^{i}n^{j}$ with invertible%
\begin{equation}
\Lambda _{j}^{i}=\sqrt{\mathrm{k}}\text{ }X_{j}^{i}
\end{equation}%
and induced $g_{\Lambda }=\mathrm{k}X^{T}\eta X$ which must have the typical
form (\ref{gl}). A particular solution for $X$ is given by the identity
matrix $X=I_{2\times 2}$. For this special solution, the induced metric
becomes $g_{\Lambda }=\mathrm{k}\eta $ which corresponds to setting $m=n=0$
and $l=\mathrm{k}$ in (\ref{gl}). Substituting into the product $\mathbf{v}_{%
\mathbf{n}}^{T}\eta \mathbf{v}_{\mathbf{n}},$ we obtain $n^{i}\left(
g_{\Lambda }\right) _{ij}n^{j}=2\mathrm{k}n^{1}n^{2}$ which is indeed an
even integer. $\left( \mathbf{2}\right) $ The sites $\mathbf{w}_{\mathbf{m}}$
in the dual $\mathbf{\Lambda }_{\mathrm{k}}^{\ast }$ have components $w_{%
\mathbf{m}}^{i}=\frac{1}{\sqrt{\mathrm{k}}}Y_{j}^{i}m^{j}$ with
\begin{equation}
\left( \Lambda ^{\ast }\right) _{j}^{i}=\frac{1}{\sqrt{\mathrm{k}}}Y_{j}^{i}
\end{equation}%
Inserting this into the form $\mathbf{v}_{\mathbf{n}}^{T}\eta \mathbf{w}_{%
\mathbf{m}},$ while requiring the duality constraint, we end up with the
condition $X^{T}\eta Y=n$ $\in $ $\mathbb{Z}$; this is easily satisfied by
choosing $Y=(X^{T}\eta )^{-1}=\eta .$ Therefore, for the particular solution
$X=I_{2\times 2}$ and $Y=\eta $, we have the characteristic matrices ($%
\Lambda ,\Lambda ^{\ast }$) and the associated induced metrics ($g_{\Lambda
},g_{\Lambda ^{\ast }}$) explicitly given by%
\begin{equation}
\begin{tabular}{lll}
$\Lambda $ & $=$ & $\sqrt{\mathrm{k}}I_{2\times 2}$ \\
$g_{\Lambda }$ & $=$ & $\mathrm{k}\eta $%
\end{tabular}%
\qquad ,\qquad
\begin{tabular}{lll}
$\Lambda ^{\ast }$ & $=$ & $\frac{1}{\sqrt{\mathrm{k}}}\eta $ \\
$g_{\Lambda ^{\ast }}$ & $=$ & $\frac{1}{\mathrm{k}}\eta $%
\end{tabular}%
\end{equation}%
with determinants $\det \Lambda =\mathrm{k}$, $\det \Lambda ^{\ast }=-\frac{1%
}{\mathrm{k}}$ and $\det g_{\Lambda }=-\mathrm{k}^{2}$ as well as $\det
g_{\Lambda ^{\ast }}=-\frac{1}{\mathrm{k}^{2}}$. For this specific solution,
one notes that $\Lambda ^{-1}=\eta \Lambda ^{\ast }$ highlighting the fact
that $\Lambda ^{\ast }$ is closely related to the inverse $\Lambda ^{-1}.$
Based on these results, we deduce the explicit forms of the intermediate
lattice $\Lambda _{\mathrm{k}\mathcal{C}}$, its dual $\Lambda _{\mathrm{k}%
\mathcal{C}}^{\ast }\sim \Lambda _{\mathrm{k}\mathcal{C}}$ and the
corresponding induced metrics $g_{\Lambda _{\mathrm{k}\mathcal{C}}}$ and $%
g_{\Lambda _{\mathrm{k}\mathcal{C}}^{\ast }}$as follows
\begin{equation}
\Lambda _{\mathrm{k}\mathcal{C}}=\left(
\begin{array}{cc}
\sqrt{\mathrm{k}} & 0 \\
0 & \frac{1}{\sqrt{\mathrm{k}}}%
\end{array}%
\right) ,\qquad \Lambda _{\mathrm{k}\mathcal{C}}^{\ast }=\left(
\begin{array}{cc}
0 & \sqrt{\mathrm{k}} \\
\frac{1}{\sqrt{\mathrm{k}}} & 0%
\end{array}%
\right) ,\qquad g_{\Lambda _{\mathrm{k}\mathcal{C}}}=\left(
\begin{array}{cc}
0 & 1 \\
1 & 0%
\end{array}%
\right)  \label{lc}
\end{equation}%
They verify the property $\Lambda _{\mathrm{k}\mathcal{C}}\eta \Lambda _{%
\mathrm{k}\mathcal{C}}^{\ast }=I_{2\times 2}$ and are unimodular in the
sense\ that $\det \Lambda _{\mathrm{k}\mathcal{C}}=1$, $\det \Lambda _{%
\mathrm{k}\mathcal{C}}^{\ast }=-1$ and $\det g_{\Lambda _{\mathrm{k}\mathcal{%
C}}}=-1.$ This unimodularity captures the self duality property of the
intermediate lattice $\mathbf{\Lambda }_{\mathrm{k}\mathcal{C}}.$

\subsection*{B.2 Higher dimensions}

Using the results obtained for the case \textrm{r=1}, the construction A for
code-CFTs with higher central charges $c_{\text{\textsc{l}/\textsc{r}}}=%
\mathrm{r\geq 2}$ extends straightforwardly by the help of tensor sums and
products. In this generalisation, the codes $\mathcal{C}$ are\textrm{\ }%
embedded in the abelian group $\mathbb{Z}_{\mathrm{k}}^{\mathrm{r}}\times
\mathbb{Z}_{\mathrm{k}}^{\mathrm{r}}$ and are associated with a 2r
dimensional lattice $\mathbf{\Lambda }_{\mathrm{k}\mathcal{C}}^{{\small
\mathrm{r,r}}}$ constrained as follows
\begin{equation}
\underbrace{\mathbf{\Lambda }_{\mathrm{k}}^{\mathrm{1,1}}\oplus ...\oplus
\mathbf{\Lambda }_{\mathrm{k}}^{\mathrm{1,1}}}_{\mathrm{r}\text{ \textrm{%
times}}}\text{ }\subset \text{ }\mathbf{\Lambda }_{\mathrm{k}\mathcal{C}}^{%
{\small \mathrm{r,r}}}\text{ }\subset \text{ }\underbrace{\mathbf{\Lambda }_{%
\mathrm{k}}^{\ast \mathrm{1,1}}\oplus ...\oplus \mathbf{\Lambda }_{\mathrm{k}%
}^{\ast \mathrm{1,1}}}_{\mathrm{r}\text{ \textrm{times}}}\text{ }\subset
\text{ }\mathbb{R}^{\mathrm{r},\mathrm{r}}  \label{csa}
\end{equation}%
When the codes $\mathcal{C}$ satisfy additional constraints, namely evenness
and self duality conditions, the associated lattices $\mathbf{\Lambda }_{%
\mathrm{k}\mathcal{C}}^{{\small \mathrm{r,r}}}$ are in turn even and self
dual, thus defining a Narain theory with central charge $\mathrm{r}$. For
the particular solution introduced earlier, the relevant matrices take the
form
\begin{equation}
\Lambda _{\mathrm{k}\mathcal{C}}^{{\small \mathrm{r,r}}}=\left(
\begin{array}{cc}
\sqrt{\mathrm{k}}I_{\mathrm{r}} & 0 \\
0 & \frac{1}{\sqrt{\mathrm{k}}}I_{\mathrm{r}}%
\end{array}%
\right) ,\quad \Lambda _{\mathrm{k}\mathcal{C}}^{\ast {\small \mathrm{r,r}}%
}=\left(
\begin{array}{cc}
0 & \sqrt{\mathrm{k}}I_{\mathrm{r}} \\
\frac{1}{\sqrt{\mathrm{k}}}I_{\mathrm{r}} & 0%
\end{array}%
\right) ,\quad g_{\Lambda _{\mathrm{k}\mathcal{C}}}^{{\small \mathrm{r,r}}%
}=\left(
\begin{array}{cc}
0 & I_{\mathrm{r}} \\
I_{\mathrm{r}} & 0%
\end{array}%
\right)
\end{equation}%
with $I_{\mathrm{r}}$ referring to the r$\times $r identity matrix. These
matrices satisfy $\Lambda _{\mathrm{k}\mathcal{C}}^{{\small \mathrm{r,r}}%
}\eta \Lambda _{\mathrm{k}\mathcal{C}}^{\ast {\small \mathrm{r,r}}}=I_{2%
{\small \mathrm{r,2r}}},$ $\det \Lambda _{\mathrm{k}\mathcal{C}}^{{\small
\mathrm{r,r}}}=1$ as well as $\det \Lambda _{\mathrm{k}\mathcal{C}}^{\ast
{\small \mathrm{r,r}}}=\left( -\right) ^{r}$.

\section{Lattices $\mathbf{\Lambda }^{\mathbf{su}_{3}},\mathbf{\Lambda }_{%
\mathcal{C}}^{\mathbf{su}_{3}},\mathbf{\Lambda }^{\ast \mathbf{su}_{3}}$ for
k\TEXTsymbol{<}3 and k\TEXTsymbol{>}3}

\label{sec:C} \setcounter{equation}{0} In this appendix, we complete the
study of section 3 by giving realisations of the triplet ($\mathbf{\Lambda }%
^{\mathbf{su}_{3}},\mathbf{\Lambda }_{\mathcal{C}}^{\mathbf{su}_{3}},\mathbf{%
\Lambda }^{\ast \mathbf{su}_{3}}$) for values of the CS level k distinct
than k=3. Particularly, we construct models with k\TEXTsymbol{>}3 and then
discuss cases having k\TEXTsymbol{<}3.

\subsection*{C.1 Lattice triplet for\emph{\ }k\TEXTsymbol{>}3}

The extension of the root \textsc{R}$_{\mathrm{3}}^{\mathbf{su}_{3}}$ and
the weight \textsc{W}$_{\mathrm{3}}^{\mathbf{su}_{3}}$ towards \textsc{R}$_{%
\mathrm{k}}^{\mathbf{su}_{3}}$ and \textsc{W}$_{\mathrm{k}}^{\mathbf{su}%
_{3}} $ are constructed using dilated roots as $\mathbf{\tilde{\alpha}}_{i}=%
\sqrt{\frac{\mathrm{k}}{3}}\mathbf{\alpha }_{i}$ and compressed weight
vectors $\mathbf{\tilde{\lambda}}_{i}=\sqrt{\frac{3}{\mathrm{k}}}\mathbf{%
\lambda }_{i},$ leading to
\begin{equation}
\begin{tabular}{lll}
$\text{\textsc{R}}_{\mathrm{k}}^{\mathbf{su}_{3}}$ & $=$ & $\mathbb{Z}%
\mathbf{\tilde{\alpha}}_{1}\oplus \mathbb{Z}\mathbf{\tilde{\alpha}}_{2}$ \\
$\text{\textsc{W}}_{\mathrm{k}}^{\mathbf{su}_{3}}$ & $=$ & $\mathbb{Z}%
\mathbf{\tilde{\lambda}}_{1}\oplus \mathbb{Z}\mathbf{\tilde{\lambda}}_{2}$%
\end{tabular}%
\qquad ,\qquad
\begin{tabular}{lll}
$L_{ij}$ & $:=$ & $\mathbf{\tilde{\alpha}}_{i}.\mathbf{\tilde{\alpha}}_{j}$
\\
$\tilde{L}^{ij}$ & $:=$ & $\mathbf{\tilde{\lambda}}^{i}.\mathbf{\tilde{%
\lambda}}^{j}$%
\end{tabular}%
\end{equation}%
with intersections $L_{ij}=\frac{\mathrm{k}}{3}K_{ij}$ and $\tilde{L}^{ij}=%
\frac{3}{\mathrm{k}}\tilde{K}^{ij}$ as well as $\mathbf{\tilde{\alpha}}_{i}.%
\mathbf{\tilde{\lambda}}^{j}=\delta _{i}^{j}$. These scaled vectors can be
solved as $\mathbf{\tilde{\lambda}}_{1}=\frac{1}{\mathrm{k}}\left( 2\mathbf{%
\tilde{\alpha}}_{1}+\mathbf{\tilde{\alpha}}_{2}\right) $ and $\mathbf{\tilde{%
\lambda}}_{2}=\frac{1}{\mathrm{k}}\left( \mathbf{\tilde{\alpha}}_{1}+2%
\mathbf{\tilde{\alpha}}_{2}\right) $ as well as $\mathbf{\tilde{\lambda}}%
_{1}+\mathbf{\tilde{\lambda}}_{2}=\frac{3}{\mathrm{k}}\left( \mathbf{\tilde{%
\alpha}}_{1}+\mathbf{\tilde{\alpha}}_{2}\right) .$ This realisation yields
the two matrices $A_{\mathrm{k}}$ and $B_{\mathrm{k}}$ that read as follows
\begin{equation}
\left( A_{\mathrm{k}}\right) _{j}^{i}=\sqrt{\frac{\mathrm{k}}{3}}\left(
\begin{array}{cc}
\sqrt{2} & -\frac{\sqrt{2}}{2} \\
0 & \frac{\sqrt{6}}{2}%
\end{array}%
\right) ,\qquad \left( B_{\mathrm{k}}\right) _{i}^{j}=\sqrt{\frac{3}{\mathrm{%
k}}}\left(
\begin{array}{cc}
\frac{\sqrt{2}}{2} & \frac{\sqrt{6}}{6} \\
0 & \frac{\sqrt{6}}{3}%
\end{array}%
\right)  \label{AKB}
\end{equation}%
with $\det A_{\mathrm{k}}=\frac{\mathrm{k}}{\sqrt{3}}$ and $\det B_{\mathrm{k%
}}=\frac{\sqrt{3}}{\mathrm{k}}.$ We then have%
\begin{equation}
\Lambda _{\mathrm{k}}^{\mathbf{su}_{3}}=\left(
\begin{array}{cc}
A_{\mathrm{k}} & 0 \\
0 & A_{\mathrm{k}}%
\end{array}%
\right) ,\qquad \Lambda _{\mathrm{k}}^{\mathbf{su}_{3}\ast }=\left(
\begin{array}{cc}
B_{\mathrm{k}} & 0 \\
0 & B_{\mathrm{k}}%
\end{array}%
\right) ,\qquad \Lambda _{\mathrm{k}\mathcal{C}}^{\mathbf{su}_{3}}=\left(
\begin{array}{cc}
A_{\mathrm{k}} & 0 \\
0 & B_{\mathrm{k}}%
\end{array}%
\right)
\end{equation}%
Site vectors \textbf{x}$_{\mathbf{n}}$ in \textsc{R}$_{\mathrm{k}}^{\mathbf{%
su}_{3}}$ are given by the expansion $\sum n^{i}\mathbf{\tilde{\alpha}}_{i}$
while the grid vectors \textbf{k}$_{\mathbf{m}}$ in \textsc{W}$_{\mathrm{k}%
}^{\mathbf{su}_{3}}$ develop like $\sum m_{i}\mathbf{\tilde{\lambda}}^{i}.$
By substituting the $\mathbf{\tilde{\lambda}}_{i}$'s in terms of the $%
\mathbf{\tilde{\alpha}}_{i}$'s, we get%
\begin{equation}
\mathbf{k}_{\mathbf{m}}=\frac{1}{\mathrm{k}}\left( 2m_{1}+m_{2}\right)
\mathbf{\tilde{\alpha}}_{1}+\frac{1}{\mathrm{k}}\left( m_{1}+2m_{2}\right)
\mathbf{\tilde{\alpha}}_{2}  \label{km}
\end{equation}%
Defining $2m_{1}+m_{2}=\mathrm{k}p+\xi $ and $m_{1}+2m_{2}=\mathrm{k}q+\zeta
$ with p,q arbitrary integers and $\xi ,\zeta =0,...,\mathrm{k-}1,$ $\func{%
mod}\mathrm{k}$ gives the vector $\mathbf{k}_{\mathbf{m}}$ as follows
\begin{equation}
\begin{tabular}{lll}
$\mathbf{k}_{\mathbf{m}}$ & $=$ & $\mathbf{k}_{\mathbf{m}}^{0}+\frac{1}{3}%
\left( 2\xi -\zeta \right) \mathbf{\tilde{\lambda}}_{1}+\frac{1}{3}\left(
2\zeta -\xi \right) \mathbf{\tilde{\lambda}}_{2}$ \\
$\mathbf{k}_{\mathbf{m}}^{0}$ & $=$ & $p\mathbf{\tilde{\alpha}}_{1}+q\mathbf{%
\tilde{\alpha}}_{2}$%
\end{tabular}%
\end{equation}%
By choosing $\zeta =2\xi $ and $\xi =\varepsilon $ with $\varepsilon
=0,...,k-1,$ $\func{mod}k$, we get%
\begin{equation}
\mathbf{k}_{\mathbf{m}}\left( \varepsilon \right) =\mathbf{k}_{\mathbf{m}%
}^{0}+\varepsilon \mathbf{\tilde{\lambda}}_{2}\qquad ,\qquad \varepsilon
=0,...,k-1
\end{equation}%
This shows that \textsc{W}$_{\mathrm{k}}^{\mathbf{su}_{3}}$ consists of k
superposed sheets (k equivalent classes) given by the values of $\varepsilon
.$ These sheets can be presented in a compact form by using the spectral
parameter $\varepsilon $ as follows
\begin{equation}
\begin{tabular}{|c|c||c|c||c|}
\hline
$2m_{1}+m_{2}$ & $m_{1}+2m_{2}$ & sites $\mathbf{k}_{\mathbf{m}}^{0}$ &
sites $\mathbf{k}_{\mathbf{m}}$ & weight lattice \textsc{W}$_{\mathrm{k}}^{%
\mathbf{su}_{3}}$ \\ \hline\hline
$\mathrm{k}p+\varepsilon $ & $\mathrm{k}q+2\varepsilon $ & $p\mathbf{\tilde{%
\alpha}}_{1}+q\mathbf{\tilde{\alpha}}_{2}$ & $\mathbf{k}_{\mathbf{m}%
}^{0}+\varepsilon \mathbf{\tilde{\lambda}}_{2}$ & \textsc{R}$_{\mathrm{k}}^{%
\mathbf{su}_{3}}\left[ \varepsilon \right] =$\textsc{R}$_{\mathrm{k}}^{%
\mathbf{su}_{3}}+{\small \varepsilon }\mathbf{\tilde{\lambda}}_{2}$ \\ \hline
\end{tabular}%
\end{equation}%
Similarly to the previous case, \textsc{W}$_{\mathrm{k}}^{\mathbf{su}_{3}}$
can be imagined as the superposition of k sheets of 2D sublattices like
\begin{equation}
\text{\textsc{W}}_{\mathrm{k}}^{\mathbf{su}_{3}}=\text{\textsc{R}}_{\mathrm{k%
}}^{\mathbf{su}_{3}}\cup \left\{ \text{\textsc{R}}_{\mathrm{k}}^{\mathbf{su}%
_{3}}+\mathbf{\lambda }_{2}\right\} \cup ...\cup \left\{ \text{\textsc{R}}_{%
\mathrm{k}}^{\mathbf{su}_{3}}+(\mathrm{k}-1)\mathbf{\lambda }_{2}\right\}
\label{u2}
\end{equation}%
Where each sheet is isomorphic to\ the hexagonal \textsc{R}$_{\mathrm{k}}^{%
\mathbf{su}_{3}}$ generated by $\mathbf{\tilde{\alpha}}_{i}$ with $(\widehat{%
\mathbf{\tilde{\alpha}}_{1},\mathbf{\tilde{\alpha}}_{2}})=2\pi /3.$ For the
CS levels k\TEXTsymbol{>}3, the discriminant \textsc{W}$_{\mathrm{k}}^{%
\mathbf{su}_{3}}/$\textsc{R}$_{\mathrm{k}}^{\mathbf{su}_{3}}$ is isomorphic
to the abelian group $\mathbb{Z}_{\mathrm{k}}$ generated by the cyclic
permutation (1,2,...,k) modulo k.

\subsection*{C.2 Lattice triplet for\emph{\ }k\TEXTsymbol{<}3}

An intriguing case arises when considering CS levels k\TEXTsymbol{<}3,
particularly for $\mathrm{k}=1$ and $\mathrm{k}=2.$ Recall that for generic
k,\textrm{\ }the scaled root and weight lattice generators $\mathbf{\tilde{%
\alpha}}_{i}$ and $\mathbf{\tilde{\lambda}}^{i}$ obey the relations: $%
\mathbf{\tilde{\alpha}}_{i}.\mathbf{\tilde{\alpha}}_{j}=\frac{\mathrm{k}}{3}%
K_{ij}$ and $\mathbf{\tilde{\lambda}}^{i}.\mathbf{\tilde{\lambda}}^{j}=\frac{%
3}{\mathrm{k}}\tilde{K}^{ij}$ as well as $\mathbf{\tilde{\alpha}}_{i}..%
\mathbf{\tilde{\lambda}}^{j}=\delta _{i}^{j}$. These imply the squared
lengths $\mathbf{\tilde{\alpha}}_{i}^{2}=\frac{2\mathrm{k}}{3}$ and $\mathbf{%
\tilde{\lambda}}_{i}^{2}=\frac{2}{\mathrm{k}}$ that lead to the ratios $%
\mathbf{\tilde{\alpha}}_{i}^{2}/\mathbf{\tilde{\lambda}}_{i}^{2}$ that
exceed one only for $\mathrm{k}\geq \sqrt{3}$ as illustrated in the table
below:%
\begin{equation}
\begin{tabular}{|c|c|c|c|}
\hline
$\mathrm{k}$ & $\quad \mathbf{\tilde{\alpha}}_{i}^{2}\quad $ & $\quad
\mathbf{\tilde{\lambda}}_{i}^{2}\quad $ & $\quad \mathbf{\tilde{\alpha}}%
_{i}^{2}/\mathbf{\tilde{\lambda}}_{i}^{2}\quad $ \\ \hline
$1$ & $\frac{2}{3}$ & $2$ & $\frac{1}{3}<1$ \\ \hline
$\frac{3}{2}$ & $1$ & $\frac{4}{3}$ & $\frac{3}{4}<1$ \\ \hline
$2$ & $\frac{4}{3}$ & $1$ & $\frac{4}{3}>1$ \\ \hline
$3$ & $2$ & $\frac{2}{3}$ & $\frac{16}{9}>1$ \\ \hline
\end{tabular}%
\end{equation}

\subsubsection*{a) Cases k=1 and k=$\protect\sqrt{3}$}

Using the relations (\ref{u1}) and (\ref{u2}), the matrices $A_{\mathrm{k}}$
and $B_{\mathrm{k}}$ for \textrm{k=1} and \textrm{k=}$\sqrt{\mathrm{3}}$
read like\footnote{%
\ Matrices of type $A_{\sqrt{\mathrm{3}}}$ and $B_{\sqrt{\mathrm{3}}}$
appeared in \textrm{\cite{1jB}; see eqs(4.5, 4.19, 4.26) there}.
\par
{}}%
\begin{equation}
\begin{tabular}{lllllll}
$A_{\mathrm{1}}$ & $=$ & $\sqrt{\frac{2}{3}}\left(
\begin{array}{cc}
1 & -\frac{1}{2} \\
0 & \frac{\sqrt{3}}{2}%
\end{array}%
\right) $ & $,\quad $ & $A_{\sqrt{\mathrm{3}}}$ & $=$ & $\frac{\sqrt{2}}{%
3^{1/4}}\left(
\begin{array}{cc}
1 & -\frac{1}{2} \\
0 & \frac{\sqrt{3}}{2}%
\end{array}%
\right) $ \\
$B_{\mathrm{1}}$ & $=$ & $\sqrt{\frac{3}{2}}\left(
\begin{array}{cc}
1 & \frac{\sqrt{3}}{3} \\
0 & \frac{2\sqrt{3}}{3}%
\end{array}%
\right) $ & $,\quad $ & $B_{\sqrt{\mathrm{3}}}$ & $=$ & $\frac{\sqrt{2}}{%
3^{1/4}}\left(
\begin{array}{cc}
\frac{1}{2} & \frac{\sqrt{3}}{6} \\
0 & \frac{\sqrt{3}}{3}%
\end{array}%
\right) $%
\end{tabular}%
\end{equation}%
with $\det A_{\mathrm{1}}=\frac{1}{\sqrt{3}}<1$\ and $\det $ $B_{\mathrm{1}}=%
\sqrt{3}>1$ while $\det A_{\sqrt{\mathrm{3}}}=\det B_{\sqrt{\mathrm{3}}}=1.$
The inequality $\det A_{\mathrm{1}}<\det $ $B_{\mathrm{1}}$ contradicts the
desired condition $\det A_{\mathrm{k}}\geq \det $ $B_{\mathrm{k}}$ expected
for the inclusion \textsc{R}$_{\mathrm{k}}^{\mathbf{su}_{3}}\subseteq $%
\textsc{W}$_{\mathrm{k}}^{\mathbf{su}_{3}}.$ This makes the case\textrm{\
k=1 }anomalous and worthy of closer scrutiny. To reconcile this, we
interpret k=1 as the integer part of $\sqrt{3}$ that is $\mathrm{k}=[\sqrt{3}%
]=1$ which allows us to retain the property $\det A_{\mathrm{k}}\geq \det $ $%
B_{\mathrm{k}}$ within the regime $\mathrm{k}\geq \sqrt{3}.$
\begin{figure}[tbph]
\begin{center}
\includegraphics[width=8cm]{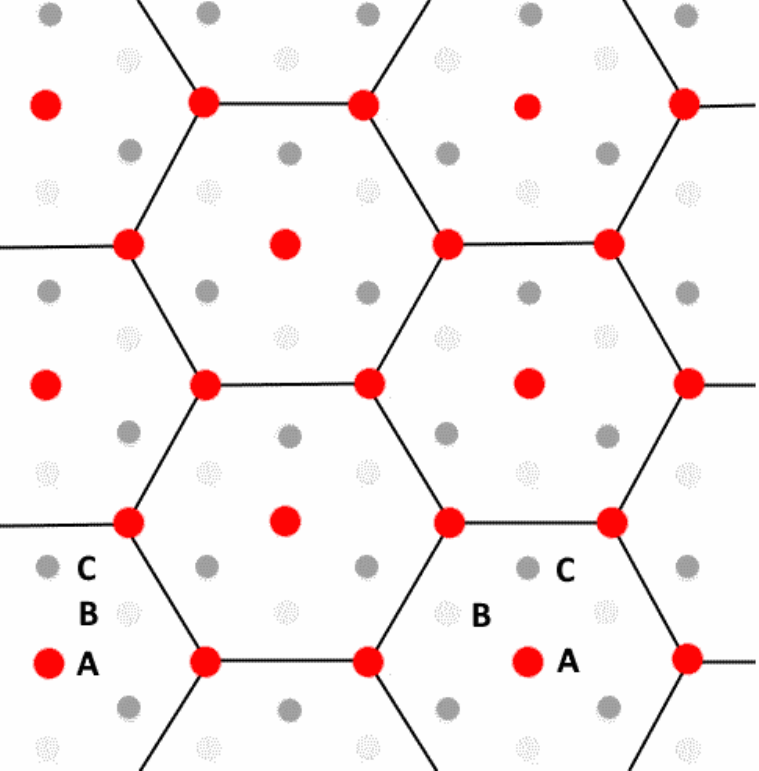}
\end{center}
\par
\vspace{-0.5cm}
\caption{the \textsc{W}$_{\mathrm{k}}^{\mathbf{su}_{3}}$\ for k=$\protect%
\sqrt{3}$ is isomorphic to the sheet $\left. \text{\textsc{R}}_{\mathrm{3}}^{%
\mathbf{su}_{3}}\left( \protect\varepsilon \right) \right\vert _{\protect%
\varepsilon =0}$ with red sites of Figure \textbf{\protect\ref{33}}.
Sublattices with light and bold grey refer to $\left. \text{\textsc{R}}_{%
\mathrm{3}}^{\mathbf{su}_{3}}\left( \protect\varepsilon \right) \right\vert
_{\protect\varepsilon =1,2}$.}
\label{34}
\end{figure}
Moreover, for $\mathrm{k}>[\sqrt{3}]$ the set \textsc{W}$_{\mathrm{k}}^{%
\mathbf{su}_{3}}$ splits into a union of k sheets. Specifically, when k$=%
\sqrt{3}$ the lattice \textsc{W}$,$ depicted in Figure \textbf{\ref{34}}, is
generated by site vectors $\mathbf{k}_{\mathbf{m}}$ labeled as $\left(
2m_{1}+m_{2}\right) \mathbf{\tilde{\alpha}}_{1}+\left( m_{1}+2m_{2}\right)
\mathbf{\tilde{\alpha}}_{2}$ with $\mathbf{m}=\left( m_{1},m_{2}\right) $
and $\mathbf{\tilde{\alpha}}_{i}=\frac{1}{3^{1/4}}\mathbf{\alpha }_{i}$
where the two $\mathbf{\alpha }_{i}$'s are the simple roots of SL$_{3}$. The
graphic of \textsc{W}$_{\sqrt{3}}^{\mathbf{su}_{3}}$ in the Figure \textbf{%
\ref{34}} is isomorphic to the red sublattice \textrm{(A-sites) }in the
three sheet lattice \textsc{W}$_{\mathrm{3}}^{\mathbf{su}_{3}}$ shown in
Figure \textbf{\ref{33}}.

\subsubsection*{b) Case k=2}

\begin{figure}[tbph]
\begin{center}
\includegraphics[width=6cm]{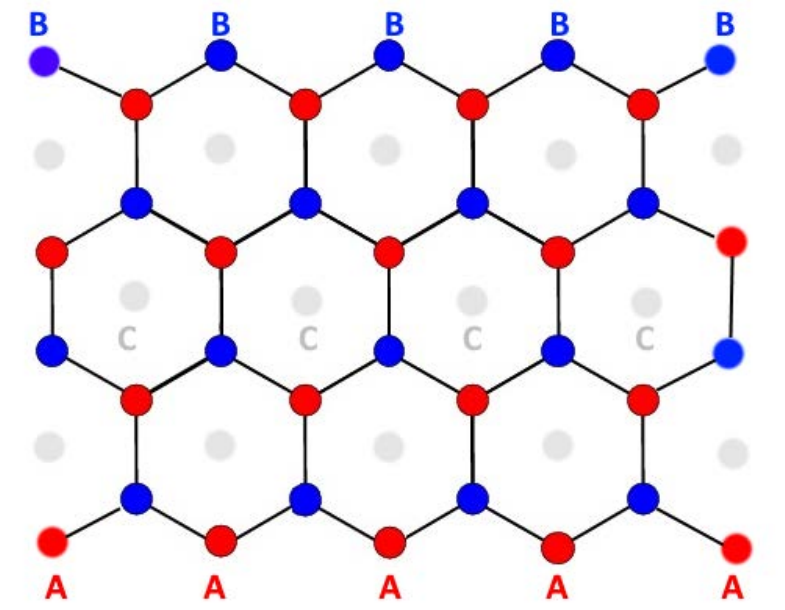}
\end{center}
\par
\vspace{-0.5cm}
\caption{The two sheets \textsc{R}$_{\mathrm{2}}^{sl_{3}}\left[ 0\right]
\cup $\textsc{R}$_{\mathrm{2}}^{sl_{3}}\left[ 1\right] $ of \textsc{W}$_{%
\mathrm{3}}^{sl_{3}}$. The sites A in \textsc{R}$_{\mathrm{2}}^{sl_{3}}\left[
0\right] $ are in red color while the site B in \textsc{R}$_{\mathrm{2}%
}^{sl_{3}}\left[ 1\right] $ are in blue. The sites C of the omitted sheet
are in grey.}
\label{35}
\end{figure}

For the CS level k=2, which lies above $\sqrt{3}$, the following hold: $%
\left( \mathbf{i}\right) $ the inclusion \textsc{R}$_{\mathrm{2}}^{\mathbf{su%
}_{3}}\subset $\textsc{W}$_{\mathrm{2}}^{\mathbf{su}_{3}}$ with \textsc{R}$_{%
\mathrm{2}}^{\mathbf{su}_{3}}$ being the hexagonal lattice generated by the
dilated $\mathbf{\tilde{\alpha}}_{i}=\sqrt{\frac{\mathrm{2}}{3}}\mathbf{%
\alpha }_{i}$. $\left( \mathbf{ii}\right) $ the weight lattice \textsc{W}$_{%
\mathrm{2}}^{\mathbf{su}_{3}}$ is generated by the compressed $\mathbf{%
\tilde{\lambda}}_{i}=\sqrt{\frac{3}{\mathrm{2}}}\mathbf{\lambda }_{i}$ and
comprises two superposed sheets \textsc{R}$_{\mathrm{2}}^{\mathbf{su}%
_{3}}\left( \varepsilon \right) $ with $\varepsilon =0,1.$ Each sheet
\textsc{R}$_{\mathrm{2}}^{\mathbf{su}_{3}}\left( \varepsilon \right) $ is
isomorphic to the hexagonal root lattice \textsc{R}$_{\mathrm{2}}^{\mathbf{su%
}_{3}}$ which can be directly inferred from eq(\ref{u2}). Analogously, for
k=3, the \textsc{W}$_{\mathrm{3}}^{\mathbf{su}_{3}}$ consists of\textrm{\ }%
three superposed sheets as depicted by the Figure \textbf{\ref{33}}. Notice
also that the coordinates $\mathbf{k}_{\mathbf{m}}^{0}$ of \textsc{R}$_{%
\mathrm{2}}^{\mathbf{su}_{3}}|_{\varepsilon =0}$ and $\mathbf{k}_{\mathbf{m}%
} $ of \textsc{R}$_{\mathrm{2}}^{\mathbf{su}_{3}}|_{\varepsilon =1}$ in
\textsc{W}$_{\mathrm{2}}^{\mathbf{su}_{3}}$ are given by%
\begin{equation}
\begin{tabular}{lll}
$\mathbf{k}_{\mathbf{m}}^{0}$ & $=$ & $\left( 2m_{1}+m_{2}\right) \mathbf{%
\tilde{\alpha}}_{1}+\left( m_{1}+2m_{2}\right) \mathbf{\tilde{\alpha}}_{2}$
\\
$\mathbf{k}_{\mathbf{m}}$ & $=$ & $\mathbf{k}_{\mathbf{m}}^{0}+\mathbf{%
\tilde{\lambda}}_{2}$%
\end{tabular}%
\end{equation}%
Here,\textrm{\ A-sites }in \textsc{R}$_{\mathrm{2}}^{\mathbf{su}%
_{3}}|_{\varepsilon =0}$ are colored red, while\textrm{\ B-sites }in \textsc{%
R}$_{\mathrm{2}}^{\mathbf{su}_{3}}|_{\varepsilon =1}$ are blue, giving rise
to the two layer honeycomb structure depicted by the Figure \textbf{\ref{35}}%
.
The corresponding matrices A$_{\mathrm{2}}$ and B$_{\mathrm{2}}$ are given
by
\begin{equation}
A_{\mathrm{2}}=\frac{2}{\sqrt{3}}\left(
\begin{array}{cc}
1 & -\frac{1}{2} \\
0 & \frac{\sqrt{3}}{2}%
\end{array}%
\right) ,\qquad B_{\mathrm{2}}=\sqrt{3}\left(
\begin{array}{cc}
\frac{1}{2} & \frac{\sqrt{3}}{6} \\
0 & \frac{\sqrt{3}}{3}%
\end{array}%
\right)  \label{2AB}
\end{equation}%
and finally the triplet ($\Lambda _{\mathrm{2}}^{\mathbf{su}_{3}},\Lambda _{%
\mathrm{2}}^{\mathbf{su}_{3}\ast },\Lambda _{\mathrm{2}\mathcal{C}}^{\mathbf{%
su}_{3}}$) read as follows%
\begin{equation}
\Lambda _{\mathrm{2}}^{\mathbf{su}_{3}}=\left(
\begin{array}{cc}
A_{\mathrm{2}} & 0 \\
0 & A_{\mathrm{2}}%
\end{array}%
\right) ,\qquad \Lambda _{\mathrm{2}}^{\mathbf{su}_{3}\ast }=\left(
\begin{array}{cc}
B_{\mathrm{2}} & 0 \\
0 & B_{\mathrm{2}}%
\end{array}%
\right) ,\qquad \Lambda _{\mathrm{2}\mathcal{C}}^{\mathbf{su}_{3}}=\left(
\begin{array}{cc}
A_{\mathrm{2}} & 0 \\
0 & B_{\mathrm{2}}%
\end{array}%
\right)  \label{c2}
\end{equation}

To complete the present study, we think it interesting to add a discussion
regarding the CS level \textrm{k=2} of the holographic dual of Narain CFT
built out of the SU(3) lattices \textsc{R}$_{\mathrm{k}}^{\mathbf{su}_{3}}$
and \textsc{W}$_{\mathrm{k}}^{\mathbf{su}_{3}}$. These lattices are given by%
\emph{\ the Figures} \textbf{\ref{34}}-\textbf{\ref{35}}. The characteristic
matrices of the three lattices $\mathbf{\Lambda }_{\mathrm{2}}^{\mathbf{su}%
_{3}},\mathbf{\Lambda }_{\mathrm{2}\mathcal{C}}^{\mathbf{su}_{3}}$ and $%
\mathbf{\Lambda }_{\mathrm{2}}^{\ast \mathbf{su}_{3}}$ are given by eqs(\ref%
{2AB},\ref{c2}). Holographically, the dual Narain CFT based on $\mathbf{%
\Lambda }_{\mathrm{2}\mathcal{C}}^{\mathbf{su}_{3}}$ is described by the
AB-action (\ref{act}) that reads in our notations like
\begin{equation}
\mathcal{S}_{\text{\textsc{cs}}}+\mathcal{S}_{\text{\textsc{bnd}}}=\frac{i}{%
2\pi }\dint\nolimits_{\mathcal{M}}Tr\left( \boldsymbol{A}\wedge d\boldsymbol{%
B}+\boldsymbol{B}\wedge d\boldsymbol{A}\right) +\dint\nolimits_{\partial
\mathcal{M}}Tr\left( \boldsymbol{A}_{z}\tilde{K}\boldsymbol{A}_{\bar{z}}+%
\boldsymbol{B}_{z}K\boldsymbol{B}_{\bar{z}}\right)
\end{equation}%
In this expression, the potentials $\boldsymbol{A}$ and $\boldsymbol{B}$\
are given by linear combinations of the old potential pairs $(\mathcal{A}%
_{1},\mathcal{A}_{2})$ and $(\mathcal{B}_{1},\mathcal{B}_{2})$ as follows
\begin{equation}
\begin{tabular}{lll}
$\boldsymbol{A}$ & $=$ & $\mathcal{A}^{1}\mathbf{\tilde{\alpha}}_{1}+%
\mathcal{A}^{2}\mathbf{\tilde{\alpha}}_{2}$ \\
$\boldsymbol{B}$ & $=$ & $\mathcal{B}_{1}\mathbf{\tilde{\lambda}}^{1}+%
\mathcal{B}_{2}\mathbf{\tilde{\lambda}}^{2}$%
\end{tabular}%
,\qquad
\begin{tabular}{lll}
$\mathbf{\tilde{\alpha}}_{i}$ & $=$ & $\sqrt{\frac{\mathrm{k}}{3}}\mathbf{%
\alpha }_{i}$ \\
$\mathbf{\tilde{\lambda}^{i}}$ & $\mathbf{=}$ & $\sqrt{\frac{3}{\mathrm{k}}}%
\mathbf{\lambda }^{i}$%
\end{tabular}%
\end{equation}%
In these relations, we have $\mathcal{A}^{i}=\mathbf{\tilde{\lambda}}^{i}.%
\boldsymbol{A}$ and $\mathcal{B}_{i}=\mathbf{\tilde{\alpha}}_{i}.\boldsymbol{%
B}$; in terms of the matrices (\~{A}$_{i}^{j}$) and (\~{B}$_{j}^{i}$) given
by eqs(\ref{vn},\ref{wn}) with \textrm{k=2,} we also have $\mathcal{A}^{i}=$(%
\~{B}$_{j}^{i}$)$\boldsymbol{A}^{j}$ and $\mathcal{B}_{i}=$(\~{A}$_{i}^{j}$)$%
\boldsymbol{B}_{j}.$ For this holographic dual, the dimension of the Hilbert
space is $\dim \mathcal{H}=$\textrm{k}$^{2}$; it is generated by the wave
functions $\Psi _{\mathbf{a,b}}\left( \xi ,\zeta \right) $ labeled by $%
\mathbf{a}=a\mathbf{\tilde{\alpha}}_{2}$ and $\mathbf{b}=b\mathbf{\tilde{%
\lambda}}_{2}$ with $a,b=0,1$ $\func{mod}$ 2$.$ These multi-holomorphic
functions are defined in terms of the path integral as
\begin{equation}
\Psi \left( \xi ,\bar{\zeta}\right) =\dint\nolimits_{\left( \xi ,\bar{\zeta}%
\right) }\left[ DADB\right] e^{-\mathcal{S}_{\text{\textsc{cs}}}+\mathcal{S}%
_{\text{\textsc{bnd}}}}
\end{equation}%
where $\left( \xi ,\bar{\zeta}\right) $ encode the boundary conditions of
the CS potentials $\mathcal{A}^{i}$ and $\mathcal{B}_{i}$ on the boundary $%
\partial \mathcal{M}=\mathbb{T}^{2}$ \textrm{\cite{1B}.} For solid torii,
their explicit expressions coincide with by the genus-one partition
functions of the $\mathbb{Z}_{2}\times \mathbb{Z}_{2}$ Narain CFTs with $%
c_{L/R}=2.$ They read like%
\begin{equation}
\Psi _{\mathbf{a,b}}\left( \xi ,\zeta \right) =\frac{1}{\left\vert \eta
\left( \tau \right) \right\vert ^{2}}\dsum\limits_{\mathbf{n,m}}e^{i\pi \tau
P_{L}^{2}-i\pi \bar{\tau}P_{R}^{2}+2i\pi \left( P_{L}\xi -P_{R}\bar{\zeta}%
\right) +\frac{\pi }{2\tau _{2}}\left( \xi ^{2}+\zeta ^{2}\right) }
\label{pab}
\end{equation}%
where $\eta \left( \tau \right) $\ is the usual Dedekind eta function and
where ($\mathrm{k=2}$) the left/right momenta $p_{L/R}$ are given by [with P$%
_{1}^{i}=(P_{L}^{i}+P_{L}^{i})/2$ and P$_{2}^{i}=(P_{L}^{i}-P_{L}^{i})/2$],
\begin{equation}
\begin{tabular}{lll}
$P_{L}^{i}$ & $=$ & $\frac{1}{\sqrt{2\mathrm{k}}}\left[ \tilde{A}%
_{j}^{i}\left( a^{j}+\mathrm{k}n^{j}\right) +\left( b_{j}+\mathrm{k}%
m_{j}\right) \tilde{B}_{i}^{j}\right] $ \\
$P_{R}^{i}$ & $=$ & $\frac{1}{\sqrt{2\mathrm{k}}}\left[ \tilde{A}%
_{j}^{i}\left( a^{j}+\mathrm{k}n^{j}\right) -\left( b_{j}+\mathrm{k}%
m_{j}\right) \tilde{B}_{i}^{j}\right] $%
\end{tabular}
\label{pa}
\end{equation}%
with $\tilde{A}_{j}^{i}$ and $\tilde{B}_{i}^{j}$ as in (\ref{2AB}). The
integers $n^{j}$ describe the KK modes and the $m_{j}$'s are windings. In
terms of the simple $\mathbf{\alpha }_{i}$ and the fundamental $\mathbf{%
\lambda }^{i}$, we have $\boldsymbol{P}_{L/R}=\frac{1}{2}[\left(
a^{i}+2n^{i}\right) \mathbf{\alpha }_{i}\pm \left( b_{i}+2m_{i}\right)
\mathbf{\lambda }^{i}].$ For the ground state for which $n^{i}=m_{i}=0,$ the
associated momenta are given by $\boldsymbol{\mathring{p}}_{L/R}=\frac{1}{2}%
(a\mathbf{\alpha }_{2}\pm b\mathbf{\lambda }_{2});$ then we have $%
\boldsymbol{P}_{L/R}=\boldsymbol{\mathring{p}}_{L/R}+\boldsymbol{p}_{L/R}$
with excitations $\boldsymbol{p}_{L/R}=n^{i}\mathbf{\alpha }_{i}\pm m_{i}%
\mathbf{\lambda }^{i}$. Moreover, the complex $\xi ^{i}$ and $\zeta ^{i},$
interpreted in terms of fugacities in conformal field theory, can be
respectively parameterised like
\begin{equation}
\begin{tabular}{lll}
$\xi ^{i}$ & $=$ & $\sqrt{\frac{\mathrm{k}}{2}}(\tilde{A}_{j}^{i}u^{j}+v_{j}%
\tilde{B}_{i}^{j})$ \\
$\zeta ^{i}$ & $=$ & $\sqrt{\frac{\mathrm{k}}{2}}(\tilde{A}_{j}^{i}\bar{u}%
^{j}-\bar{v}_{j}\tilde{B}_{i}^{j})$%
\end{tabular}%
\end{equation}%
For k=3, the manifestation of the SU(3) symmetry is given by $K=A^{T}A$ and $%
\tilde{K}=BB^{T}$ where K is the Cartan matrix and $\tilde{K}$ its inverse ($%
K\tilde{K}=I$). Further developments in this direction and in link with
lattice topological matter with Dirac points will be reported in a future
occasion.%
\begin{equation*}
\end{equation*}

\end{document}